\documentclass[final]{siamart1116}
\usepackage{amsfonts}

\usepackage{float}  
\usepackage{graphicx,epstopdf}
\usepackage{subfig}

\usepackage{dsfont}

\newsiamthm{claim}{Claim}
\newsiamremark{remark}{Remark}
\newsiamremark{example}{Example}
\newsiamremark{hypothesis}{Hypothesis}
\newsiamremark{algor}{Algorithm}
\crefname{hypothesis}{Hypothesis}{Hypotheses}

\numberwithin{theorem}{section}
\DeclareMathAlphabet{\bi}{OML}{cmm}{b}{it}
\newcommand{\rme}{\mathrm{e}}
\newcommand{\rmi}{\mathrm{i}}
\newcommand{\rmd}{\mathrm{d}}
\newcommand{\case}[2]{{\textstyle\frac{#1}{#2}}}

\newcommand{\Ip}{\mbox{$\mathcal{I}^{\scriptscriptstyle{\rm p}}$}}
\newcommand{\Ia}{\mbox{$\mathcal{I}^{\scriptscriptstyle{\rm a}}$}}

\newcommand{\Aarr}{\mbox{$A_{\scriptscriptstyle{\rm arr}}$}}
\newcommand{\larr}{\mbox{$l_{\rm arr}$}}
\newcommand{\za}{\mbox{$z_{\rm a}$}}

\newcommand{\fmax}{f_{\rm max}}
\newcommand{\fmin}{f_{\rm min}}

\newcommand{\Gh}{\mbox{$\widehat{G}$}}
\newcommand{\Go}{\mbox{$\widehat{G}^{\scriptscriptstyle{0}}$}}
\newcommand{\GR}{\mbox{$\widehat{G}^{\scriptscriptstyle{R}}$}}
\newcommand{\GRn}{\mbox{$\widehat{G}_n^{\scriptscriptstyle{R}}$}}
\newcommand{\GRm}{\mbox{$\widehat{G}_m^{\scriptscriptstyle{R}}$}}
\newcommand{\Gon}{\mbox{$\widehat{G}_n^{\scriptscriptstyle{0}}$}}
\newcommand{\Gom}{\mbox{$\widehat{G}_m^{\scriptscriptstyle{0}}$}}
\def\ipart{{\rm Im}\,}
\newcommand{\xb}{\mbox{$\vec{\bi{x}}$}}
\newcommand{\yb}{\mbox{$\vec{\bi{y}}$}}

\newcommand{\ybs}{\mbox{$\vec{\bi{y}}^{\,\rm s}$}}

\newcommand{\Qq}{\mbox{$\mathbb{Q}$}} 

\def\Hnods{\stackrel{\scriptscriptstyle 0}{H^s}}
\def\ov{\overline}

\xdefinecolor{darkblueacmac}{RGB}{42, 107, 135}
\xdefinecolor{blueacmac}{RGB}{55, 146, 179}
\xdefinecolor{ltblueacmac}{RGB}{230, 230, 244}
\xdefinecolor{redacmac}{RGB}{152, 0, 52}

\xdefinecolor{linkcol}{named}{blueacmac}
\xdefinecolor{citecol}{named}{redacmac} 

\title{Imaging extended reflectors in a terminating waveguide}

\author{Chrysoula Tsogka%
  \thanks{Applied Math Unit,
  University of California, Merced, 5200 North Lake Road, Merced, CA
  95343 (\email{ctsogka@ucmerced.edu}).}%
  \and
  Dimitrios A. Mitsoudis%
  \thanks{Department of Energy Technology Engineering, Technological Educational Institute of Athens 
                  \& IACM/FORTH, Heraklion 70013, Greece.
    (\email{dmits@teiath.gr}).}
  \and
  Symeon Papadimitropoulos%
   \thanks{Department of Mathematics \& Applied Mathematics, University of
  Crete \& IACM/FORTH, Heraklion 70013, Greece. (\email{spapadem@tem.uoc.gr}).}%
}

\begin{document}
\maketitle

\begin{abstract}
We consider the problem of imaging extended reflectors in  terminating waveguides. 
We form the image by back-propagating the array response matrix projected on the waveguide's non evanescent modes. 
The projection is adequately defined for any array aperture size covering fully or partially the waveguide's vertical cross-section.  
We perform a resolution analysis of the imaging method and show that the resolution is determined by the central frequency
while the image's signal-to-noise ratio improves as the bandwidth increases.
The robustness of the imaging method is assessed with fully 
non-linear scattering data in terminating waveguides with complex geometries. 
\end{abstract}

%
%
%
\section{Introduction}           \label{sec:Intro}

We consider in this paper the problem of imaging extended reflectors in terminating acoustic waveguides with complex geometries as the one depicted in  \cref{fig:WG_setup}. Specifically, we assume that the waveguide $\Omega \subset {\mathbb R}^2$  consists of a semi-infinite strip 
$\Omega_{L^-} = (-\infty,L) \times (0,D)$ in which the speed of propagation may only depend on the cross-range variable $x$, i.e., $c=c(x)$, and a bounded domain $\Omega_{L^+}$ in which it may be fully inhomogeneous, i.e., $c=c(z,x)$, and contains the reflector that we wish to image. Although we restrict our presentation in the two-dimensional case, the proposed imaging methodology can be extended in a straightforward manner to a three-dimensional waveguide with bounded cross-section. We illustrate this with some numerical results  in the three dimensional case. 

Our data is the array response matrix for the scattered field collected on an array of transducers which can play the dual role of emitters and receivers. This is a three-dimensional data structure $[\widehat{\Pi}]_{srl}$ that depends on the location of the source $\xb_s$, $s=1,\ldots,N_s$, the receiver $\xb_r$, $r=1,\ldots N_r$, and the frequency $\omega_l$, $l=1,\ldots,N_f$. The element $\widehat{\Pi}(\xb_s,\xb_r;\omega_l)$  denotes the response recorded at $\xb_r$ when a unit amplitude signal at frequency $\omega_l$ is send from a point source at $\xb_s$. Furthermore, we consider that the array is located in $\Omega_{L^-}$, it has an equal number of sources and receivers $N_s=N_r=N$ and may span fully or partially the vertical cross-section of the waveguide. 

Imaging in waveguides is of particular interest in underwater acoustics \cite{BGWX_2004, JKPS_04, XML_00, IMN_2004, Fink1, Prada2, DMcL_2006, BL_2008} where one wants to characterize sound speed inhomogeneities in shallow ocean environments  with applications in sonar, marine ecology, seabed imaging, etc. Moreover, imaging in waveguides finds also applications in inspections of underground pipes using acoustic waves \cite{MS_2012, PAHWTBLS_07} as well as in non-destructive evaluation of materials where elastic wave propagation should be considered \cite{BLL_2011}.  In any case, this is a challenging inverse scattering problem since in a waveguide geometry the wave field may be  decomposed in a finite number of propagating modes and an infinite number of evanescent modes. The evanescent part of the wave field is in general not available in the measured data because it decays exponentially fast with the propagation distance. Let us denote $(\mu_n,X_n)$ the eigenvalues and corresponding orthonormal eigenfunctions of the vertical eigenvalue problem for the negative Laplacian ($-\partial^2_x$ in the 2D-case) in a vertical cross-section of $\Omega_{L^-}$ and let $M$ be the number of propagating modes. 

We propose and analyze in this paper an imaging method that relies only on the propagating modes in the waveguide. The idea of formulating the inverse scattering problem in terms of the propagating modes has been considered by several authors in the past; indicatively we refer to the relatively recent works \cite{DMcL_2006, PR_07, BL_2008}. In \cite{DMcL_2006} the problem of reconstructing weak inhomogeneities located in an infinite strip is addressed and the solution of the linearized inverse scattering problem is obtained using the spectral decomposition of the 
far-field matrix. We note that in this case the measurements consist of both the transmitted and the reflected (backscattered) field. In \cite{PR_07} the problem of selective focusing on small scatterers in two dimensional acoustic waveguides is considered and the spectral decomposition of the time-reversal operator is analyzed in this setting.  In \cite{BL_2008} the authors establish a modal formulation for the Linear Sampling Method (LSM) \cite{CK_96} for imaging extended reflectors in waveguides. The extension to the case of anisotropic scatterers that may touch the waveguide boundaries is carried out in \cite{MS_2012} where both the LSM  and the Reciprocity Gap Method (RGM) \cite{CH_05} are studied theoretically and numerically. The case of imaging cracks in acoustic waveguides is considered in \cite{BL_2012} using LSM and the factorization method \cite{K_98}. In all the aforementioned works the waveguide geometry is infinite in one-dimension. 

The case of a semi-finite, terminating waveguide as the one considered here was first studied in our knowledge in \cite{BN_2016} for electromagnetic waves in three dimensions. In particular in \cite{BN_2016} the forward data model was derived using Maxwell's equations and two imaging methods were formulated: reverse time migration (phase conjugation in frequency domain) that is obtained by applying the adjoint of the forward operator to the data and an $l_1$-sparsity promoting optimization method. 

Our imaging approach is also inspired by phase conjugation and consists in back-propagating the array response matrix projected on the $M$ waveguide's propagating modes. It is important to note that the projection on the propagating modes is not an obvious procedure when the array does not span the whole aperture of the waveguide. Following our previous work \cite{TMP_16} we define this modal projection adequately using the eigenvalue decomposition of the matrix $\Aarr \in \mathbb{C}^{M\times M}$ whose $mn$-th component is the integral over the array aperture of the product $X_n (\cdot) X_m(\cdot)$. The orthonormality of $X_n$ implies that $\Aarr$ reduces to the identity matrix when the array spans the entire waveguide depth. The properties of the eigenvalues and eigenvectors of $\Aarr$ were analyzed in detail in \cite{TMP_16} for the partial aperture case. We show in particular that there is no loss of information and therefore no change in the image as long as the minimal eigenvalue of $\Aarr$ remains above a threshold value $\epsilon$ which depends on the noise level in the data or equals the machine precision in the noiseless case. As the array aperture decreases the number of the eigenvalues that fall below $\epsilon$ increases and consequently the quality of the image deteriorates (see \cite{TMP_16}).    

To analyze the resolution of the proposed imaging method we consider the case of a point reflector and prove that the single frequency point-spread function equals the square of the imaginary part of the Green's function.  This is established using the Kirchhoff-Helmholtz identity which we derive for the terminating waveguide configuration. Furthermore, for the simple geometry of a semi-infinite strip in two dimensions a detailed resolution analysis is carried out. This determines the resolution of the imaging method which depends only on the central frequency and equals half the wavelength in both directions. Although the bandwidth does not affect the resolution, it does play an important role as it significantly improves the signal-to-noise ratio of the image. This is shown theoretically and is also confirmed by our numerical simulations. 

Imaging in the terminating waveguide geometry allows for an improvement in the reconstructions compared to the infinite waveguide case. This is because multiple-scattering reflections that bounce off the terminating boundary of the waveguide provide multiple views of the reflector that are not available in the infinite waveguide case. To benefit from this multipathing we need to know or determine the boundary of the waveguide prior to imaging the reflector. In this work we considered that the waveguide boundary is known. We refer to \cite{BGT_2015} for a study of source imaging in waveguides with random boundary perturbations where it is shown that uncertainty in the location of the boundaries can be mitigated using filters that imply a somewhat reduced resolution. Moreover, it is shown in \cite{BGT_2015} that there is an optimal trade-off between robustness and resolution which can be adaptively determined during the image formation process.

The robustness of the proposed imaging method is assessed with fully non-linear scattering data obtained using the Montjoie software \cite{Montjoie}. The use of this software allows us to model wave propagation in waveguides with complicated geometries and study the reconstruction of diverse reflectors. For all the examples considered we have obtained a significant improvement in the reconstruction in the terminating waveguide geometry as compared to the infinite case. We have also studied the robustness of the method for different array apertures ranging from full to one fourth of the waveguide's vertical cross-section. The quality of the image deteriorates as we decrease the array aperture but our imaging results remain very satisfactory even with an array-aperture equal to one fourth of the full one.
In most of the examples we consider that the multistatic array response matrix is available. However, the same method can be also applied to synthetic array data obtained with a single transmit/receive element. We obtain good reconstructions for this reduced data modality as well but for larger array apertures that cover at least half of the waveguide's width in the vertical direction.

The paper is organized as follows. In \cref{sec:Form} we present the formulation of the problem. In \cref{sec:Im} we describe our imaging methodology inspired by phase conjugation for both the passive imaging configuration which concerns imaging a source, as well as the active setup that refers to imaging a reflector. The resolution analysis is carried out in \cref{sec:Resol} for single and multiple frequency imaging. Finally, in \cref{sec:Numer} we illustrate the performance of our approach with numerical simulations in two and three  dimensions.

%
%
%
%
\section{Formulation of the problem} \label{sec:Form}
\begin{figure}[h]
\begin{center}
\includegraphics[width=0.8\linewidth]{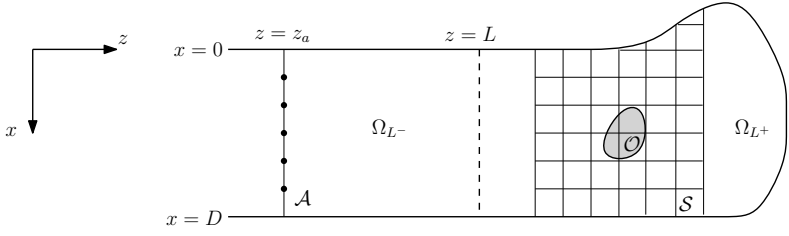}
\caption{Schematic representation of the semi-infinite waveguide.}
\label{fig:WG_setup}
\end{center}
\end{figure}

In this work, we study the problem of imaging extended reflectors in a two-dimensional terminating waveguide, as shown 
in \cref{fig:WG_setup}.  The  reflector is illuminated by an active vertical array $\mathcal{A}$, composed of $N$ transducers 
that act as sources and receivers. The array may span the whole vertical cross-section of the waveguide or part of it. 
The array transducers are assumed to be  distributed uniformly, and densely enough, that is the inter-element distance $h$ 
is considered to be small, typically a fraction  of the wavelength $\lambda$. 
The term extended indicates that the reflectors are comparable in size to $\lambda$.  

We also assume that the array measurements can be cast in the form of the so-called array response matrix, denoted by 
$\widehat{\Pi}$. This is an $N \times N$ complex matrix whose $(r,s)$ entry is the Fourier transform of the time traces of the 
echoes recorded at the $r$-th receiver when the $s$-th source emits a signal. 
In particular, we shall use the array response matrix for the scattered field that is due to the presence of
an extended reflector $\mathcal{O}$ located somewhere in the bounded part of the waveguide delimited by the cross-section 
at $z = L$ (see \cref{fig:WG_setup}). As usual, the scattered field is determined by subtracting the incident field from the total field.

Specifically, we consider a Cartesian  coordinate system  $(z,x)$, where $z$ denotes the main direction of propagation 
called hereafter range, and $x$ the cross-range direction  taken to be positive downwards. Our terminating waveguide 
$\Omega$ consists of two subdomains: the semi-infinite strip $\Omega_{L^-} = (-\infty,L) \times (0,D)$ and a bounded 
domain in  ${\mathbb R}^2$ denoted by $\Omega_{L^+}$. 
Let us also assume that all the inhomogeneities of the medium are contained in $\Omega_{L^+}$ while the medium is 
homogeneous in the semi-infinite strip $\Omega_{L^-}$,  i.e. the wave speed may depend on range and cross-range in 
$\Omega_{L^+}$, and varies smoothly to the constant value that has for $z\le L$. Note that the assumption of a constant 
wave speed in $\Omega_{L^-}$ may be relaxed by requiring the speed to depend 
on the cross-range variable $x$. However, to facilitate the presentation in this paper we will consistently assume 
that $\Omega_{L^-}$ is filled with a homogeneous medium.


The total field for our waveguide in the presence of a scatterer  solves the scalar wave equation
\begin{equation} \label{eq:ptot}
\Delta p^{\rm tot} (t,\xb) - \frac{1}{c(\xb)^2} \frac{\partial^2 p^{\rm tot} (t,\xb)}{\partial t^2} = -f(t,\xb),
\end{equation}
where  $\xb=(z,x) \in \Omega$  and the source term $f(t,\xb)$ models a point-like source with time-harmonic dependence. 
\Cref{eq:ptot} is supplemented by homogeneous Dirichlet conditions on the boundary of $\Omega$. The scatterer 
is modeled as an acoustically hard scatterer with a homogeneous Neumann condition on $\partial \mathcal{O}$ and a suitable 
outgoing radiation condition is assumed as  $z\rightarrow -\infty$. 
Moreover, we assume that the medium is quiet for $t\leq  0$,  i.e. $p^{\rm tot} (t,\xb) = 0$ for $t\le 0$. 
Note that the scalar wave equation that we consider here is used quite often instead of the full Maxwell's or elastic wave equations 
since it captures the main features of the scattering problem. 

By applying the Fourier transform
\begin{equation} \label{eq:FT}
\widehat{p}^{\rm tot}(\omega,\xb) = \int \rme^{i\omega t} p^{\rm tot} (t,\xb) dt
\end{equation}
on \cref{eq:ptot}, we obtain the Helmholtz equation for the total field
\begin{equation} \label{eq:HE}
-\Delta\widehat{p}^{\rm tot}(\omega,\xb) - k^2 \eta(\xb) \,\widehat{p}^{\rm tot}(\omega,\xb) = \widehat{f}(\omega,\xb),\quad \xb\in\Omega,
\end{equation}
where  $\omega$ is the angular frequency, $k=\omega/c_0$ is the (real) wavenumber and 
$\eta(\xb)=c^2_0/c^2(\xb)$ is the index of refraction. (Notice that $\eta(\xb)=1$ for all $\xb\in\Omega_{L^-}$.)

Let also $\Gh(\xb,\xb_s;\omega)$ denote the Green's function  for the Helmholtz operator 
(and the associated boundary conditions) due to a point source located at $\xb_s = (z_s,x_s) \in \Omega$ 
and for a single frequency $\omega$, \textit{i.e.} $\Gh(\xb,\xb_s;\omega)$ is the solution of 
\begin{equation}   \label{eq:Green1}
   -\Delta \Gh(\xb,\xb_s;\omega) - k^2 \eta(\xb)\Gh(\xb,\xb_s;\omega) = \delta(\xb-\xb_s).
\end{equation} 

Finally, let $(\mu_n,X_n)$ be the eigenvalues and corresponding orthonormal eigenfunctions of the following vertical 
eigenvalue  problem in the homogeneous part of the waveguide $\Omega_{L^-}$:
\begin{equation} \label{eq:VE}
X_n''(x) + \mu_n X(x) = 0, \qquad X_n(0) = X_n(D) = 0.
\end{equation}
Henceforth we shall assume that there exists an index $M$ such that the constant value $k^2$ 
of the wavenumber  satisfies in $\Omega_{L^-}$:
\begin{equation*}
\lambda_M < k^2 < \lambda_{M+1}.
\end{equation*}
In other words, $M$ is the number of \textit{ propagating modes} in $\Omega_L^-$.
Let us also denote the horizontal wavenumbers in $\Omega_L^-$ by
\begin{equation}    \label{eq:betas}
  \beta_n = \left\{\begin{array}{ll}
                  \sqrt{k^2 - \mu_n},      & 1 < n < M,\\
                  \rmi \sqrt{\mu_n - k^2}, & n> M+1 .\end{array}\right.
\end{equation}

In what follows we will assume that the problems for the incident and the total fields, which are governed by the 
Helmholtz equation and satisfy the boundary conditions in the perturbed semi-infinite cylinder described before, are well-posed.
Notice that for the incident field it is known, \cite{Jones_1953}, that the problem is well-posed except for a set of values of $k^2$ 
that are equal to point eigenvalues of the negative Laplacian associated with zero Dirichlet conditions on the boundary.  
This set is at most countable, it has no finite accumulation point and in many cases it is empty. For the total field there 
are examples in infinite waveguides that suggest existence of the so-called trapped modes, i.e. nonzero localized solutions of the
associated homogeneous problem, see e.g. \cite{EP_1998}.
%
%
%
%
\section{Imaging}   \label{sec:Im}
Our main objective in this work is to form images of extended reflectors that lie somewhere in a terminating 
waveguide like the one described in the previous section. The usual steps that one may follow to this end, is to first 
identify a search domain $\mathcal{S}$ (see \cref{fig:WG_setup}), discretize it using a grid, and then compute 
the value of an appropriate imaging functional in each grid point in $\mathcal{S}$. It is expected that these 
values, when they are graphically displayed in the search domain, should exhibit peaks that indicate the presence 
of the reflector. 
\subsection{Imaging with a full-aperture array}  \label{ssec:Full}
We shall first consider the easier case where the array spans the whole vertical cross-section of the waveguide.
Moreover, although we are interested in imaging extended reflectors we will first examine the so-called passive imaging
problem in order to motivate the use of the imaging functional that we will introduce next.
\subsubsection{Passive Imaging}
\begin{figure}[ht]
\begin{center}
\includegraphics[width=0.7\textwidth]{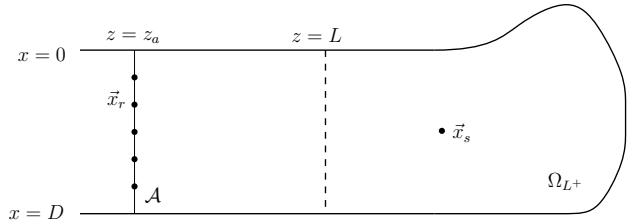}\\
\caption{Passive imaging setup in a terminated waveguide.}
\label{fig:term_passive}
\end{center}
\end{figure}

So, let us assume that a point source of unit strength, located at the point $\xb_s = (z_s,x_s)\in\Omega$, emits  a signal
that is recorded on a vertical array $\mathcal{A}$ located in $\Omega_{L^-}$. Moreover, we assume that the array 
$\mathcal{A}=\{\xb_r = (z_a,x_r)\}_{r=1}^N$, ($z_a < L$), spans the whole vertical cross-section of the waveguide as 
illustrated in \cref{fig:term_passive}. Our aim is to find the location of the source.
In this case the array response matrix $\widehat{\Pi}$ at frequency $\omega$ reduces to a $N\times 1$ vector, whose 
$r$-th component equals the Green's function evaluated at receiver $\xb_r$ due to the source $\xb_s$, i.e.
\begin{equation}    \label{eq:src_respmatr}
   \widehat{\Pi}(\xb_r;\omega) = \Gh(\xb_r,\xb_s;\omega).
\end{equation}
In what follows we consider a monochromatic source and to simplify the notation we suppress parameter $\omega$ 
from the imaging functional and the Green's function. The dependence on $\omega$ will be recalled in \cref{sec:mf} 
where imaging with multiple frequency data is considered. 

The imaging functional that we propose to use is based on the concept of phase conjugation, which may be physically
interpreted by virtue of the Huygen's principle. As pointed out in \cite{JD_91}, Huygen's principle states that a
propagating wave may be viewed as superposition of wavelets reemitted from a fictitious surface with amplitudes
proportional to those of the original wave. In phase conjugation, which may be seen as the equivalent of time reversal
in the frequency domain, the reemitted wavelets' amplitudes are proportional to the complex conjugate of the corresponding 
ones in the original wave.  These remarks lead naturally one to define the following classical phase conjugation imaging functional
\begin{equation}  \label{eq:PC_simple}
\mathcal{I}^{\scriptscriptstyle{\rm pc}}(\yb^s) = \int_{\mathcal{A}} \ov{\Gh(\xb_r,\xb_s)} \Gh(\yb^s,\xb_r) dx,
\end{equation}
where $\xb_r = (z_a,x)\in \mathcal{A}$ and $\yb^s \in \mathcal{S}$.
However, if we assume for a moment that apart from recording the value of the field on the array we would be able to record  
its normal derivative as well, then we may define the following imaging functional, which as we will show next has
very nice theoretical properties. So let
\begin{equation}   \label{eq:PC_comp}
\mathcal{I}(\yb^s) :=      \int_{\mathcal{A}} \Big(\ov{\Gh(\xb_r,\xb_s)} \nabla\Gh(\xb_r,\yb^s) 
                           -  \Gh(\xb_r,\yb^s) \ov{\nabla\Gh(\xb_r,\xb_s)}~\Big)\cdot \bi{\nu} \, dx, 
\end{equation}
where $\bi{\nu}$ is the outward-pointing unit normal vector to $\mathcal{A}$. Of course this functional is more
complicated than phase conjugation but  the following proposition shows that in order to compute $\mathcal{I}(\yb^s)$
in a terminating waveguide it is required to know only the values of the wave field on the array  and not its derivatives.

%
%
\begin{proposition}[Kirchhoff-Helmholtz identity]    \label{pr:HK_termwg}
Assume that a point source is located in the terminating waveguide that we have described in \cref{sec:Form} 
(see, also, \cref{fig:term_passive}), and that a vertical array $\mathcal{A}$,  which spans the whole vertical cross-section 
of the waveguide,  is located in  $\Omega_{L^-}$. Then, the imaging functional that we have defined in (\ref{eq:PC_comp}) 
satisfies the following Kirchhoff-Helmholtz identity:
\begin{equation}  \label{eq:HK_termwg1}      
   \mathcal{I}(\yb^s) = \Gh(\yb^s,\xb_s) - \ov{\Gh(\yb^s,\xb_s)} = 2i \ipart \Gh(\ybs,\xb_s).
\end{equation}
Moreover, we can show that,
\begin{equation}    \label{eq:HK_termwg2} 
  \mathcal{I}(\yb^s) =  2i   \sum_{n=1}^{M} \beta_n\, \overline{\Gh_n(z_a,\xb_s)}\, \Gh_n(z_a,\yb^s),                    
\end{equation}
where $ \Gh_n(z_a,\cdot)$, $n = 1,\ldots,M$,  denote the first $M$ Fourier coefficients of the Green's function 
(which correspond to the propagating modes) with respect to the orthonormal basis of $L^2(0,D)$ that is 
formed by the vertical eigenfunctions $X_n$, i.e.
\begin{equation} \label{eq:GRn}
 \Gh_n(z_a,\cdot) = \int_0^D \Gh((z_a,x'),\cdot)\, X_n(x')\, dx'.
\end{equation}  
\end{proposition}
{\bf Proof.} See \cref{sec:KHpr}.\ \hfill $\Box$
\paragraph{The passive imaging functional}
Motivated by \cref{pr:HK_termwg} we define here our imaging functional for the passive case. Assuming that the array elements 
are dense enough, so that we may think of the array as  being continuous, we define 
\begin{equation} \label{eqPnpassive}
   \widehat{\mathbb{Q}}_n = \int_0^D  \widehat{\Pi}\big(\xb_r;\omega\big) X_n(x) \, dx, \quad n = 1\ldots,M,
\end{equation}
to be the projection of the recorded field on the first $M$ eigenfunctions $X_n$, $n= 1\ldots,M$, of the vertical eigenvalue 
problem \cref{eq:VE}.  
Notice that using \cref{eq:src_respmatr}, $\mathbb{Q}_n$ may be written as
\begin{align*}
   \widehat{\mathbb{Q}}_n &= \int_0^D   \GR\big((z_a,x),\xb_s\big) X_n(x) \, dx = \GRn(z_a,\xb_s).
\end{align*}
In view of \cref{eq:HK_termwg2}  we define our imaging functional as:
\begin{equation}  \label{eq:Ip_gen}
   \Ip(\yb^s) := \sum_{n=1}^{M} \beta_n \overline{\widehat{\mathbb{Q}}_{n}} \, \Gh_n(z_a,\yb^s).
\end{equation}

Note that the evaluation of $\Ip(\yb^s)$, for $\yb^s\in\mathcal{S}$, requires only recordings of the wave field. 
Moreover, \cref{eq:HK_termwg1} and \cref{eq:HK_termwg2} ensure that 
\begin{equation}  \label{eq:Ip_Gipart}
  \Ip(\yb^s) = \ipart \Gh(\ybs,\xb_s).
\end{equation}
This last equation is a very interesting result, and says that the quality of the focusing in the image is determined by the 
 imaginary part of the Green's function in our waveguide. Therefore, a resolution analysis for  $\Ip$ will entail 
 the study of the behaviour of $\ipart \Gh$.

\begin{figure}[ht]
\begin{center}
\includegraphics[width=0.34\linewidth]{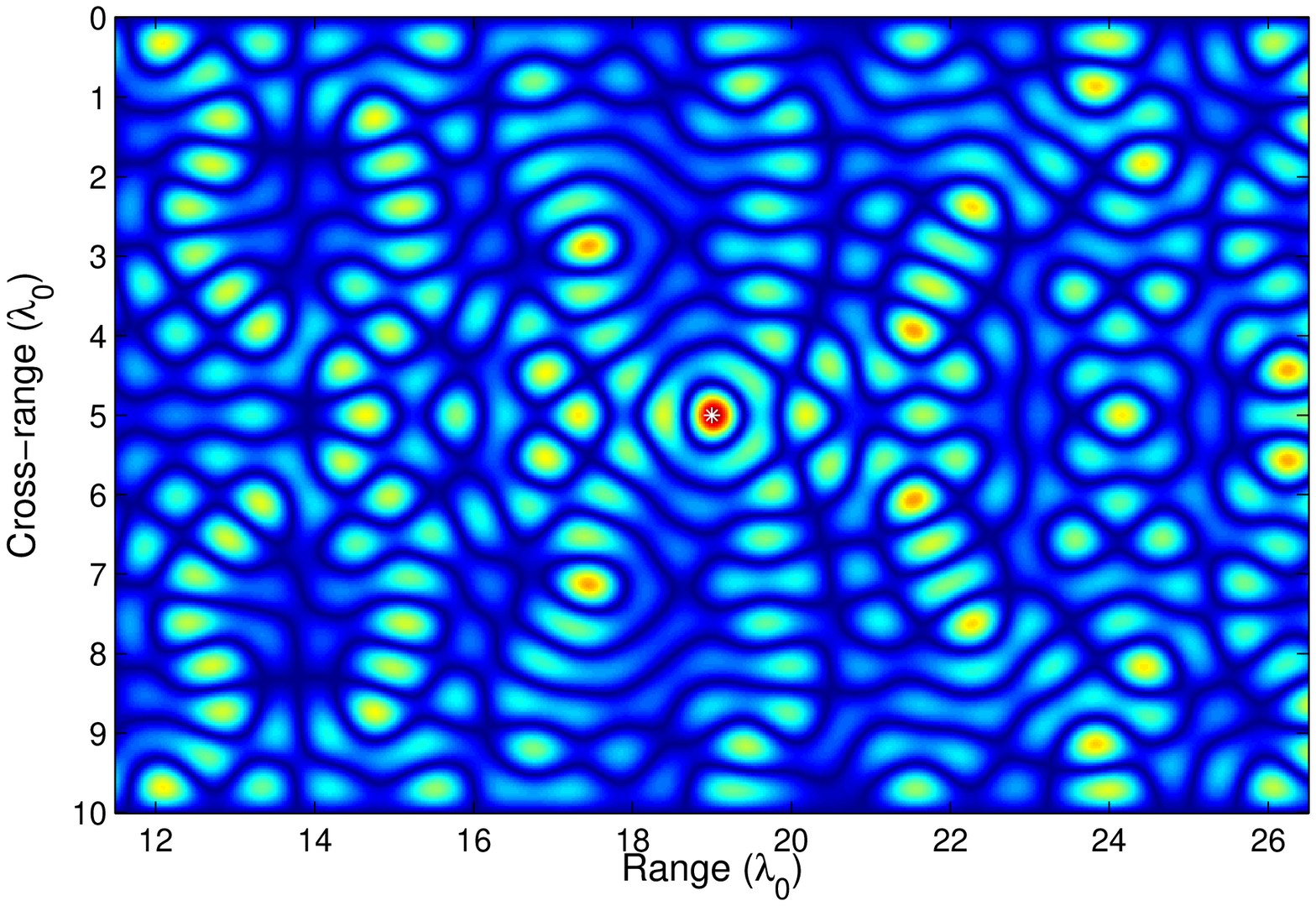}
\includegraphics[width=0.3\linewidth]{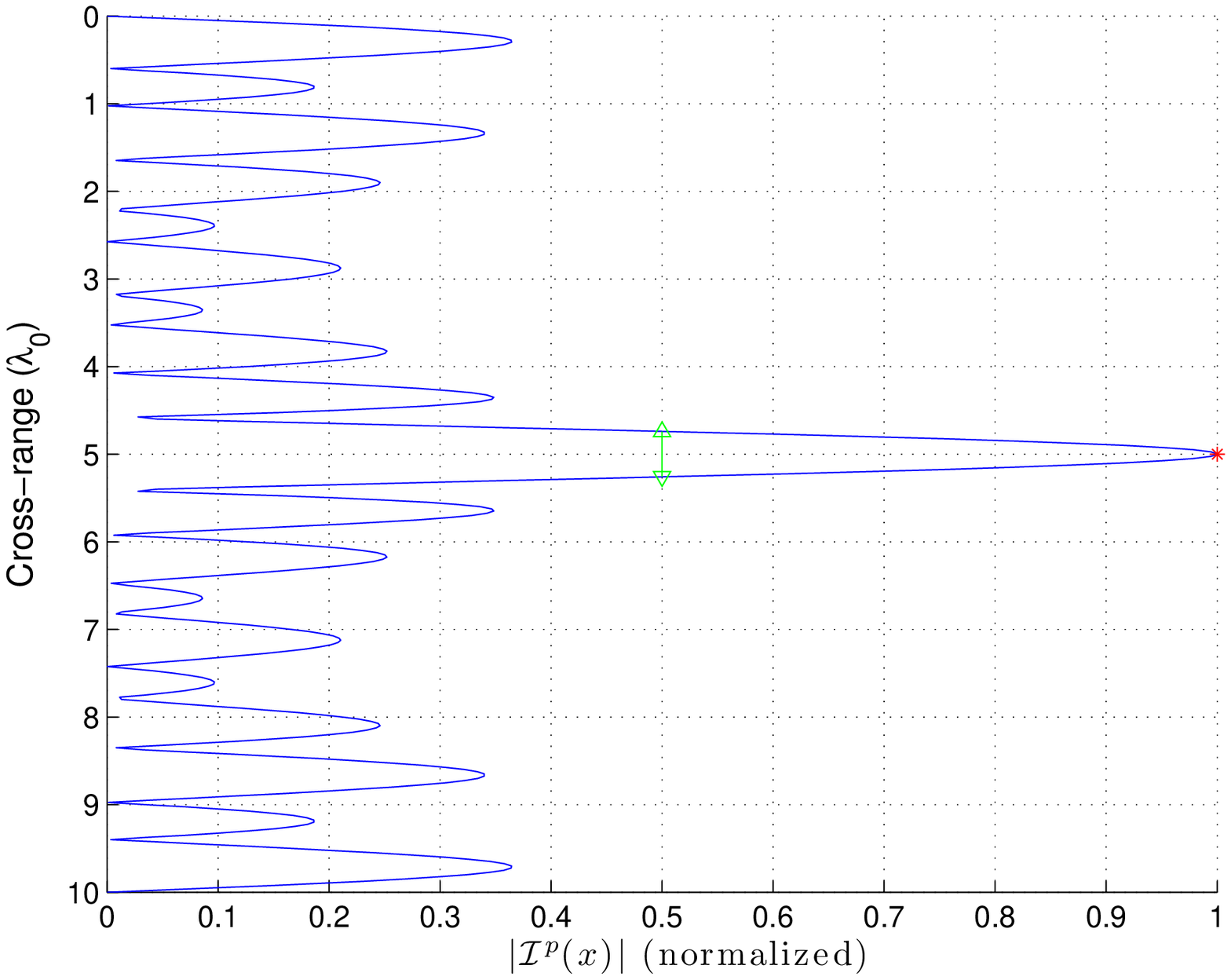}
\includegraphics[width=0.3\linewidth]{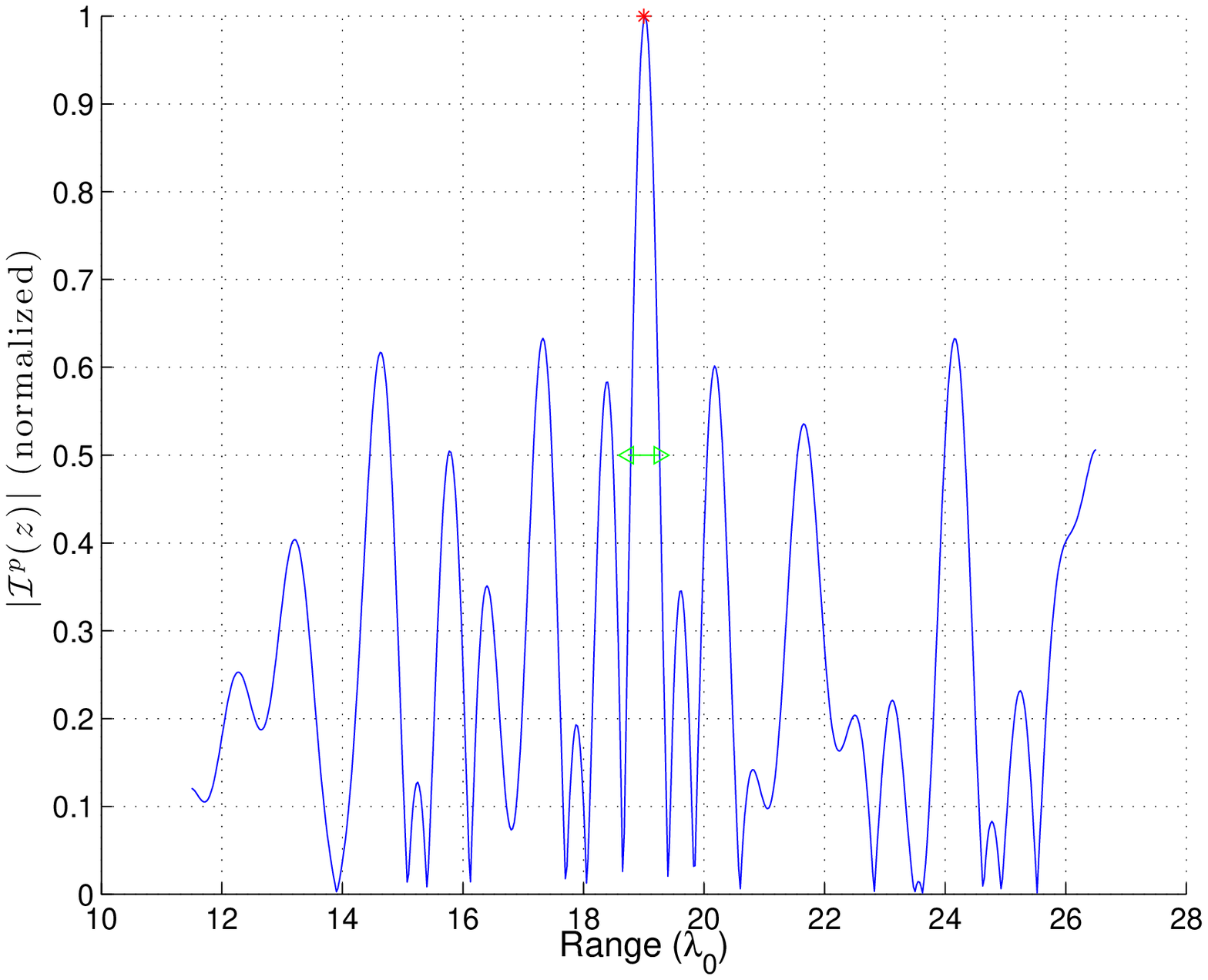}
\caption{Normalized modulus of $\Ip$ for a point source located at $\xb_s = (19,5)\,\lambda_0$ and for a single frequency 
corresponding to $k = 0.973k_0$. Imaging on the whole search domain (left), for search points fixed at the
correct range $z=z_s$ (middle) and at the correct cross-range $x=x_s$ (right). The green arrowed segment 
indicates length equal to $\lambda/2$ and a red asterisk points to the location of the source.}
\label{fig:Ip}
\end{center}
\end{figure}

\example[Imaging a point source]     \label{ex:PointSource}
In order to provide to the reader a sense of how $\Ip(\ybs)$ behaves we consider the simple case of imaging 
a source in a homogeneous terminating waveguide that forms a semi-infinite strip, i.e. $\Omega = (-\infty,R)\times (0,D)$.  
We assume a reference wavenumber $k_0 = \pi/10$ that corresponds to a reference wavelength $\lambda_0$, and take
$D = 10\lambda_0$ while the vertical (terminating) boundary is placed at $R=27.5 \lambda_0$.
In \cref{fig:Ip}, we plot the modulus of equation \cref{eq:Ip_gen}, for a source placed at $\xb_s=(19,5)\lambda_0$ 
(shown in the plot as a white asterisk)  and for a single frequency $f$ that corresponds to a wavenumber $k = 0.973k_0$. 
This results to a number of propagating modes $M=19$. Finally our search domain is $\mathcal{S} = [11.5,26.5]  \times [0,10]$,
where all distances are expressed in terms of the reference wavelength $\lambda_0$.  

We observe that the $\Ip(\ybs)$ image, despite a relatively high noise level, displays a clear peak around $\xb_s$, 
which is a key property for an imaging functional.

\subsubsection{Active Imaging}
\label{sec:active}
\begin{figure}[ht]
\begin{center}
\includegraphics[width=0.7\textwidth]{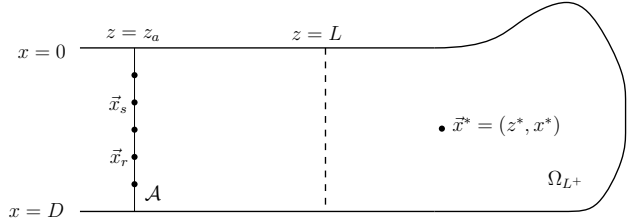}\\
\caption{Active imaging setup in a terminated waveguide.}
\label{fig:termWG3}
\end{center}
\end{figure}

As a step forward to the general case of an extended scatterer, we will now deal with the active imaging problem
where we are interested in locating a single point scatterer of unit reflectivity that is situated at 
$\xb^\ast = (z^\ast,x^\ast)$ while the array $\mathcal{A}$ is like the one in the passive imaging case as illustrated  
in \cref{fig:termWG3}.

Then, the $(s,r)$ entry of the array response matrix: 
\begin{align*}     
   \widehat{\Pi}(\xb_s,\xb_r;\omega) = k^2 \Gh(\xb^\ast,\xb_s;\omega)\Gh(\xb_r,\xb^\ast;\omega),
\end{align*}
corresponds to the scattered signal received at $\xb_r$ when the point reflector at $x^\ast$ is illuminated by 
a unit amplitude signal emitted at frequency $\omega$ from a point source located at $\xb_s$. 
In what follows we suppress the multiplicative constant $k^2$, hence we assume that
\begin{equation}      \label{eq:sct_respmatr}
   \widehat{\Pi}(\xb_s,\xb_r;\omega) = \Gh(\xb^\ast,\xb_s;\omega)\Gh(\xb_r,\xb^\ast;\omega).
\end{equation}
In the mutiple-frequency case we can also remove this factor by rescaling the data matrix $\widehat{\Pi}(\xb_s,\xb_r;\omega)$
to be equal to $k^{-2}\widehat{\Pi}(\xb_s,\xb_r;\omega)$.

Assuming again that the array is continuous we define the projected response matrix $\widehat{\mathbb{Q}}$ as
\begin{equation}      \label{eq:sct_respmatr_proj}
   \widehat{\mathbb{Q}}_{nm} = \int_0^D\int_0^D  \widehat{\Pi}(\xb_s,\xb_r;\omega) X_n(x_s) X_m(x_r) \, dx_s \, dx_r, 
   \quad n,m = 1\ldots,M,
\end{equation}
where $X_n$, $n= 1\ldots,M$, are the first $M$ eigenfunctions of problem \cref{eq:VE} as before.
\paragraph{The active imaging functional}
A natural  generalization of the imaging functional that we have proposed in the passive case is the following active imaging functional 
\begin{equation}   \label{eq:Ia}
   \Ia(\ybs) := \sum_{n=1}^{M}\sum_{m=1}^{M}  \beta_n \beta_m \ov{\widehat{\mathbb{Q}}_{nm}} \, \Gh_n(z_a,\ybs)\, \Gh_m(z_a,\ybs),
\end{equation}
defined for each point $\ybs$ in the search domain $S$.

Note that by replacing \cref{eq:sct_respmatr} into \cref{eq:sct_respmatr_proj} and using the expression of 
$\Gh_n$ given in \cref{eq:GRn},  it is easy to show that 
\begin{equation}     \label{eq:Qnm}
   \widehat{\mathbb{Q}}_{nm} = \Gh_n(z_a,\xb^\ast)\,\Gh_m(z_a,\xb^\ast).
\end{equation}
In turn, \cref{eq:Ia} now becomes,
\begin{eqnarray*}
   \Ia(\ybs)  = \sum_{n=1}^{M}  \beta_n\, \ov{\Gh_n(z_a,\xb^\ast)}\,\Gh_n(z_a,\ybs) 
                    \sum_{m=1}^{M} \beta_m\,\ov{\Gh_m(z_a,\xb^\ast)}\,\Gh_m(z_a,\ybs),
\end{eqnarray*}
and \cref{pr:HK_termwg} ensures that 
\begin{equation}     \label{eq:IaHK}
     \Ia(\ybs)  = \Big(\ipart \Gh(\ybs,\xb^\ast)\Big)^2 .   
\end{equation}

Thus we deduce that the imaging functional \cref{eq:Ia} for a point scatterer behaves like the square of the imaginary 
part of the Green's  function.

\example[Imaging a point scatterer]    \label{ex:PointScatt} 
To illustrate how $\Ia(\ybs)$ behaves we consider a point scatterer in the homogeneous terminating waveguide that we have described 
in \cref{ex:PointSource}. The scatterer is placed at $\xb^\ast = (19,5)\,\lambda_0$ while all the other parameters
are the same as in the previous example.
\begin{figure}[ht]
\begin{center}
\includegraphics[width=0.34\linewidth]{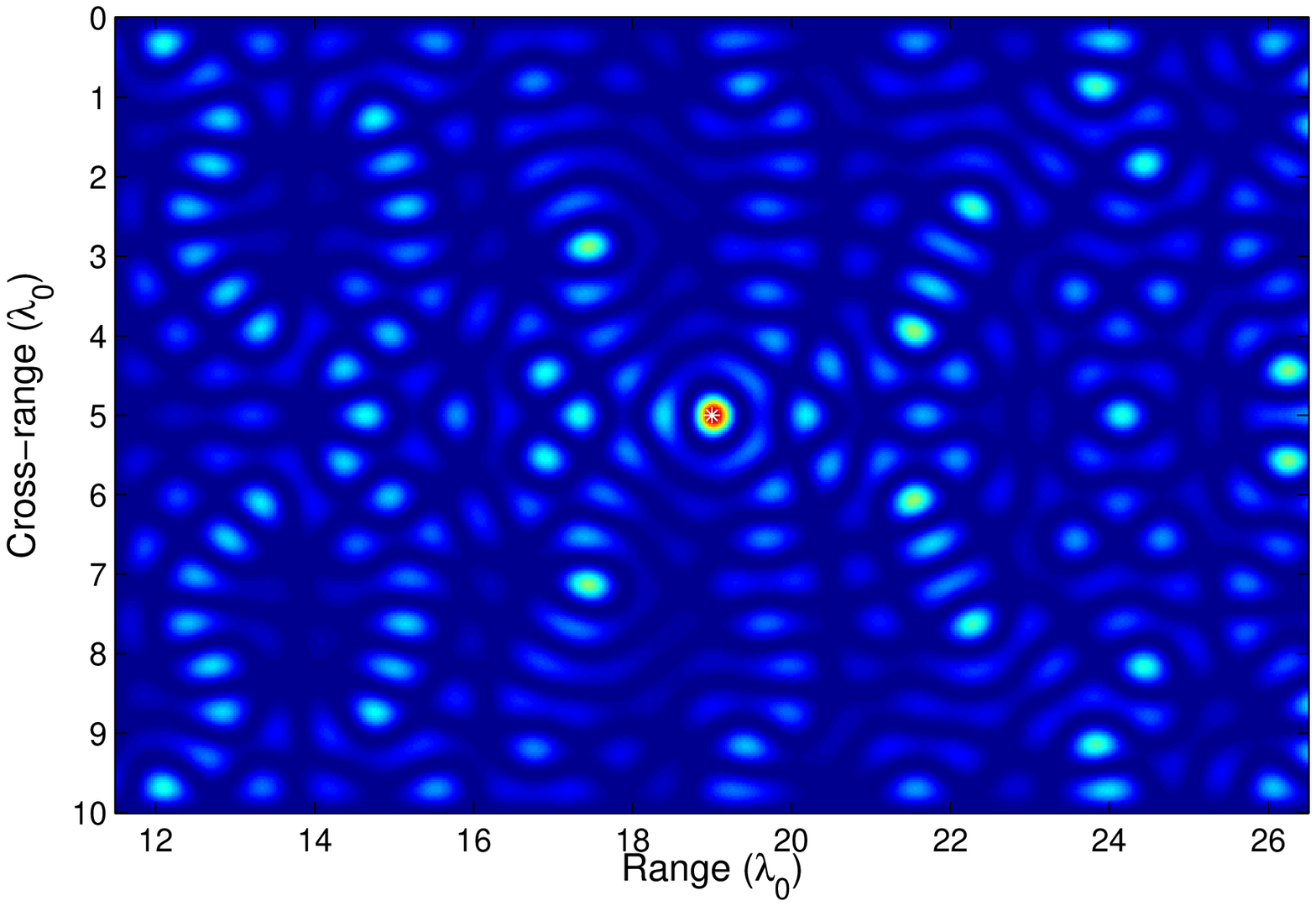}
\includegraphics[width=0.3\linewidth]{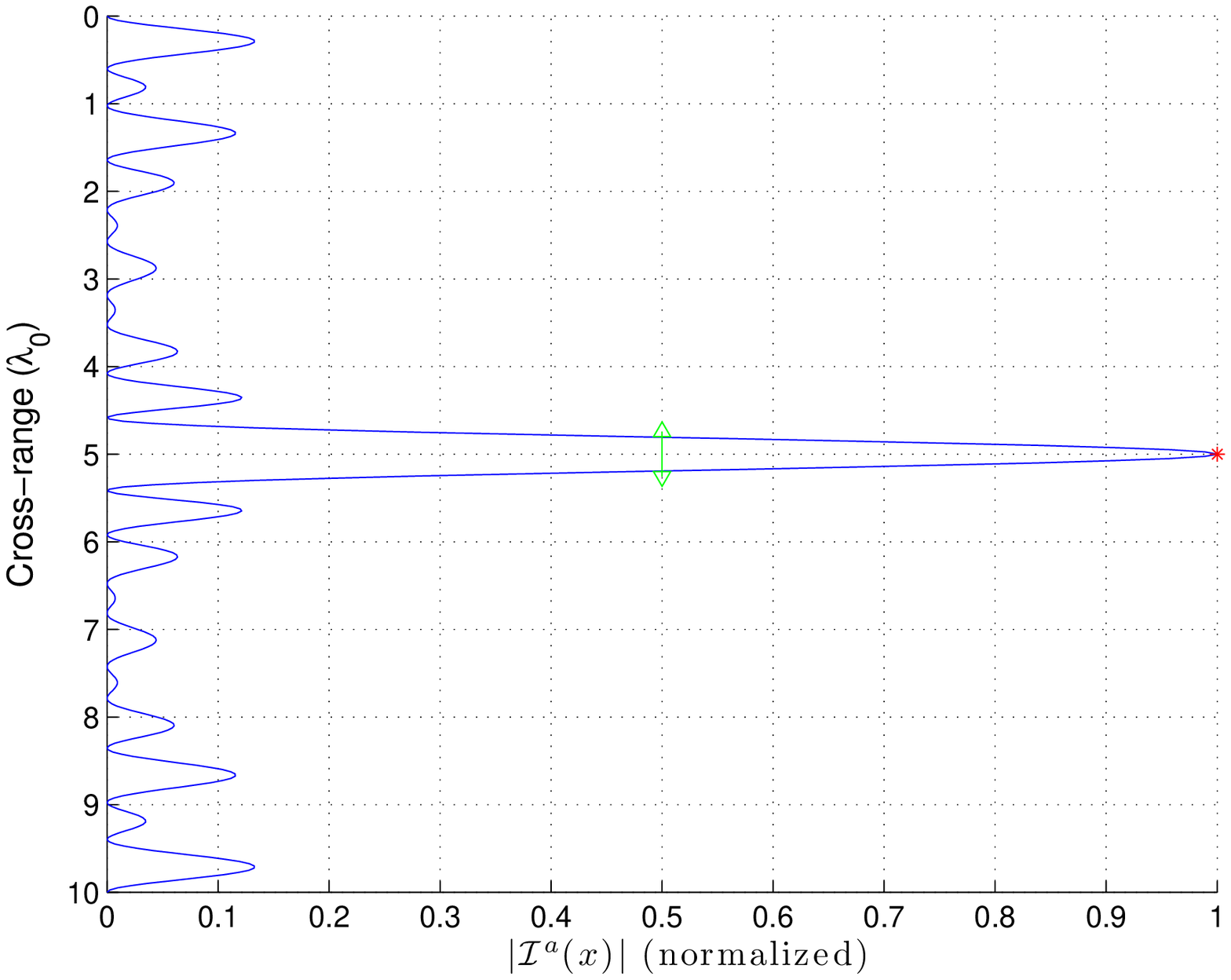}
\includegraphics[width=0.3\linewidth]{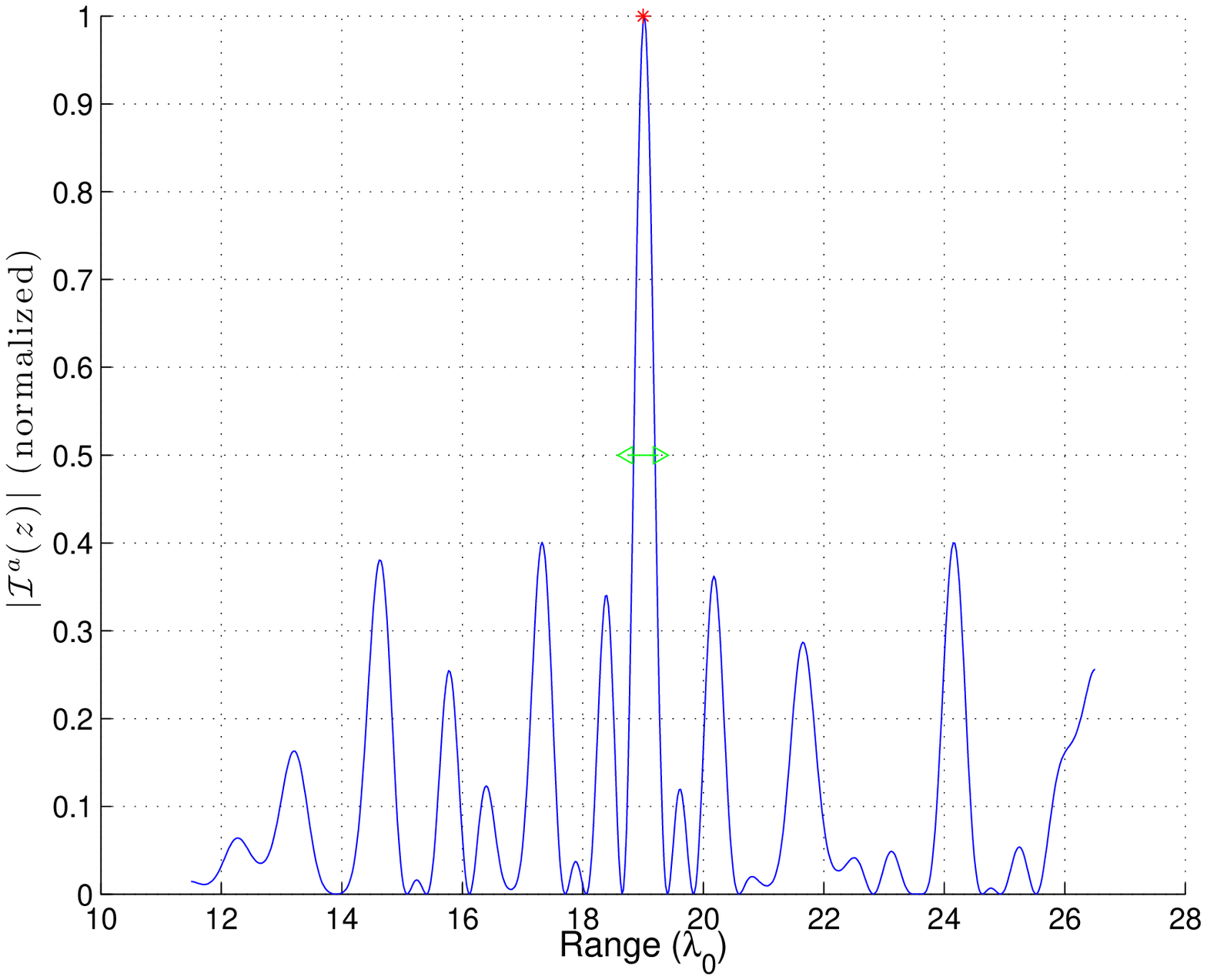}
\caption{Normalized modulus of $\Ip$ for a point scatterer located at $\xb^\ast = (19,5)\,\lambda_0$ and for a single frequency 
corresponding to $k = 0.973k_0$. Imaging on the whole search domain (left), for search points fixed at the
correct range $z=z^\ast$ (middle) and at the correct cross-range $x=x^\ast$ (right).
The green arrowed segment indicates length equal to $\lambda/2$ and a red asterisk points to the location of the scatterer.}
\label{fig:Ia}
\end{center}
\end{figure}
In \cref{fig:Ia} we plot the modulus of (\ref{eq:Ia}). As one may immediately verify this image has better signal-to-noise (SNR) 
ratio than the one shown in  \cref{fig:Ip}. This is something to be expected since $\Ia$ is just the square of $\Ip$.

We conclude noting that so far we have proven that the point spread functions in the passive and in 
the active imaging case are determined by the the imaginary part of the Green's  function when the array spans the whole vertical cross-section of the waveguide. Next, we consider the partial-aperture array case and present the modifications that have to be carried 
out in the previous approach in order to extend its applicability in this more challenging setup.
\subsection{Imaging with a partial-aperture array}  \label{ssec:Partial}
We now turn our attention to the case where the array does not span the whole vertical cross-section of the waveguide.
In \cite{TMP_16} we have presented a way to construct a projection of the array response matrix that is well suited to
that case.  Here we briefly describe the basic idea, necessary notation and main results that appear in \cite{TMP_16} 
in order to apply them in our current setup.

Let $\Aarr$ be the $M\times M$ matrix with entries 
\begin{equation}    \label{eq:Aarr}
   (\Aarr)_{mn} = \int_{\mathcal A} X_m(x) X_n(x) \rmd x,\quad m,n=1,\ldots,M,
\end{equation}
where $M$ is the number of propagating modes in $\Omega_{L^-}$.
We have shown that  $\Aarr$ is a real, symmetric Toeplitz-minus-Hankel matrix and its eigenvalues
$\nu_j$, $j=1,\ldots,M$, are clustered near $0$ and $1$; in fact if $\larr$ is the length of the array 
$\mathcal{A}$, then approximately $[\larr/(\lambda/2)]$ of the $\nu_i$'s lie near $1$ and the rest
$M-[\larr/(\lambda/2)]$ are approaching zero. 
Let, also, ${\bi w}^j = (w^j_1, w^j_2,\ldots,w^j_M)^T$ be the corresponding orthonormal eigenvectors, 
which turn out to be discrete prolate (or prolate-like) spheroidal sequences, and $W$ be the $M\times M$
orthogonal matrix  $W = (\bi{w}^1,\bi{w}^2,\ldots ,\bi{w}^M)$. Then we introduce the trigonometric polynomials
\begin{equation}   \label{eq:trigp}
  s_j(x) =  \sum_{i = 1}^M w_i^j X_i(x), \quad j = 1,2,\ldots,M.
\end{equation}
Next, we project $\widehat{\Pi}$ onto the first $M$ trigonometric polynomials $s_n$, $n=1,\ldots,M$ instead of 
projecting onto the eigenfunctions $X_n$. Specifically, we define  $\widehat{\mathbb S}$ to be the $M\times M$ 
matrix with entries 
\begin{equation}    \label{eq:Smatrx}
   \widehat{\mathbb S}_{mn} = \frac{1}{\nu_m \nu_n}\,\int_{\mathcal A} \int_{\mathcal A} 
   \widehat{\Pi}( \xb_s,\xb_r,\omega) \, s_m(x_s)\, s_n(x_r)\, d x_s\, d x_r,\  m,n=1,\ldots,M.
\end{equation}
It is easy to check that
\begin{equation}    \label{eq:SXWrel}
    \int_{\mathcal A} s_k(x) X_m(x) \, dx = \nu_k \,w_m^k,\quad k,m=1,\ldots,M.
\end{equation}
Finally, we define $\widehat{\mathbb{Q}}$ as
\begin{equation}    \label{eq:Qpartial}
    \widehat{\mathbb{Q}} = W \,\widehat{\mathbb S} \,W^T.
\end{equation}
For implementation aspects of this approach and for an extensive discussion on its performance we refer to \cite{TMP_16}.
Notice that in the case of an array with full aperture the matrix $\Aarr$ is just the identity matrix,
$s_j(x) = X_j(x)$ and $W = I_{M}$, thus we recover the definition of $\mathbb{Q}$ given in \cref{eq:sct_respmatr_proj}.

The following proposition shows that in the case of a point scatterer (active imaging) when we use \cref{eq:Smatrx,eq:Qpartial}
to construct the matrix  $\widehat{\mathbb{Q}}$ we practically recover the same data as if we were 
working with a full-aperture array.
\begin{proposition}   \label{pr:PartialPointScatt}
Let a single point scatterer of unit reflectivity be located at $\xb^\ast = (z^\ast,x^\ast)\in\Omega_{L^+}$ (see \cref{fig:termWG3}). 
We assume that the array $\mathcal{A}$ is at range $z = z_a \ll L$, so that the evanescent part of the 
wave field may be neglected. Then the projected array response matrix $\widehat{\mathbb{Q}}$ defined by \cref{eq:Smatrx,eq:Qpartial} 
for a partial aperture array is equal to the projected matrix $\widehat{\mathbb{Q}}$ for an array that spans the whole vertical cross-section $[0,D]$.
\end{proposition}
{\bf Proof.} 
For a single point scatterer the $(s,r)$ entry of the array response matrix $\widehat{\Pi}$ is given by
$\widehat{\Pi}(\xb_s,\xb_r) = \Gh(\xb^\ast,\xb_s)\Gh(\xb_r,\xb^\ast)$, see \cref{eq:sct_respmatr}.
Moreover, note that for each $\xb = (z,x) \in\Omega_{L^-}$, that satisfies the assumption $z \ll L$,  the Green's function  
may be written as   
\begin{equation}    \label{eq:Greensexp}
  \Gh(\xb,\xb^\ast) \approx \sum_{i=1}^{M} C_i X_i(x),
\end{equation}
since the evanescent modes can be neglected for large propagation distances.
Obviously $C_i = \Gh_i(z,\xb^\ast)$, 
where $\Gh_i(\cdot,\xb^\ast)$ is defined in \cref{eq:GRn}.

In the full-aperture array case the orthonormality of the $X_n$'s immediately implies that
$\widehat{\mathbb{Q}}_{nm} = C_n C_m$, see also \cref{eq:Qnm}.

If the array has partial aperture, then \cref{eq:Smatrx} implies that
\begin{align*}
   \widehat{\mathbb S}_{nm} &= \frac{1}{\nu_n \nu_m}\,\int_{\mathcal A} \int_{\mathcal A} 
   \Gh(\xb^\ast,\xb_s)\, \Gh(\xb_r,\xb^\ast) \, s_n(x_s)\, s_m(x_r)\, dx_s dx_r \\
   &= \frac{1}{\nu_n \nu_m}\, \sum_{k=1}^{M} \sum_{l=1}^{M} C_k C_l 
        \int_{\mathcal A} s_n(x_s) X_k (x_s) dx_s \int_{\mathcal A} s_m(x_r) X_l (x_r) dx_r \\
   &\stackrel{\cref{eq:SXWrel}}{=}     \sum_{k=1}^{M} C_k w_k^n\sum_{l=1}^{M}  C_l w_l^m . 
\end{align*}
Hence
\begin{align*}
  \widehat{\mathbb{Q}}_{nm} &\stackrel{\cref{eq:Qpartial}}{=}  (W \,\widehat{\mathbb S} \,W^T)_{nm} 
  = \sum_{j=1}^{M} w_n^j \sum_{i=1}^{M} s_{ji} w_m^i
  = \sum_{j=1}^{M} w_n^j \sum_{i=1}^{M}  \sum_{k=1}^{M} \sum_{l=1}^{M} C_k C_l w_k^j w_l^i w_m^i \\
  &= \sum_{k=1}^{M} \sum_{l=1}^{M} C_k C_l  \sum_{j=1}^{M} w_n^j w_k^j \sum_{i=1}^{M} w_l^i w_m^i 
    =  \sum_{k=1}^{M} \sum_{l=1}^{M} C_k C_l  (W W^T)_{nk} (W W^T)_{lm} \\
  &= C_n C_m,
\end{align*}
where the last equality holds since $W$ is orthogonal. \hfill $\Box$

\begin{remark} \label{remark2}~\\[-5pt]
\begin{enumerate}
\item[(a)] An analogous result to that stated in  \cref{pr:PartialPointScatt} is expected to hold also for extended scatterers 
              under the linearized Born approximation.
\item[(b)] \Cref{pr:PartialPointScatt} essentially  says that when the array has partial aperture and we construct $\widehat{\mathbb{Q}}$ 
by means of   \cref{eq:Smatrx,eq:Qpartial}, then we do not lose any information and the resulting image is expected to be as
good as if the array was spanning the whole vertical cross-section of the waveguide. Of course, this holds ideally assuming, 
e.g., that the array is continuous and that all necessary computations are performed `exactly', i.e. with infinite precision.
However, our analysis in \cite{TMP_16} suggests that the minimum eigenvalue $\nu_{\rm min}$ of $\Aarr$ decays to zero
as the length of the array decreases. As a result, numerical instabilities occur when the length of the array is decreased so that 
some of the smaller in magnitude eigenvalues drop below some small-valued threshold $\epsilon$, which may depend on the 
noise level in the data or on  the machine epsilon in the noiseless case. In such case, we propose to filter the matrix 
$\widehat{\mathbb S}$ by setting $1/\nu_i = 0$ for those indices $i$ that correspond to eigenvalues $\nu_i$ that 
satisfy $\nu_i < \epsilon$. For the details we refer to \cite{TMP_16}.
\item[(c)]In the case of reflectors that are in the vicinity of the array we 
may include a number of evanescent terms in the expansion  \cref{eq:Greensexp} and adjust appropriately the size of the matrices $\Aarr$, 
$\widehat{\mathbb S}$, and $\widehat{\mathbb Q}$. 
\end{enumerate}
\end{remark}

In the case of passive imaging with a partial-aperture array our methodology is modified as follows:
We first construct the vector $\widehat{\mathbb S}$ with entries 
\begin{equation}    \label{eq:Svect} 
    \widehat{\mathbb S}_n = \frac{1}{\nu_n }\,\int_{\mathcal A}  \widehat\Pi(\xb_r) \,  s_n(x_r)\,  dx_r,\ \  
    n = 1,\ldots,M,
\end{equation} 
and then we define the vector 
\begin{equation}    \label{eq:Qvect} 
     \widehat{\mathbb Q} =  W \widehat{\mathbb S},
\end{equation}      
where the matrix $W$ is as before. It is straightforward to show that the projected array response vector 
$\widehat{\mathbb{Q}}$ for a continuous array that spans the whole vertical cross-section $[0,D]$ is equal 
to the vector defined by \cref{eq:Svect,eq:Qvect} in the case of a  partial-aperture array.

We conclude this section by proposing the following imaging algorithms for imaging one or more extended sources or scatterers 
located in $\Omega_{L^+}$.

\begin{algor}[Passive imaging]     \label{alg:Passive}
\begin{enumerate}
\item[(a)] Given the $N\times 1$ array response vector $\widehat{\Pi}$ we compute the $M\times 1$ projected 
         vector $\widehat{\mathbb{Q}}$ by means of \cref{eq:Svect,eq:Qvect} .
\item[(b)] Next, we compute the imaging functional $\Ip$ given in (\ref{eq:Ip_gen}) for each point of a predefined
         search domain $S$ and we display graphically the modulus of these values.
\end{enumerate}
\end{algor}

\begin{algor}[Active imaging]      \label{alg:Active}
\begin{enumerate}
\item[(a)] Given the $N\times N$ array response matrix $\widehat{\Pi}$ we compute the $M\times M$ projected 
         matrix $\widehat{\mathbb{Q}}$ by means of \cref{eq:Smatrx,eq:Qpartial}.
\item[(b)] Next, we compute the imaging functional $\Ia$ given in (\ref{eq:Ia}) for each point of a predefined
         search domain $S$ and we display graphically the modulus of these values.
\end{enumerate}
\end{algor}
%
%
%
%
\section{Resolution analysis}    \label{sec:Resol}
In this section we present a detailed resolution analysis for the imaging functionals
$\Ip$ and $\Ia$ defined in \cref{eq:Ip_gen} and \cref{eq:Ia}, respectively.
As usual, this amounts in studying the behaviour of the point spread function (PSF)
which is the  imaging functional for a point source (passive case)
or a point scatterer (active case). In fact, we are going to examine only the 
case of a point source since the results of the previous section ensure that the 
PSF for a point scatterer is just the square of the PSF for a point source.

Specifically, we restrict ourselves in the simple case of a homogeneous waveguide 
($\eta(\xb) = 1$) which forms the semi-infinite strip $(-\infty,R) \times (0,D)$. The Green's function in this waveguide,
hereafter denoted by $\GR$, may be found analytically; the derivation is given in \cref{sec:GMI}. 
We have that for each $\ybs = (z,x) \in\Omega$,
\begin{equation}    \label{eq:green_split}
  \GR(\yb^s,\xb_s) 
   = \left\{ \begin{array}{lc}
                  \displaystyle  \sum_{m=1}^\infty \frac{1}{\beta_m}\,
                  \rme^{i\beta_m (R - z_s)}  \, \sin\beta_m(R - z) \, X_m(x) \, X_m(x_s), &  z > z_s \\
                  \displaystyle  \sum_{m=1}^\infty \frac{1}{\beta_m}\,
                  \rme^{i\beta_m (R - z)}  \, \sin\beta_m(R - z_s) \, X_m(x) \, X_m(x_s), & z < z_s   
    \end{array} \right.  ,                    
\end{equation}
where the point source is located at  $\xb_s = (z_s , x_s)$, the vertical eigenpairs
$(\mu_n,X_n)$ are equal to
\begin{equation}   \label{eq:eigs}
  \mu_n = (n\pi / D)^2, \quad X_n(x) = \sqrt{2/D} \sin (\sqrt{\mu_n}x),
  \ n = 1,2,\ldots,
\end{equation}
and the horizontal wavenumbers $\beta_n$ are defined in \cref{eq:betas}.

Then, as \cref{eq:Ip_Gipart} suggests, the PSF for a point source is
\begin{align}
   \Ip(\ybs) &= \ipart \GR(\ybs,\xb_s) 
   = \sum_{n=1}^M \frac{1}{\beta_n} \sin{\beta_n (R - z_s)}  \, \sin\beta_n(R - z) \, X_n(x) \, X_n(x_s)  \nonumber\\
  &= \frac{1}{2} \sum_{n=1}^M \frac{1}{\beta_n} 
        \Big(\cos{\beta_n (z - z_s)}  - \cos\beta_n(2R - z - z_s)\Big) \, X_n(x) \, X_n(x_s).   \label{eq:PSF_passive} 
\end{align}

\subsection{Single frequency} 
The analysis in this subsection will be carried out for a monochromatic source. 
The following two propositions provide analytical estimates of the PSF when we fix range or cross-range to
that of the point source and look at a cross-section in the other direction.
%
%
\begin{proposition}[Cross-range resolution]   \label{pr:crr}
Assume that  the search point is located at the correct range, \textit{i.e.}, $\ybs = (z_s,x)$. Then
\begin{equation}    \label{eq:Ip_form_cr}
     \Ip(z_s,x) \approx \case{1}{4}\left[ \big( J_0(\alpha_x) - J_0(\beta_x) \big)  - 
                                \big( J_0(\sqrt{\alpha_x^2 + \gamma_x^2}) -  J_0(\sqrt{\beta_x^2 + \gamma_x^2}) \big)\right],     
\end{equation}   
where 
\begin{equation}      \label{eq:coef1}
      \alpha_x = \frac{2\pi (x - x_s)}{\lambda}, \quad \beta_x = \frac{2\pi (x + x_s)}{\lambda}, \quad 
      \gamma_x = \frac{4\pi}{\lambda}(R - z_s) .
\end{equation}
\end{proposition}
{\it Proof}. For $\ybs = (z_s,x)$, and in view of \cref{eq:eigs}, \cref{eq:PSF_passive} becomes
\begin{equation}   \label{eq:Ip_sum_cr} 
   \Ip(z_s,x) =   \frac{1}{D} \sum_{n=1}^M \frac{1}{\beta_n} 
        \Big( 1  - \cos(2\beta_n(R - z_s))\Big)  \, \sin \frac{n\pi x}{D}\sin \frac{n\pi x_s}{D}.     
\end{equation}

Letting  $\xi_n = {n \lambda}/{(2D)}$ we may view the right-hand side of \cref{eq:Ip_sum_cr} as a 
Riemann sum approximation of  the integral 
$$
   \frac{1}{\pi} \int_0^1 \frac{1}{\sqrt{1-\xi_n^2}} 
                   \Big( 1  - \cos\big(\frac{4\pi}{\lambda}(R - z_s)\sqrt{1-\xi_n^2}\big)\Big)     
                  \sin \Big (\frac{2\pi x}{\lambda} \xi_n\Big)\, \sin \Big(\frac{2\pi x_s}{\lambda} \xi_n\Big) d\xi_n.
$$
Hence, using the simple trigonometric identity $\sin A \sin B = \case{1}{2}(\cos(A-B) - \cos(A+B))$ we may
approximate $\Ip$ as:
\begin{align*}
   \Ip(x) &\approx \frac{1}{\pi} \int_0^1 \frac{1}{\sqrt{1-\xi_n^2}} 
                   \Big( 1  - \cos\big(\frac{4\pi}{\lambda}(R - z_s)\sqrt{1-\xi_n^2}\big)\Big) \\
            &\phantom{\int_0^1 888}
                  \times \frac{1}{2}\Big(\cos\big (\frac{2\pi (x - x_s)}{\lambda} \xi_n\big) - 
                                            \cos\big (\frac{2\pi (x + x_s)}{\lambda} \xi_n\big)\Big)\,  d\xi_n,
\end{align*}  
where we have slightly extended the notation and used here $\Ip$ as a function of a single variable (cross-range).
Now, with $\alpha_x,\beta_x$ and $\gamma_x$ given by (\ref{eq:coef1}), $\Ip$ can be written as
\begin{align}  \label{eq:Ip_int_cr}
  \Ip(x) &\approx  \frac{1}{2\pi} \int_0^1 \frac{1}{\sqrt{1-\xi_n^2}} 
  \, \Big( 1  - \cos\big(\gamma_x\sqrt{1-\xi_n^2}\big)\Big) \Big(\cos(\alpha_x\xi_n) - \cos(\beta_x \xi_n)\Big)\,  d\xi_n \nonumber\\
  &=  \frac{1}{2\pi} \int_0^1 \frac{1}{\sqrt{1-\xi_n^2}} \cos(\alpha_x\xi_n) \, d\xi_n 
    -      \frac{1}{2\pi} \int_0^1 \frac{1}{\sqrt{1-\xi_n^2}} \cos(\beta_x\xi_n) \, d\xi_n \nonumber\\
  &+ \frac{1}{2\pi}\int_0^1 \frac{1}{\sqrt{1-\xi_n^2}}\, \cos(\alpha_x\xi_n) \cos\big(\gamma_x\sqrt{1-\xi_n^2}\big)\,  d\xi_n\nonumber\\
  &- \frac{1}{2\pi}\int_0^1 \frac{1}{\sqrt{1-\xi_n^2}}\, \cos(\beta_x\xi_n) \cos\big(\gamma_x\sqrt{1-\xi_n^2}\big)\,  d\xi_n \nonumber\\
  &=: I_1 - I_2 + I_3 - I_4. 
\end{align}
The integrals $I_i$, $i = 1,\ldots,4$, in (\ref{eq:Ip_int_cr}) may be evaluated analytically. We look at each term separately. 
For example, it is known,  \cite[(3.753.2)]{gradshteyn2007}, that
$$
    \int_0^1 \frac{1}{\sqrt{1-\xi_n^2}}\, \cos(\alpha_x\xi_n)\,  d\xi_n = \frac{\pi}{2} J_0(\alpha_x),
$$
where $J_0(\cdot)$ is the Bessel function of the first kind of order 0. Therefore, 
$$
    I_1 = \frac{1}{4}J_0(\alpha_x) \quad\mbox{ and } \quad I_2 = \frac{1}{4}J_0(\beta_x).
$$
Next, in order to evaluate $I_3$ we change variables, letting $\theta = \arcsin \xi_n$, and use the fact that 
$J_{-1/2}(t) = \sqrt{\frac{2}{\pi t}} \cos(t)$ 
(see, e.g., \cite[(10.16.1)]{NIST:DLMF}) to write $I_3$  as
$$
   I_3 = \frac{1}{4}\sqrt{\alpha_x \gamma_x} \int_0^{\pi/2} J_{-1/2}(\alpha_x \sin\theta) J_{-1/2}(\gamma_x \cos\theta)
   (\sin\theta)^{1/2} (\cos\theta)^{1/2} d\theta.
$$
Then, according to \cite[(6.683.2)]{gradshteyn2007}, $I_3 = \frac{1}{4} J_0(\sqrt{\alpha_x^2 + \gamma_x^2})$.
Similarly, $I_4 = \frac{1}{4} J_0(\sqrt{\beta_x^2 + \gamma_x^2})$ and the proof is complete. \hfill $\Box$

\medskip

%
%
\begin{proposition}[Range resolution]   \label{pr:rr}
Assume that  the search point is located at the correct cross-range, \textit{i.e.}, $\ybs = (z,x_s)$. Then
\begin{equation}    \label{eq:Ip_form_r}
     \Ip(z,x_s) \approx \case{1}{4}\left[ \big( J_0(\alpha_z) - J_0(\beta_z) \big)  - 
                                \big( J_0(\sqrt{\alpha_z^2 + \gamma_z^2}) -  J_0(\sqrt{\beta_z^2 + \gamma_z^2}) \big)\right],     
\end{equation}   
where now
\begin{equation}      \label{eq:coef2}
     \alpha_z = \frac{2\pi (z - z_s)}{\lambda}, \quad \beta_z = \frac{2\pi (2R - z - z_s)}{\lambda}, \quad 
     \gamma_z = \frac{4\pi x_s}{\lambda}.
\end{equation}
\end{proposition}
{\it Proof}. Since $\yb^s$ is placed at the correct cross-range we now let $x = x_s$ in (\ref{eq:PSF_passive}). Thus,
by a slight abuse of notation, $\Ip(z)$ as a function of the range variable equals to 
\begin{equation}     \label{eq:Ip_sum_rr} 
   \Ip(z)  = \frac{1}{D} \sum_{n=1}^M \frac{1}{\beta_n} 
        \Big(\cos{\beta_n (z - z_s)}  - \cos\beta_n(2R - z - z_s)\Big) \, \sin^2 \frac{n\pi x_s}{D}.     
\end{equation}
As in the proof of Proposition~\ref{pr:crr}, we let $\xi_n = {n \lambda}/{(2D)}$ and approximate 
the right-hand side of (\ref{eq:Ip_sum_cr}) by an integral. Specifically, if 
$\alpha_z, \beta_z, \gamma_z$ are as in (\ref{eq:coef2}), and if we use that $\sin^2 A = \frac{1}{2}(1 - \cos 2A)$, 
we may deduce that $\Ip(z)$ is approximated as
\begin{align} 
      \Ip(z)   &\approx  \frac{1}{2\pi} \int_0^1 \frac{1}{\sqrt{1-\xi_n^2}} 
                   \big(\cos(\alpha_z)\sqrt{1-\xi_n^2})   -  \cos(\beta_z\sqrt{1-\xi_n^2})\big)   \nonumber \\
                  &\times \big(1 - \cos(\gamma_z \xi_n)\big)\, d\xi_n .                                   \label{eq:Ip_int_rr}               
\end{align} 
The integral of the various terms of \cref{eq:Ip_int_rr} are of the same type as those in \cref{eq:Ip_int_cr}
and they can be evaluated analytically resulting to \cref{eq:Ip_form_r}. \hfill $\Box$

\medskip

\begin{figure}[h]
\begin{center}
\includegraphics[width=0.45\linewidth]{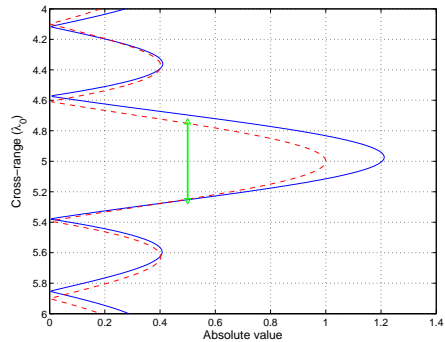} 
\caption{Comparison between \cref{eq:Ip_form_cr}  multiplied by 4 (blue line) and $J_0(a_x)$ (red dashed line), 
for a source located at $(z_s,x_s)=(19, 5)\, \lambda_0$ and for a single frequency corresponding to $k = 0.973k_0$,
where the reference wavenumber is $k_0 = \pi/10$. The green arrowed segment indicates length equal to $\lambda/2$.  }
\label{fig:CR_bess_test}
\end{center}
\end{figure}

A first remark is that the approximate formulas \cref{eq:Ip_form_cr} and \cref{eq:Ip_form_r} for the PSF 
when range or cross-range, respectively, is fixed at the correct location of the point-source suggest that 
the term that mainly contributes in defining the resolution in the vicinity of the source is $J_0(\alpha_x)$
or $J_0(\alpha_z)$, respectively. To illustrate this, in  \cref{fig:CR_bess_test} we superimpose  the graphs of  
\cref{eq:Ip_form_cr} multiplied by 4  (typed in blue) and of  $J_0(a_x)$ (red dashed line), 
for a source located at $(z_s,x_s)=(19, 5)\, \lambda_0$ and for a single frequency corresponding to $k = 0.973k_0$.  
The reference wavenumber is $k_0 = \pi/10$.

Hence, if we define the resolution to be the width of the PSF at its half maximum, it is immediate to 
check that both cross-range and range resolution are approximately equal to $\lambda/2$.

\begin{figure}[h]
\begin{center} 
\includegraphics[width=0.45\linewidth]{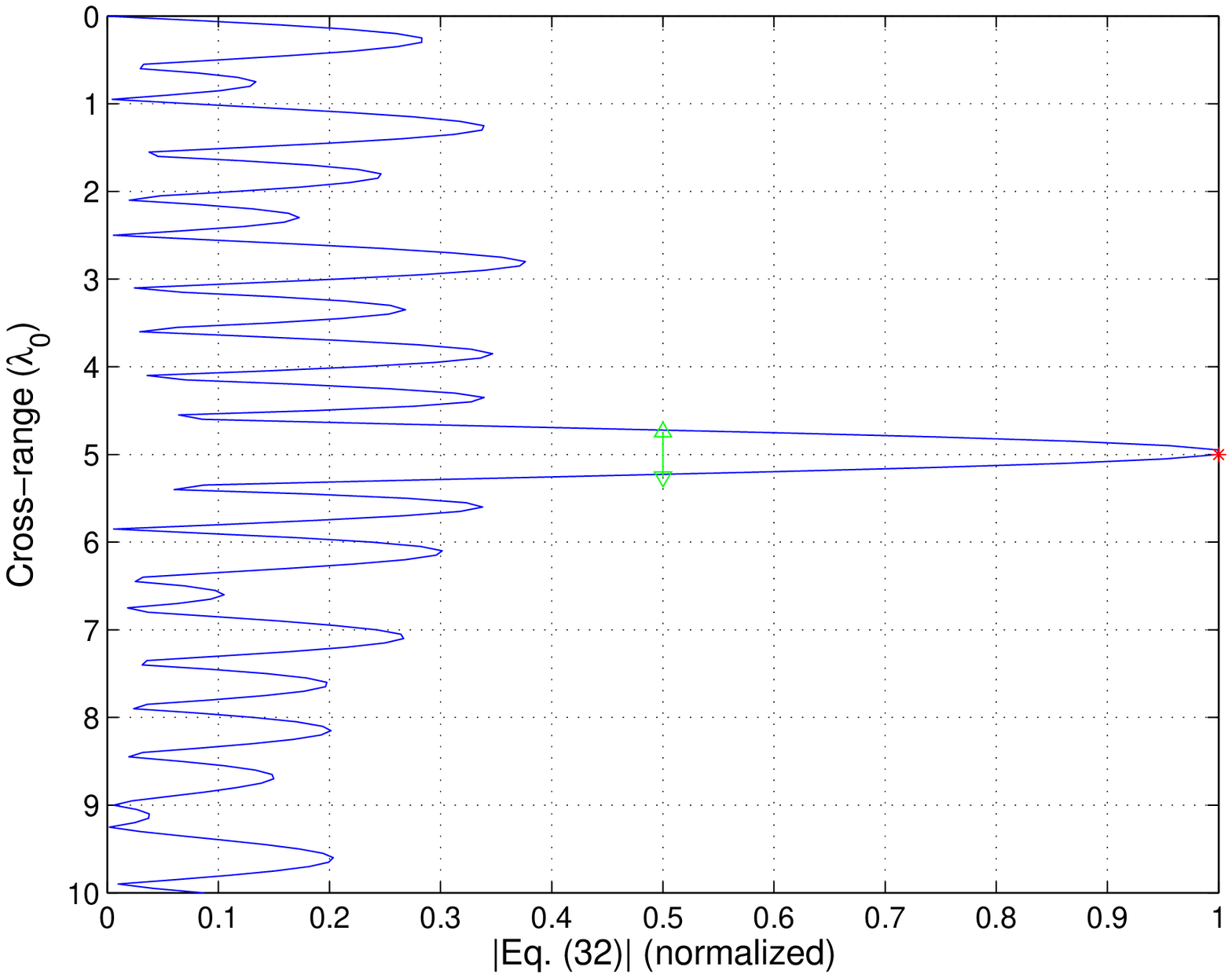} 
\includegraphics[width=0.43\textwidth]{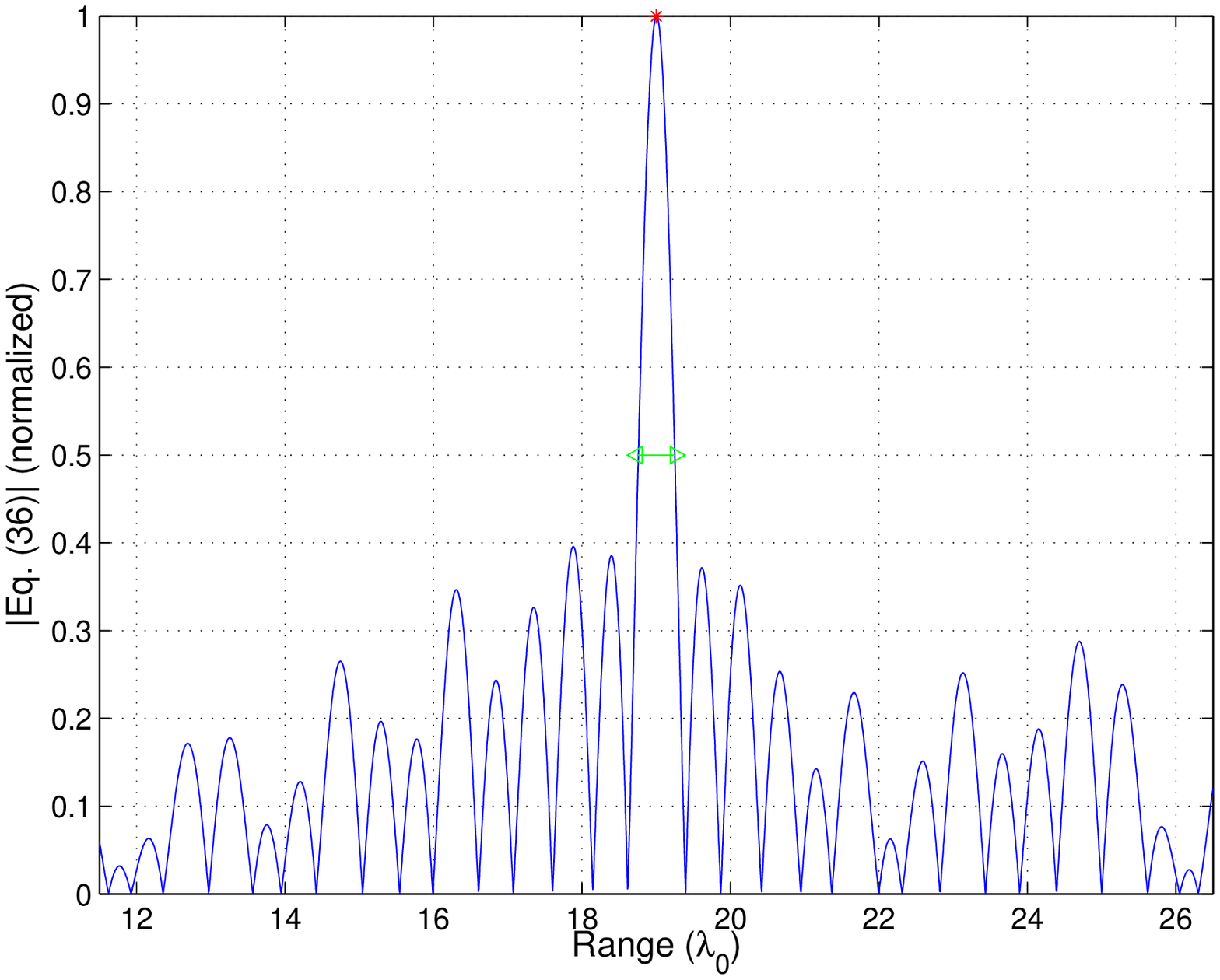}  
\caption{ Normalized absolute value of \cref{eq:Ip_form_cr} versus cross-range (left subplot) and 
              normalized absolute value of \cref{eq:Ip_form_r} versus range (right subplot) 
              for a point source located at $(z_s,x_s)=(19, 5)\, \lambda_0$. 
              Here the reference wavenumber is $k_0 = \pi/10$  and results are shown for a single frequency 
              that corresponds to $k = 0.973 k_0$. The green arrowed segment indicates length equal to $\lambda/2$ 
              and a red asterisk points to the location of the source.}
\label{fig:Res_f73p125}
\end{center}
\end{figure}

Next, we plot in \cref{fig:Res_f73p125}, the absolute values of \cref{eq:Ip_form_cr} (left subplot) 
and \cref{eq:Ip_form_r} (right subplot) for a point source located at $(z_s,x_s)=(19, 5)\, \lambda_0$. 
As before, the reference wavenumber is $k_0 = \pi/10$  and results are shown for a single frequency 
that corresponds to $k = 0.973 k_0$.  The analytical expressions we have derived for the cross-range 
and range resolution capture the behaviour of the imaging functional as we may check by comparing the plots in 
\cref{fig:Res_f73p125} with the two rightmost subplots of \cref{fig:Ip}. 
The images shown in \cref{fig:Res_f73p125,fig:Ip}, 
peak at the right position of $\xb_s$ and as predicted by the theoretical analysis they have a resolution of $\lambda/2$ in both range and cross-range directions. We observe however that they are quite oscillatory and their SNR is not
very satisfactory.

%
%
\subsection{Multiple frequencies} \label{sec:mf}
In this subsection we will show that the SNR in our images can be significantly improved using multiple frequencies. For most practical purposes this is something feasible since in many applications
sources are not monochromatic but they rather emit pulses. The multiple frequency version of the imaging functional defined in \cref{eq:Ip_gen} is simply the summation over frequencies of the corresponding monochromatic one
\begin{equation}  \label{eq:Ip_genf}
   \Ip(\yb^s) :=  \sum_{l=1}^{N_f} \Ip(\yb^s;f_l) = \sum_{l=1}^{N_f} \sum_{n=1}^{M_l} \beta_n(f_l) \overline{\widehat{\mathbb{P}}_{n}}(f_l) \, \Gh_n(z_a,\yb^s;f_l)
\end{equation}
where $f_l$, $l=1,\ldots,N_f$ are the discrete frequencies that span the available frequency interval
$[\fmin,\fmax]$ in our data. Note that $M_l$ depends on the index $l$ since the number of propagating modes depends on the frequency $f_l$.  The definition of the corresponding active imaging functional for multiple frequencies follows similarly.  
 
Let us first look at the cross-range direction.
To investigate the PSF behaviour with multiple frequencies in the ideal setting that we have examined thus far, 
we integrate \cref{eq:Ip_form_cr} with respect to frequency $f$ over an interval with bandwidth $B$. 
Specifically, letting $\Psi(x;B)$ denote the PSF for multiple frequencies at the correct range, we have
\begin{align}    \label{eq:int_multifreq}
      \Psi(x;B) &=\int_{\fmin}^{\fmax} \Ip(z_s,x;f) df  \nonumber \\
      &\approx  \case{1}{4}\int_{\fmin}^{\fmax}
                       \left[ \big( J_0(\alpha_x) - J_0(\beta_x) \big)  - 
                                \big( J_0(\sqrt{\alpha_x^2 + \gamma_x^2}) -  J_0(\sqrt{\beta_x^2 + \gamma_x^2}) \big)\right]\,df
            \nonumber\\
     &\approx  \case{1}{4}\int_{\fmin}^{\fmax}  J_0(\alpha_x) \,df ,
\end{align}
where now the parameters $\alpha_x,\beta_x$ and $\gamma_x$ (given in \cref{eq:coef1}) are written in terms of the 
frequency $f$ as
$$
      \alpha_x = \frac{2\pi}{c_0}(x - x_s)f, \quad \beta_x = \frac{2\pi}{c_0}(x + x_s)f, \quad 
      \gamma_x = \frac{4\pi(R - z_s)}{c_0}f ,
$$
where $c_0$ is the constant wave speed, $f_c$ is the central frequency, 
and $[\fmin,\fmax] = [f_c - \frac{B}{2},f_c + \frac{B}{2}]$. Note that we have numerically verified the validity of the last 
approximation in \cref{eq:int_multifreq} at least in the frequency range that we have examined. 
Now, let  $\zeta_x :=  \frac{2\pi}{c_0}(x - x_s)$. Then, \cite[11.1.7]{AbramowitzStegun}, 
\begin{equation}    \label{eq:PSF_multfreq}
   \Psi(x;B) \approx  \case{1}{4}\int_{\fmin}^{\fmax}  J_0(\zeta_x f) \,df 
   = \frac{1}{4\zeta_x} \Big( \Lambda_0\big(\zeta_x \fmax \big) - \Lambda_0\big(\zeta_x \fmin \big)\Big),
\end{equation}
where 

\begin{equation}    \label{eq:Lambda0}
   \Lambda_0(s) := s J_0(s) +\frac{\pi s}{2} \Big(J_1(s) {\bf H}_0(s) - J_0(s) {\bf H}_1(s) \Big),
\end{equation}

$J_n(\cdot)$ is the Bessel function of the first kind of order $n$, and ${\bf H}_n(\cdot)$, is 
the Struve function of order $n$, respectively. 
(For the definition of the Struve function see e.g. \cite[Ch. 12]{AbramowitzStegun}.) 

\begin{figure}[h]
\begin{center}
\includegraphics[width=0.5\linewidth]{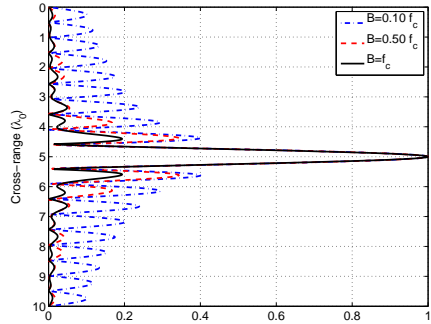}
\caption{Modulus of \cref{eq:PSF_multfreq} for bandwidth equal to  $B=0.10 f_c$ (dash-dot blue line), 
$B = 0.50 f_c$ (dashed red line) and $B = 1.00 f_c$ (solid black line). The point source  is placed at 
$(z_s,x_s) = (19,5)\lambda_0$, the reference wavenumber is $k_0 = \pi/10$ and the central frequency 
$f_c$ corresponds to $k = 0.973k_0$. }
\label{fig:PSF_cr_multifreq}
\end{center}
\end{figure}

Let us consider a specific example. Assume a point source located at  $(z_s,x_s) = (19,5)\lambda_0$, a reference 
wavenumber equal to $k_0 = \pi/10$ and a central frequency $f_c$ corresponding to $k_c= 0.973k_0$.
In \cref{fig:PSF_cr_multifreq} we superimpose the modulus of the right-hand side of (\ref{eq:PSF_multfreq}) for 
three different bandwidths that are equal to $B=10\%, 50\%$ and $100\%$ of the central frequency $f_c$.
All three are normalized with respect to their maximum value which, as may be immediately inferred
from \cref{eq:PSF_multfreq}, is equal to $B/4$. Moreover, we observe that resolution is determined by the 
central frequency while SNR is improved as the bandwidth increases.
Specifically, when $B = 0.10 f_c$ the SNR seems to 
be of the same order as in the single frequency case (compare the blue dashed-dotted line with the 
one shown in the left plot in \cref{fig:Res_f73p125}), it is slightly improved  when $B = 0.50 f_c$
and it is considerably improved by a factor of $2$ for the larger bandwidth $B = f_c$. Let us quantify
these observations. Obviously the global maximum of $|\Psi(x;B)|$ is attained at $x = x_s$ and is equal
to $B/4$. Then, for the bandwidths considered in the example referring to \cref{fig:PSF_cr_multifreq}, SNR is determined
as the ratio of the maximum value to the second taller peak; assume that the latter is attained at $\rho(B)$.
Hence \cref{eq:PSF_multfreq} and \cref{eq:Lambda0} imply that $\zeta_\rho := \frac{2\pi}{c_0}(\rho(B) - x_s)$ 
satisfies the following equation
\begin{eqnarray}     \label{eq:extr}
 &&\fmax \Big(J_1(\zeta_\rho\fmax) {\bf H}_0(\zeta_\rho\fmax) - J_0(\zeta_\rho\fmax) {\bf H}_1(\zeta_\rho\fmax) \Big) \nonumber\\
 && -  \fmin \Big(J_1(\zeta_\rho\fmin) {\bf H}_0(\zeta_\rho\fmin) - J_0(\zeta_\rho\fmin) {\bf H}_1(\zeta_\rho\fmin) \Big) = 0.
\end{eqnarray}
Moreover, it is immediate to check that since $\zeta_\rho$ is a root of (\ref{eq:extr}) then 
$$
    \Psi(\rho(B);B) =  \frac{1}{4}\big(\fmax J_0(\zeta_\rho\fmax) - \fmin J_0(\zeta_\rho\fmin)\big),
$$
hence ${\rm SNR} = B/|\fmax J_0(\zeta_\rho\fmax) - \fmin J_0(\zeta_\rho\fmin)|$.

We compute numerically $\rho(B)$ for the various bandwidths reported above and our results
are summarized in \cref{tbl:SNR_cr}.

\begin{table}[h]
\caption{SNR in cross-range for various bandwidths.}
\label{tbl:SNR_cr}
\begin{center}
\begin{tabular}{ccc} \hline
   $B$            &  $\rho(B)$                    & SNR       \\ \hline
   $0.10 f_c$ & $5.62521 \lambda_0$ & $2.4981$ \\
   $0.50 f_c$ & $5.61865 \lambda_0$ & $2.9097$ \\
   $1.00 f_c$ & $5.59433 \lambda_0$ & $5.1513$ \\ \hline
\end{tabular}
\end{center}
\end{table}

The situation in the range direction is completely similar so we do not present it here.

\begin{figure}[h]
\begin{center}
\includegraphics[width=0.32\textwidth]{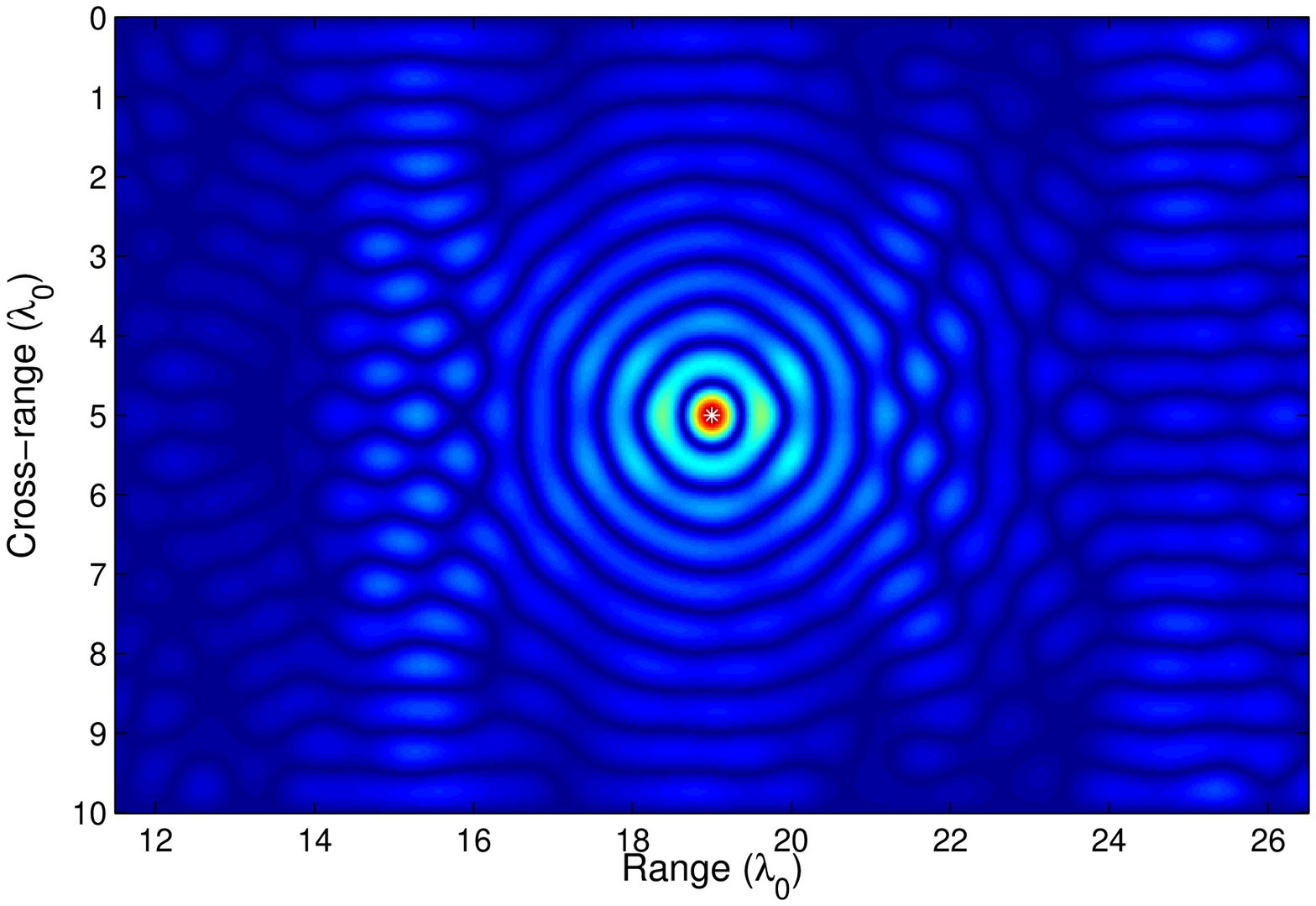} 
\includegraphics[width=0.32\textwidth]{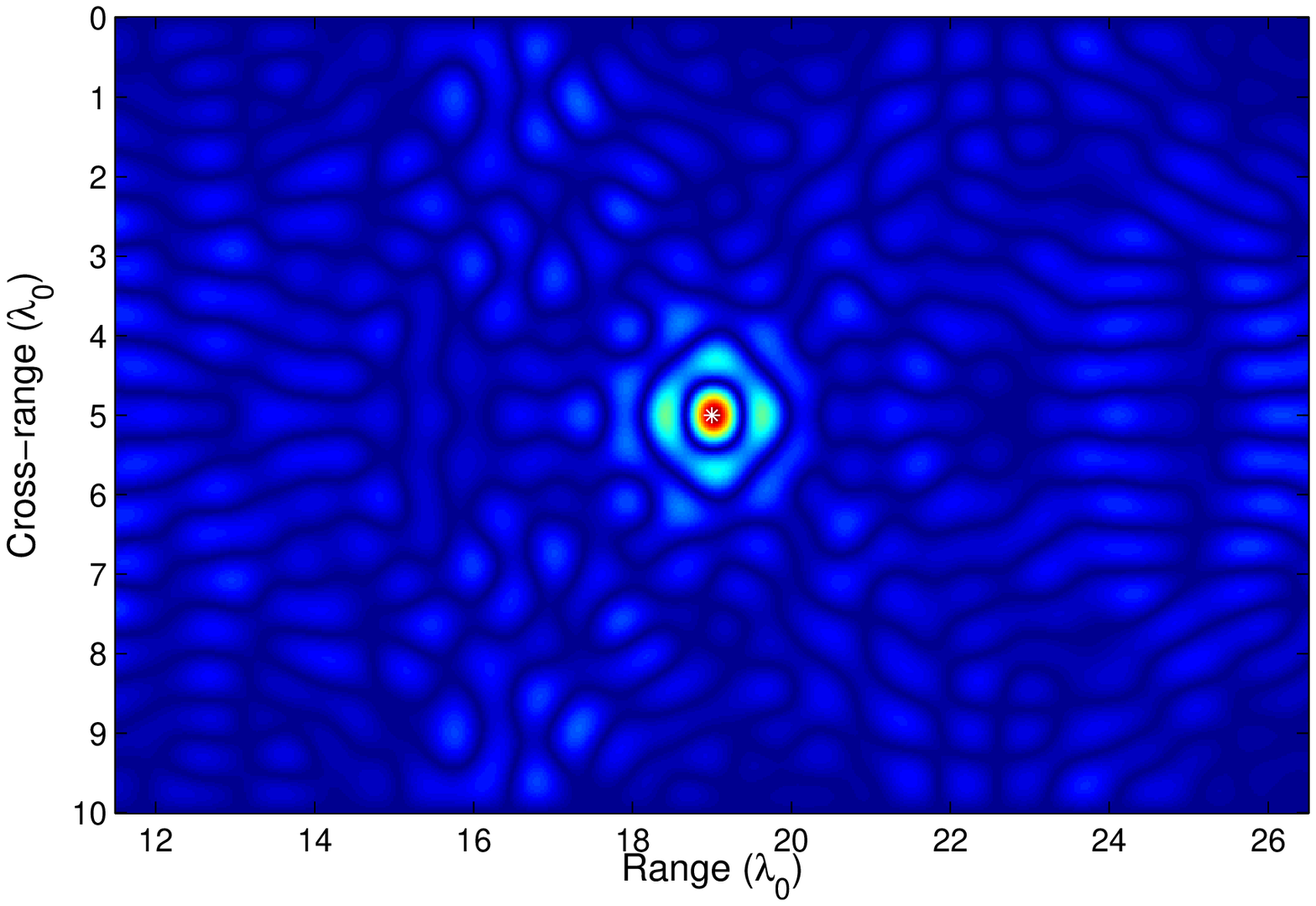} 
\includegraphics[width=0.32\textwidth]{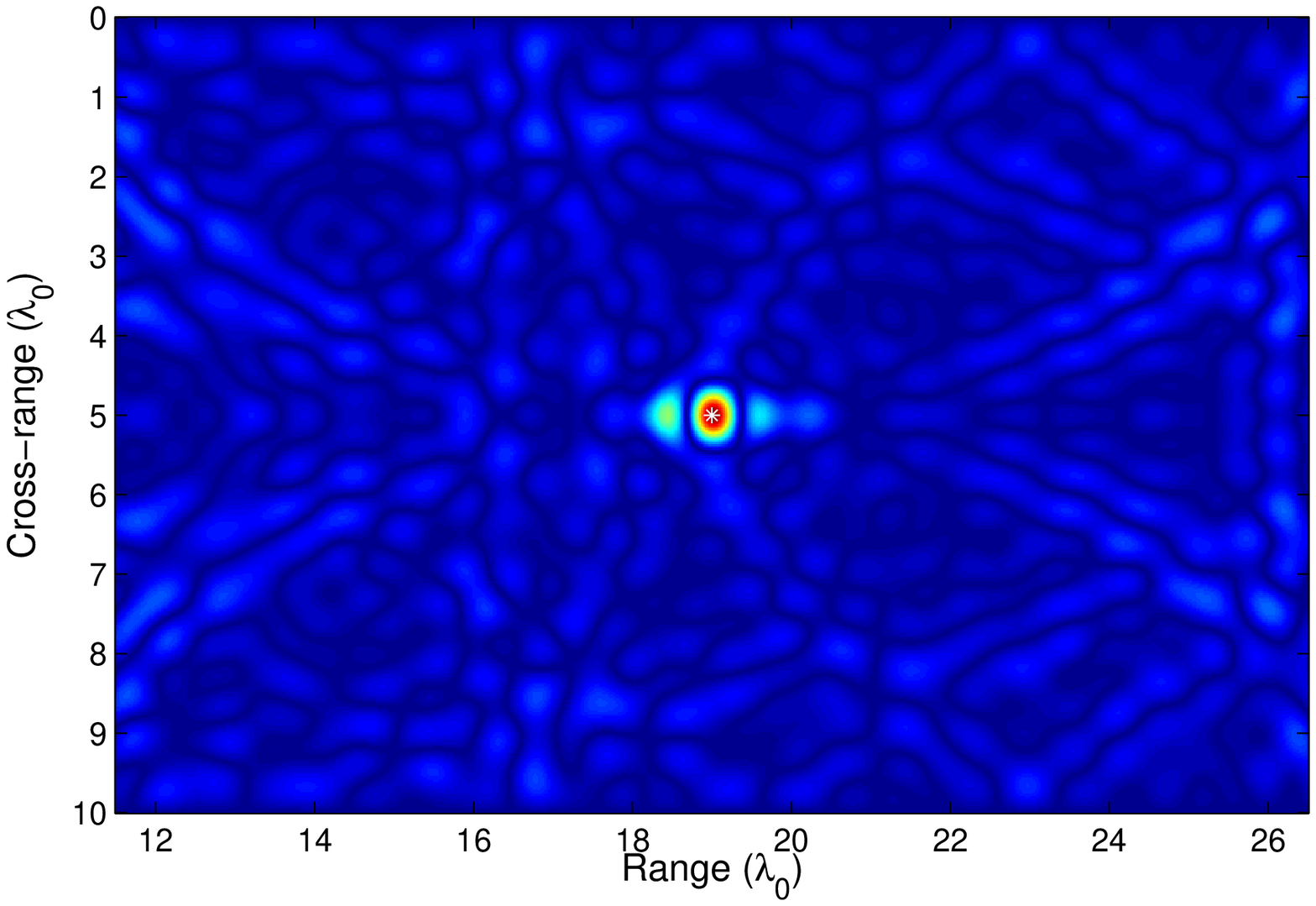} 
\caption{Imaging with $\Ip$ for multiple frequencies for a point source placed at $(z_s,x_s) = (19,5)\lambda_0$.
The reference wavenumber equals $k_0 = \pi/10$ and the central frequency $f_c$ corresponds to $k_c = 0.973k_0$.
Left image: Bandwidth $B = 0.15 f_c$, Middle image: $B = 0.51 f_c$, Right image: $B = 0.92 f_c$. }
\label{fig:Ip_2D}
\end{center}
\end{figure}

We observe the same behaviour when we work with the actual imaging functional $\Ip$.
For example, in \cref{fig:Ip_2D} we plot the modulus of $\Ip(\ybs)$ for a point source
located (as before) at  $(z_s,x_s) = (19,5)\lambda_0$, a reference wavenumber equal to $k_0 = \pi/10$ 
and a central frequency $f_c$ that corresponds to $k_c= 0.973k_0$. The left image is obtained 
when the bandwidth $B \approx 0.15 f_c$, in the middle one $B\approx 0.51 f_c$, and the one on the
right corresponds to $B=0.92 f_c$. 
The advantage of using multiple frequencies is evident when we 
compare these images with the one shown in the left plot of \cref{fig:Ip}. Moreover, 
using a bandwidth of the same order as the central frequency greatly improves the SNR in the image.

\begin{figure}[h]
\begin{center}
\includegraphics[width=0.32\textwidth]{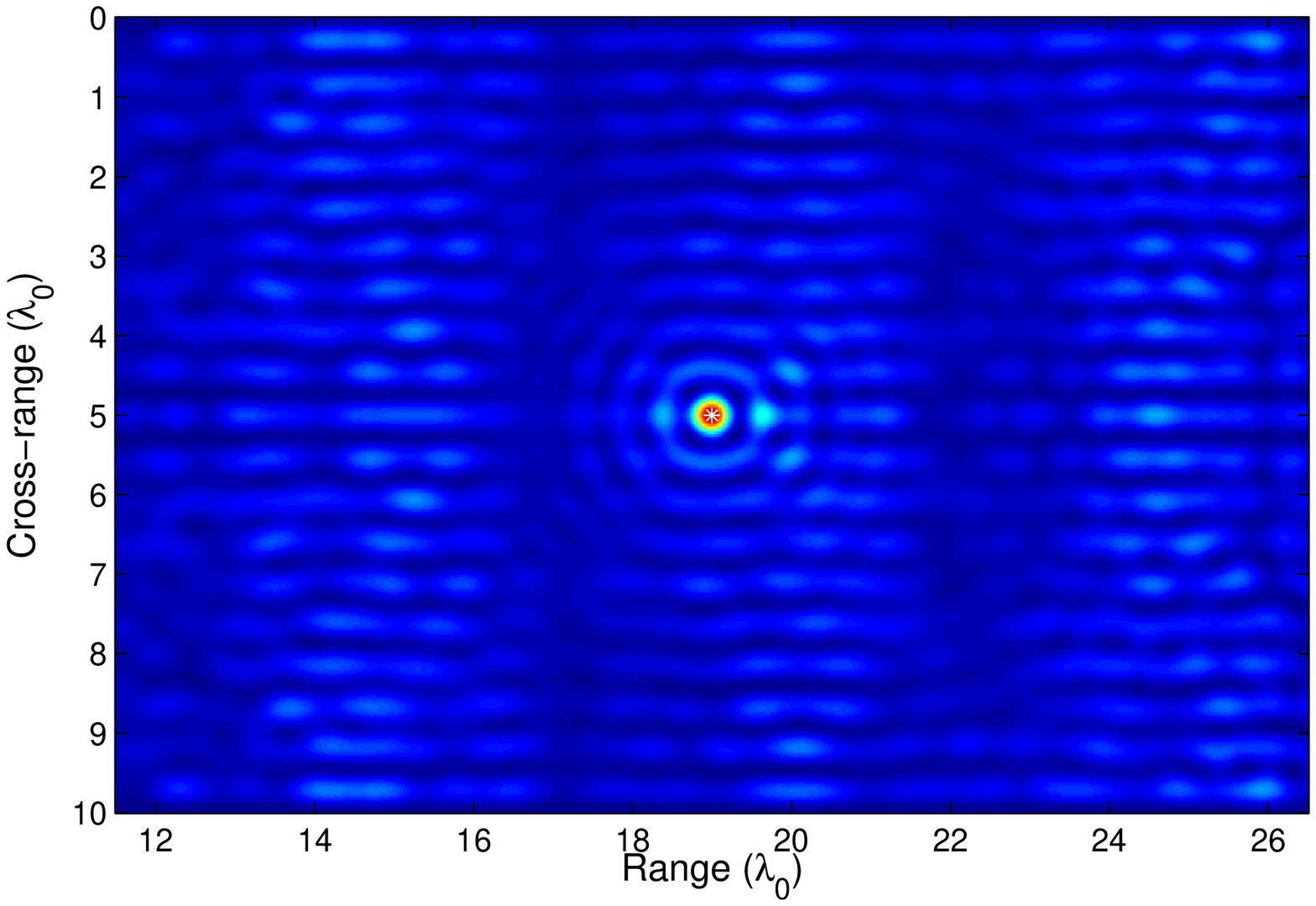} 
\includegraphics[width=0.32\textwidth]{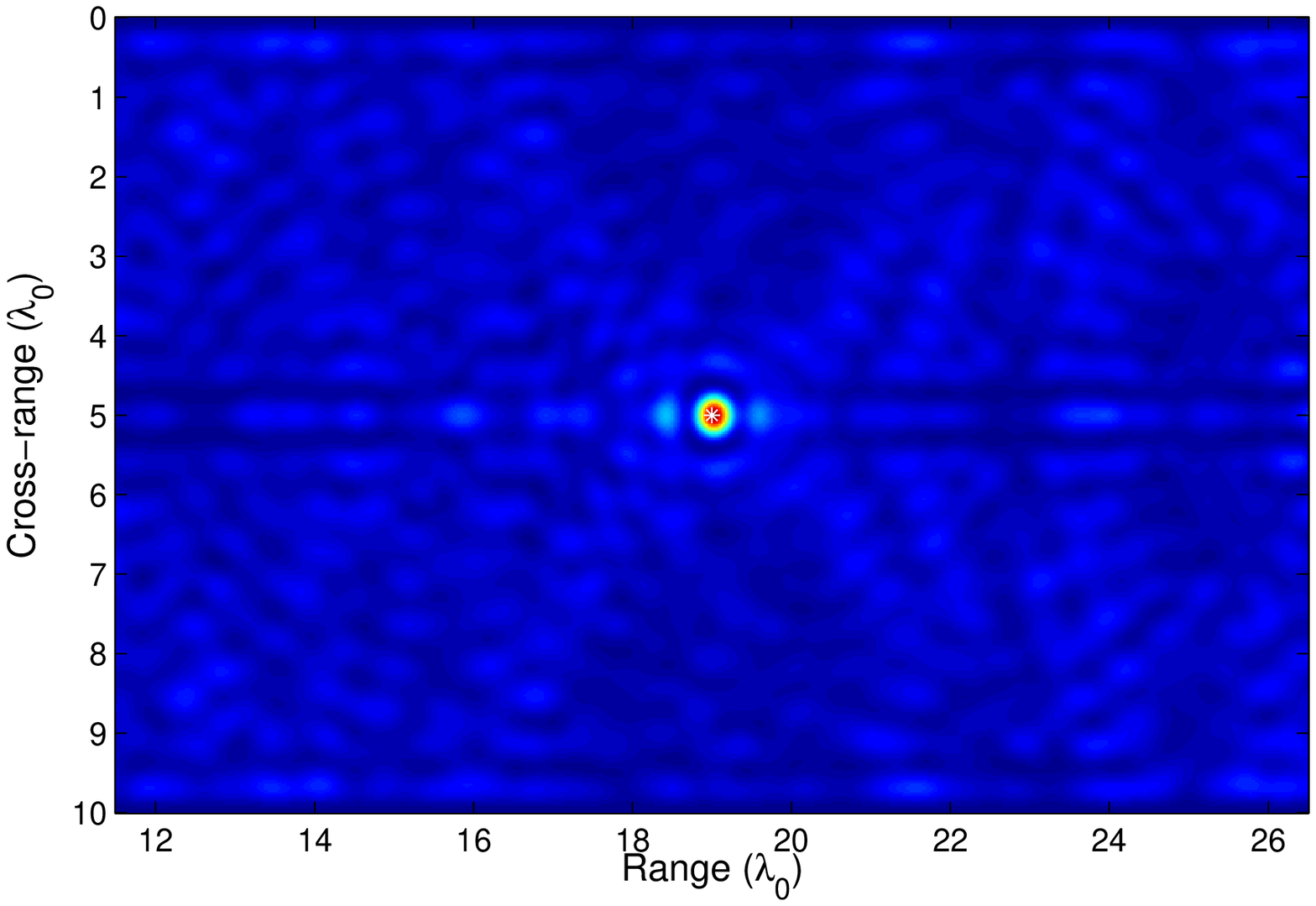} 
\includegraphics[width=0.32\textwidth]{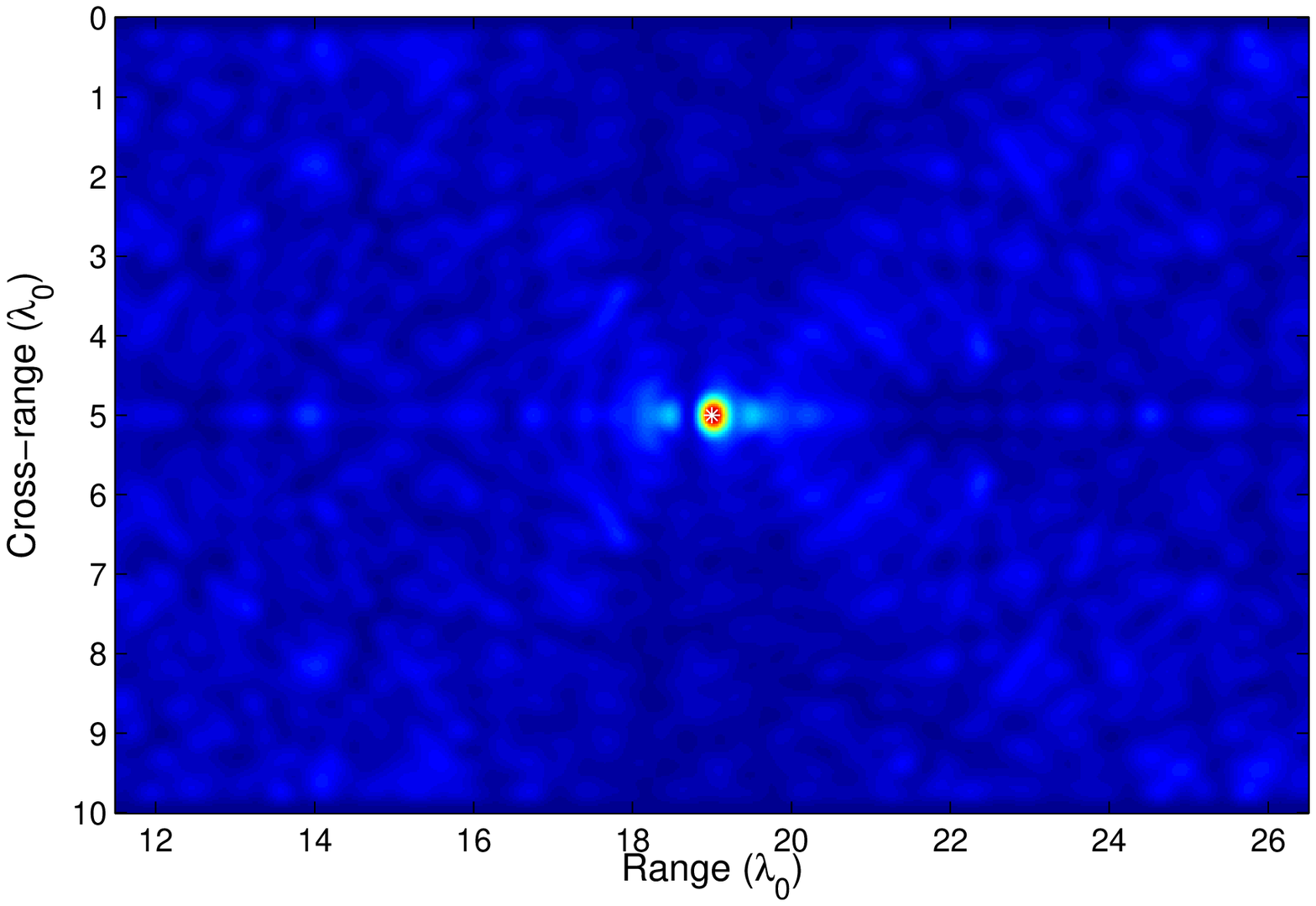} 
\caption{Imaging with $\Ia$ for multiple frequencies for a point scatterer placed at $(z_s,x_s) = (19,5)\lambda_0$.
The central frequency $f_c$ is the same as in \cref{fig:Ip_2D}.
Left image: Bandwidth $B = 0.15 f_c$, Middle image: $B = 0.51 f_c$, Right image: $B = 0.92 f_c$. }
\label{fig:Ia_2D}
\end{center}
\end{figure}

Finally, in \cref{fig:Ia_2D}, we plot the $(\ipart \GR(\ybs;
\xb_s))^2$ which, as (\ref{eq:IaHK}) suggests, is equal to  $\Ia(\ybs)$.
We examine the same cases as in \cref{fig:Ip_2D} and, while the noise levels are lower 
even at the single frequency case (compare with the left plot in \cref{fig:Ia}), 
again we have a clear SNR improvement  as the  bandwidth increases. As we will see in the next section, 
this effect is of greater importance when one deals with extended scatterers.

To summarize, in this section we  derived analytical formulas that approximate the PSF for a point source 
in cross-range and range. We have concluded that both range  and cross-range resolution equal to 
$\lambda/2$ in the monochromatic case. 
Moreover, we addressed the improvement in SNR that brings upon the images the use 
of multiple frequencies,  
and we have shown that the resolution in the multiple frequency case is
$\lambda_c/2$, where $\lambda_c$ is the wavelength that corresponds to the central frequency 
of the available bandwidth. Let us note that the resolution analysis carries over to the partial aperture case at least for array apertures such that the minimum eigenvalue $\nu_{\rm min}$ of $\Aarr$  is larger than $\epsilon$ (see \cref{remark2}).

%
%
%
%
%
\section{Numerical experiments}     \label{sec:Numer}
In this section we focus on the active imaging case and assess the performance of $\Ia$ for imaging extended reflectors in terminating waveguides. We start with a model problem for which the scattered data are computed using the linearized Born approximation and then consider several extended reflectors for which 
the scattered data are computed by solving the full wave equation.  In all cases we will show imaging results obtained 
using $\Ia$ with multiple frequencies.   We first show numerical results for a full aperture array and then consider the more challenging case of a partial aperture array.
\subsection{Linearized Born scattered data}
We consider a one-dimensional scatterer  $\mathcal{C}$, which is a semicircle placed in a homogeneous waveguide with flat horizontal boundaries and a vertical terminating boundary at $z=R$, i.e. $\Omega = (-\infty,R)\times (0,D)$. The response matrix is computed using the Born approximation and is given by
\begin{equation}     \label{eq:Pisec5}
   \widehat{\Pi}(\xb_s,\xb_r;\omega) = \int_\mathcal{C} \GR(\xb,\xb_s;\omega)\GR(\xb_r,\xb;\omega) d\xb,
\end{equation}
with $\GR(\xb,\yb;\omega)$ as in  \cref{eq:green_split}. Note that in \cref{eq:Pisec5} we have suppressed the 
multiplicative factor $k^2$ that usually appears in its right-hand side. As already mentioned in \cref{sec:active}, this can be performed 
in practice by rescaling the data matrix $\widehat{\Pi}(\xb_s,\xb_r;\omega)$ as $k^{-2}\widehat{\Pi}(\xb_s,\xb_r;\omega)$.    

Recall also that our imaging functional $\Ia$ is given by
\begin{equation} \label{eq:Iamf}
   \Ia(\ybs) = \sum_{l=1}^{N_f} \sum_{n=1}^{M_l}\sum_{m=1}^{M_l}  \beta_n(f_l) \beta_m(f_l) \ov{\widehat{\mathbb{Q}}_{nm}(f_l)} \, \GRn(z_a,\ybs;f_l)\, \GRm(z_a,\ybs;f_l),
\end{equation}
where $\widehat{\mathbb{Q}}$ is the projected array response matrix (see \cref{eq:Smatrx,eq:Qpartial}) and 
$\GRn$, $n=1,\ldots,M_l$ is the projection of the Green's function on the first 
$M_l$ vertical eigenfunctions, cf. \cref{eq:GRn}. 
To demonstrate the effect of the terminating boundary of the waveguide on imaging, we compare the results obtained 
when the same reflector is placed in a terminating and in an open-ended (infinite-strip) waveguide. For both the open ended 
and the terminating waveguide the array is placed at $z_a=0$ and spans the whole vertical cross-section of the 
waveguide. The semicircular scatterer $\mathcal{C}$ is centered at $(z^*,x^*) = (19,5)\lambda_0$ with 
diameter $b=2\lambda_0$ and we use frequencies $f \in [f_c-B/2, f_c+B/2]$, 
where the central frequency $f_c$ corresponds to the wavenumber $k_c=0.975 k_0$, the reference wavenumber, as before, equals $k_0=\pi/10$, and the bandwidth is equal to $B=0.92 f_c$. For the terminating waveguide, the vertical boundary is placed at $R=27.5 \lambda_0$. 

To compute the data and the image for the open ended waveguide, we simply replace $\GR(\yb,\xb_s;\omega)$  
in \cref{eq:Pisec5,eq:Iamf} by the Green's function for the infinite waveguide, hereafter denoted by
$\Go(\yb,\xb_s;\omega)$. Recall that  $\Go$  is given by (see e.g. \cite{JKPS_04})
%
\begin{equation} \label{eq:Gopen}
   \Go(\yb,\xb_s;\omega) 
   =   \displaystyle  \frac{ i}{2} \sum_{m=1}^\infty \frac{1}{\beta_m}\,
                  \rme^{i\beta_m | z - z_s| }  \, X_m(x) \, X_m(x_s),                    
\end{equation}
where $\yb = (z,x) \in\Omega$ and  $\xb_s = (z_s , x_s)$, the vertical eigenpairs
$(\mu_n,X_n)$ are as in \cref{eq:eigs}, and the horizontal wavenumbers $\beta_n$ are defined in \cref{eq:betas}.  

\begin{figure}[ht]
 \begin{minipage}{0.49\linewidth}
\begin{center}
$B=0.51f_c$ \\
\begin{minipage}{.45\linewidth}
\begin{center}
infinite \\
\includegraphics[width=\linewidth]{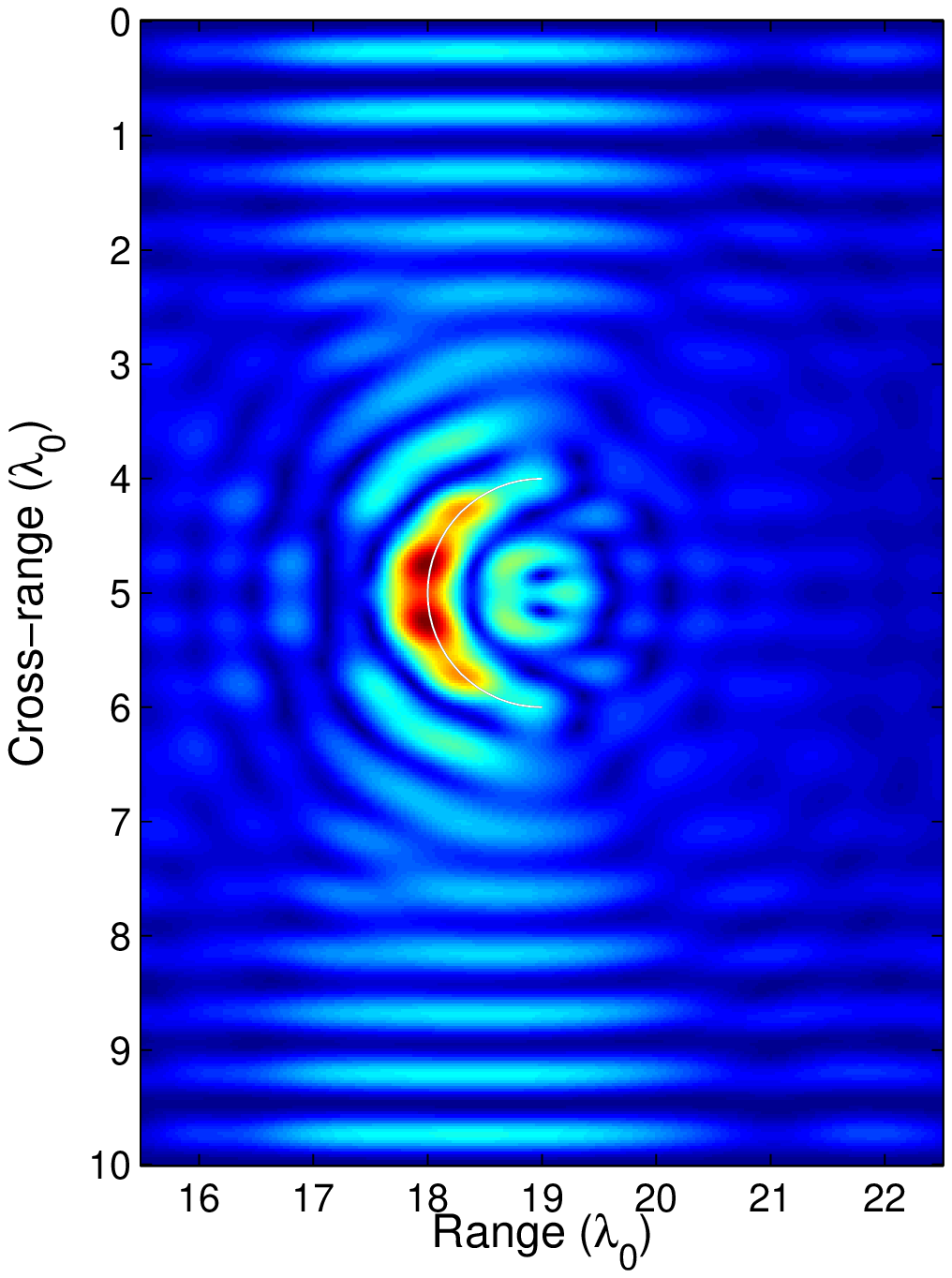} \\
\includegraphics[width=\linewidth]{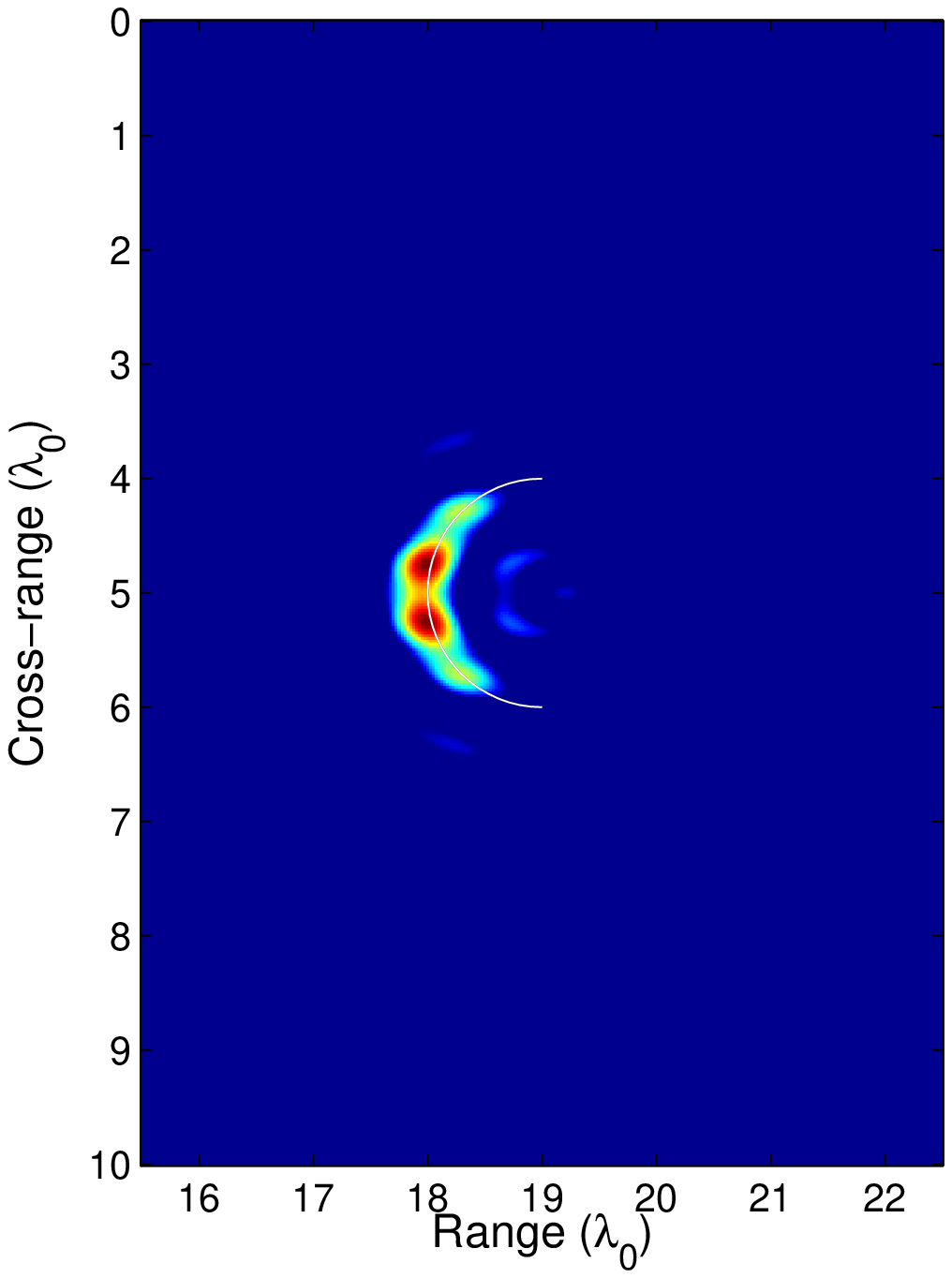}
\end{center}
\end{minipage}
\begin{minipage}{.45\linewidth}
\begin{center}
terminating \\
\includegraphics[width=\linewidth]{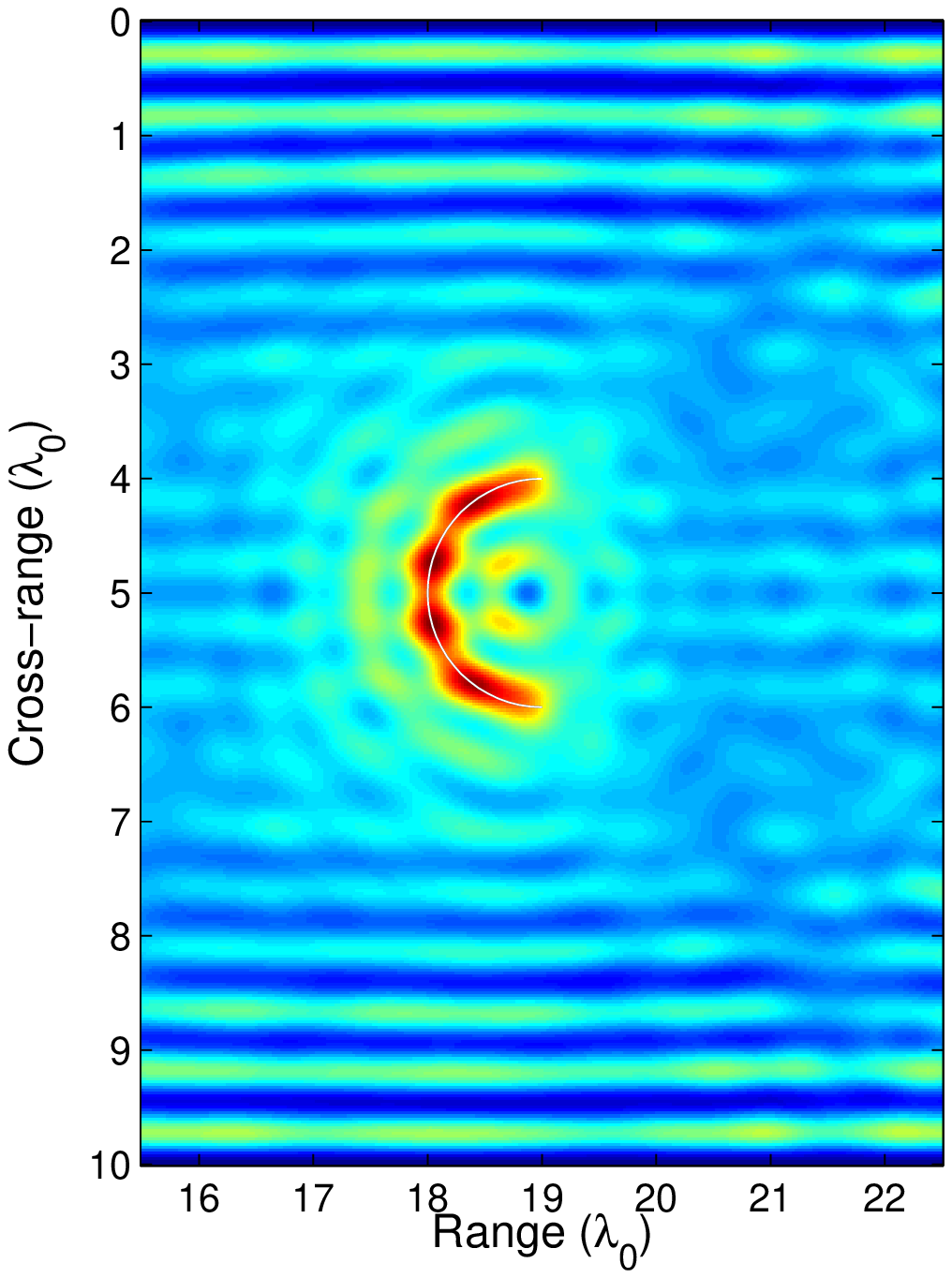} \\
\includegraphics[width=\linewidth]{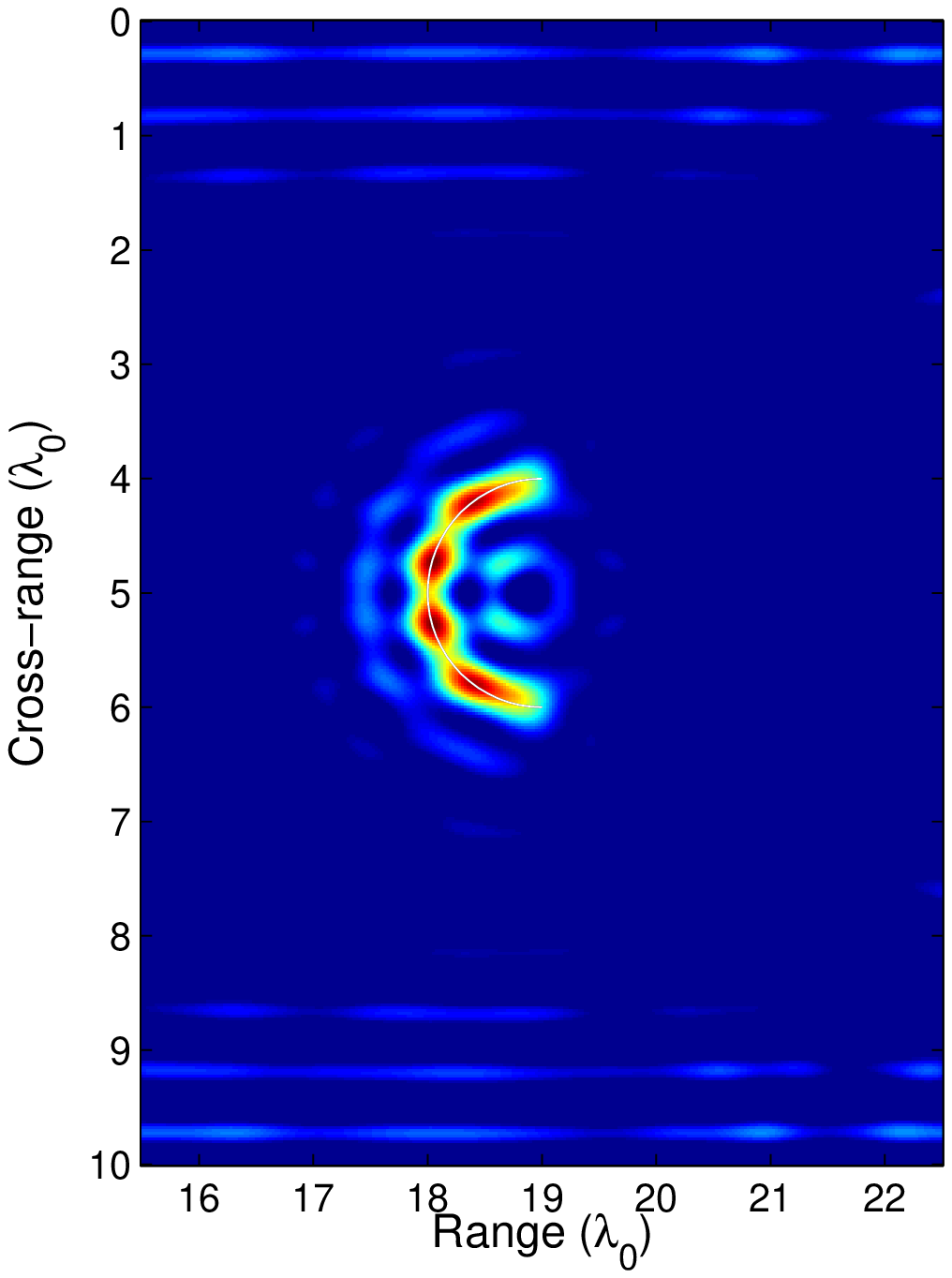}
\end{center}
\end{minipage}
\end{center}
\end{minipage}
 \begin{minipage}{0.49\linewidth}
\begin{center}
$B=0.92f_c$ \\
\begin{minipage}{.45\linewidth}
\begin{center}
infinite \\
\includegraphics[width=\linewidth]{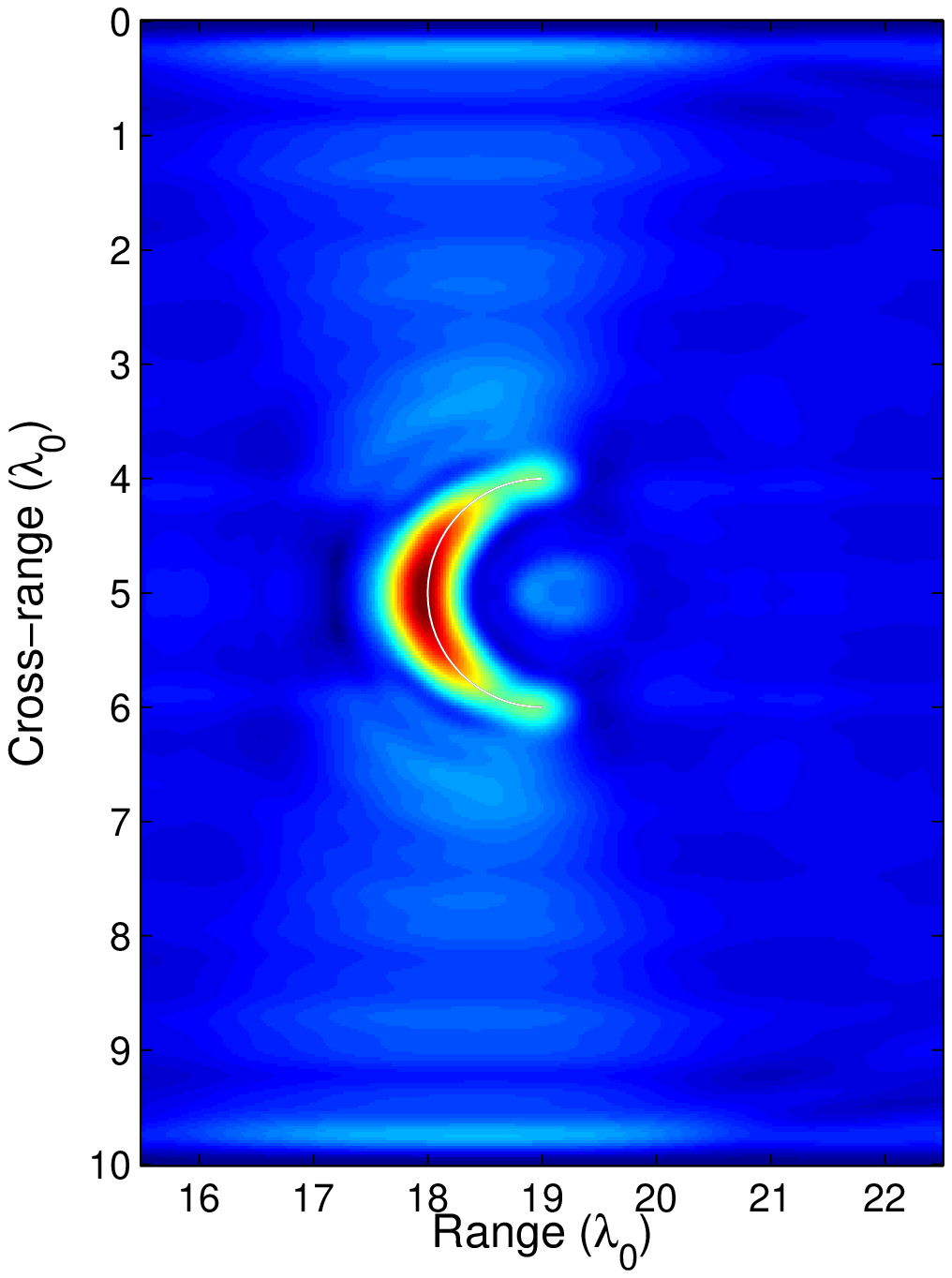}  \\
\includegraphics[width=\linewidth]{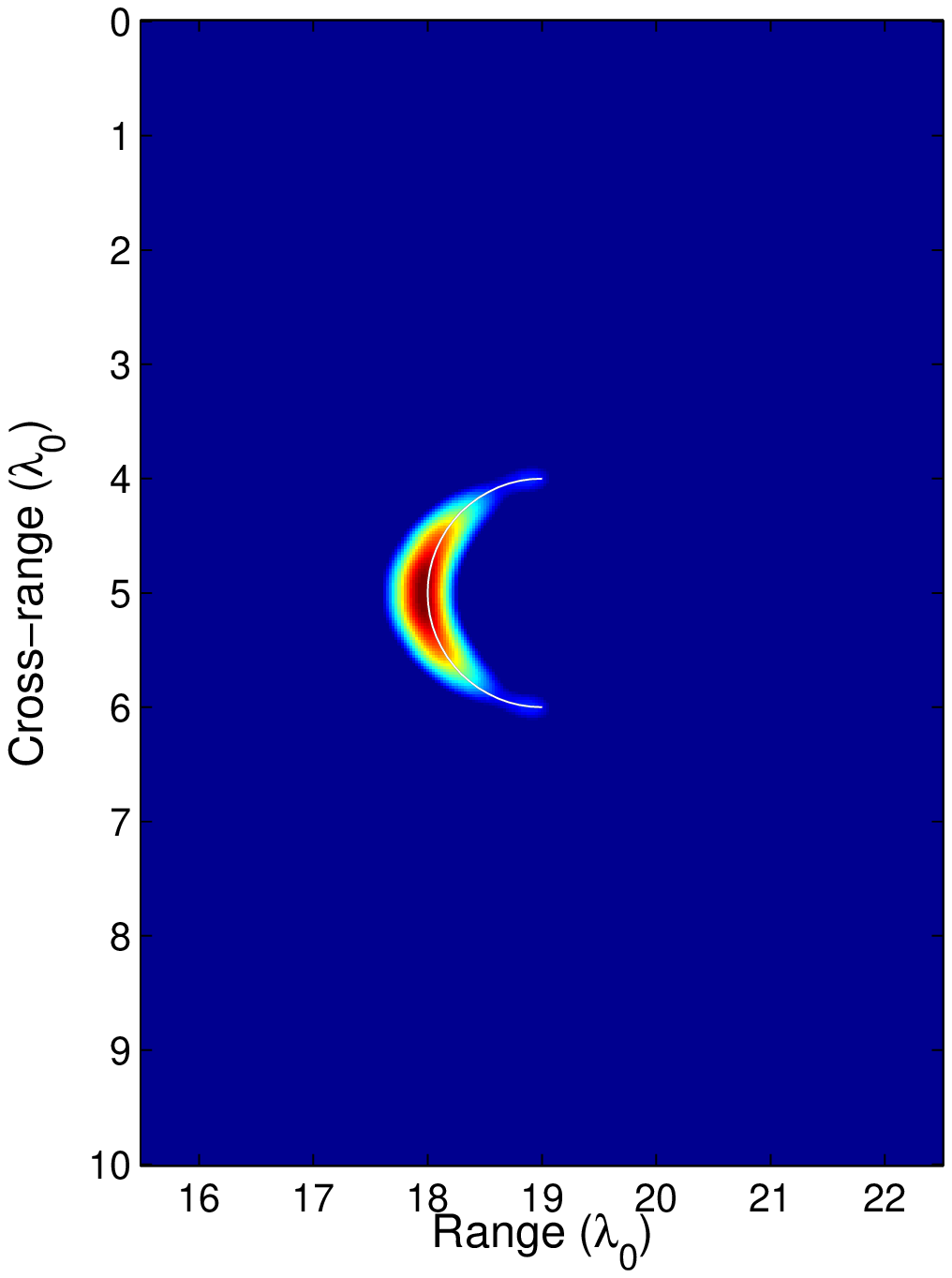} 
\end{center}
\end{minipage}
\begin{minipage}{.45\linewidth}
\begin{center}
terminating~\\
\includegraphics[width=\linewidth]{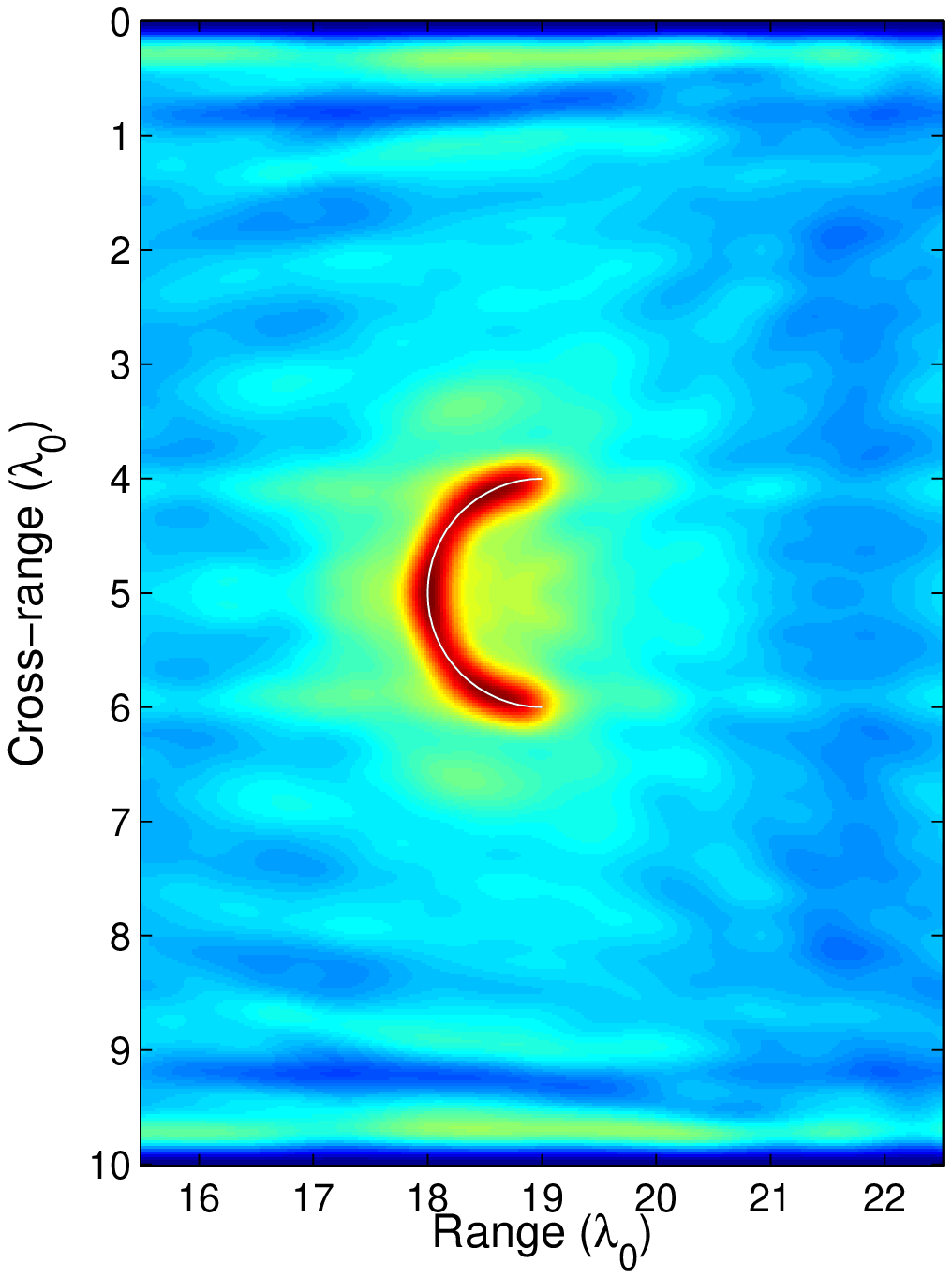}  \\
\includegraphics[width=\linewidth]{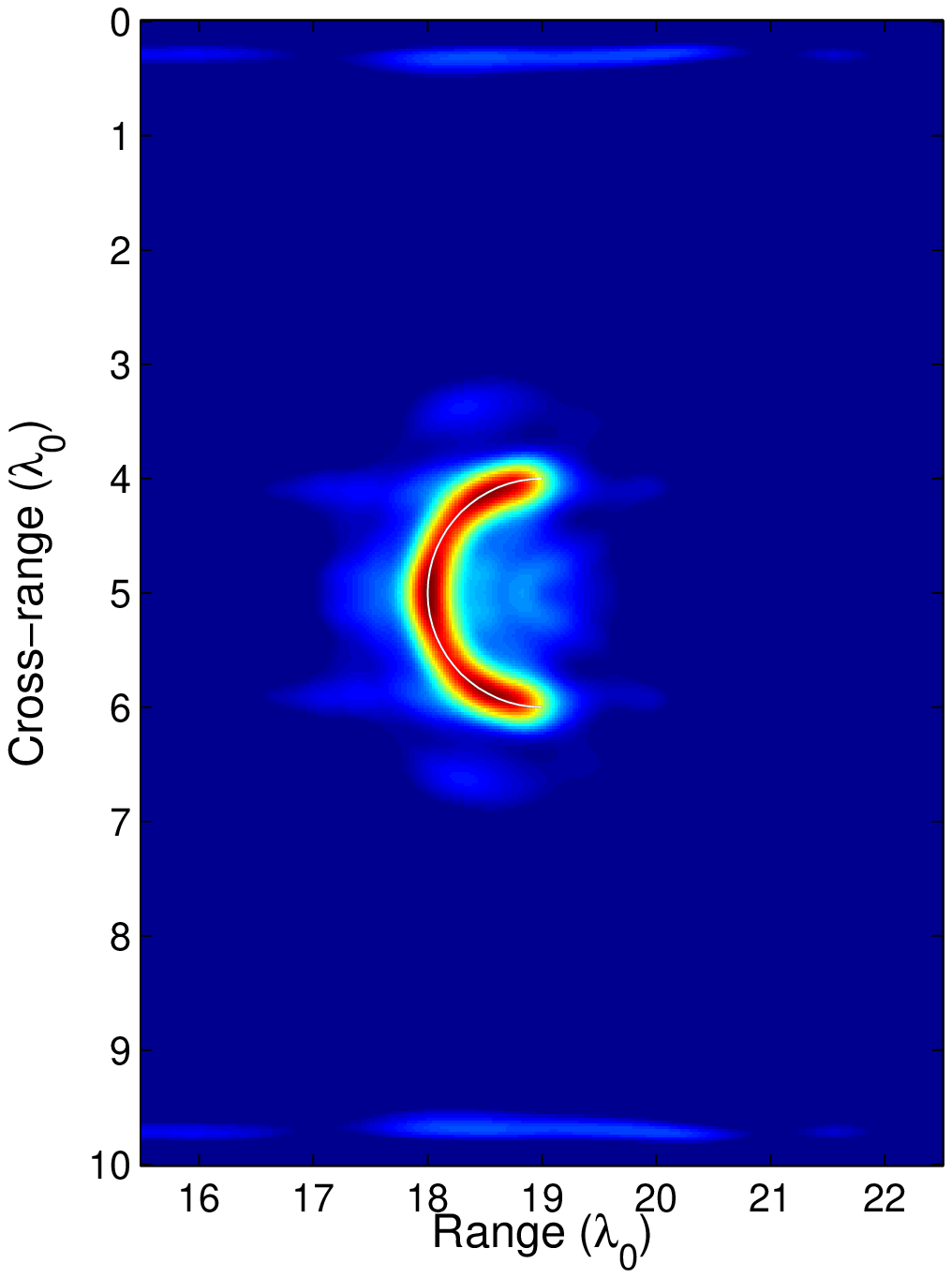} 
\end{center}
\end{minipage}
\end{center}
\end{minipage}
\caption{Multiple frequency imaging with $\Ia$ of a semicircular reflector centered at $(z^*,x^*)=(19,5)\lambda_0$. 
Specifically, $f \in [f_c-B/2, f_c+B/2]$ with $k_c=0.975 k_0$, $k_0=\pi/10$. For the first two columns the bandwidth is 
equal to $B=0.51 f_c$, while a larger bandwidth $B=0.92f_c$ is used for 
the two columns on the right. The images shown in the first and and in the third column correspond to the
open-ended  waveguide while those depicted in the second and fourth to the terminating waveguide. 
On the top row we plot the modulus of the image normalized by its maximum value, while on the bottom row we 
use a threshold that sets to zero the values of the image with normalized modulus less than $\ell=0.4$.}
\label{fig:IKMa_inf_term}
\end{figure}

In \cref{fig:IKMa_inf_term} we plot the modulus of $\Ia$ for the case of an open-ended waveguide (plots shown in the first and third
columns), and a terminating waveguide (second and fourth columns). We have used two different bandwidths.
The images shown in the  first two columns were obtained with bandwidth $B=0.51 f_c$, while $B$ was taken equal to
$B=0.92f_c$ for the images shown in the third and fourth columns. These results are in perfect agreement with our 
theoretical analysis that suggests that SNR improves as we increase the bandwidth. 
In the remaining part of this section we fix the bandwidth to $B=0.92 f_c$. In all plots, the image is normalized with respect to its maximum value. 
Looking carefully at the images shown in \cref{fig:IKMa_inf_term}, we observe that these in the open ended waveguide 
exhibit a lower noise compared to the corresponding ones in the terminating waveguide, while the latter offer a better 
reconstruction of the entire scatterer shape compared to those in the infinite waveguide which focus mainly around the midpoint 
of the semicircle.  This can be seen more clearly in the images displayed on the bottom row, where we threshold the 
normalized modulus of the values of the image that are less than $\ell = 0.4$. From now on we will refer to this process 
as thresholding with parameter $\ell$.

\subsection{Full wave scattered data}
\begin{figure}[h]
  \centerline{
  \includegraphics[width=0.80\textwidth]{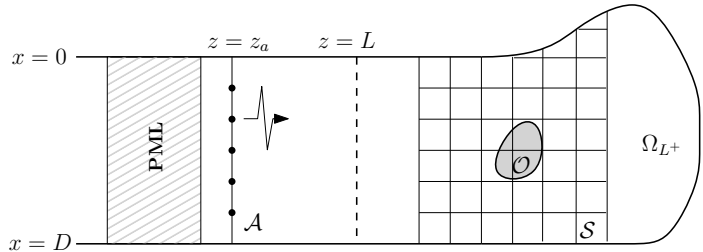}}
 \caption{Sketch of a waveguide that is truncated near  the source with a PML.}
  \label{fig:WG_PML}     
 \end{figure}

Next, we want to test our approach in imaging extended scatterers without using any simplifying approximation 
for the forward model. To this end, we now construct the array response matrix $\widehat{\Pi}$ by solving 
the wave equation  \cref{eq:ptot} numerically, with the aid of the high-order finite element C{\small ++} code Montjoie \cite{Montjoie},
which was developed  at INRIA. The originally semi-infinite waveguide is truncated with a perfectly matched layer (PML), as 
shown in  Figure~\ref{fig:WG_PML}, that ranges between $-5\lambda_0$ and $0$, a width sufficient to absorb waves propagating  to 
$-\infty$. We discretize the finite computational domain using quadrangles, in which we use 
$\mathbb{Q}_{12}$ polynomials ($\mathbb{Q}_n={\rm span}\{x^l y^m,\; 0\leq l,m\le n\}$), while we use a fourth-order leapfrog 
scheme for the time domain discretization.

\begin{figure}[ht]
\centering
\subfloat[No threshold used.]{
\includegraphics[width=0.22\linewidth]{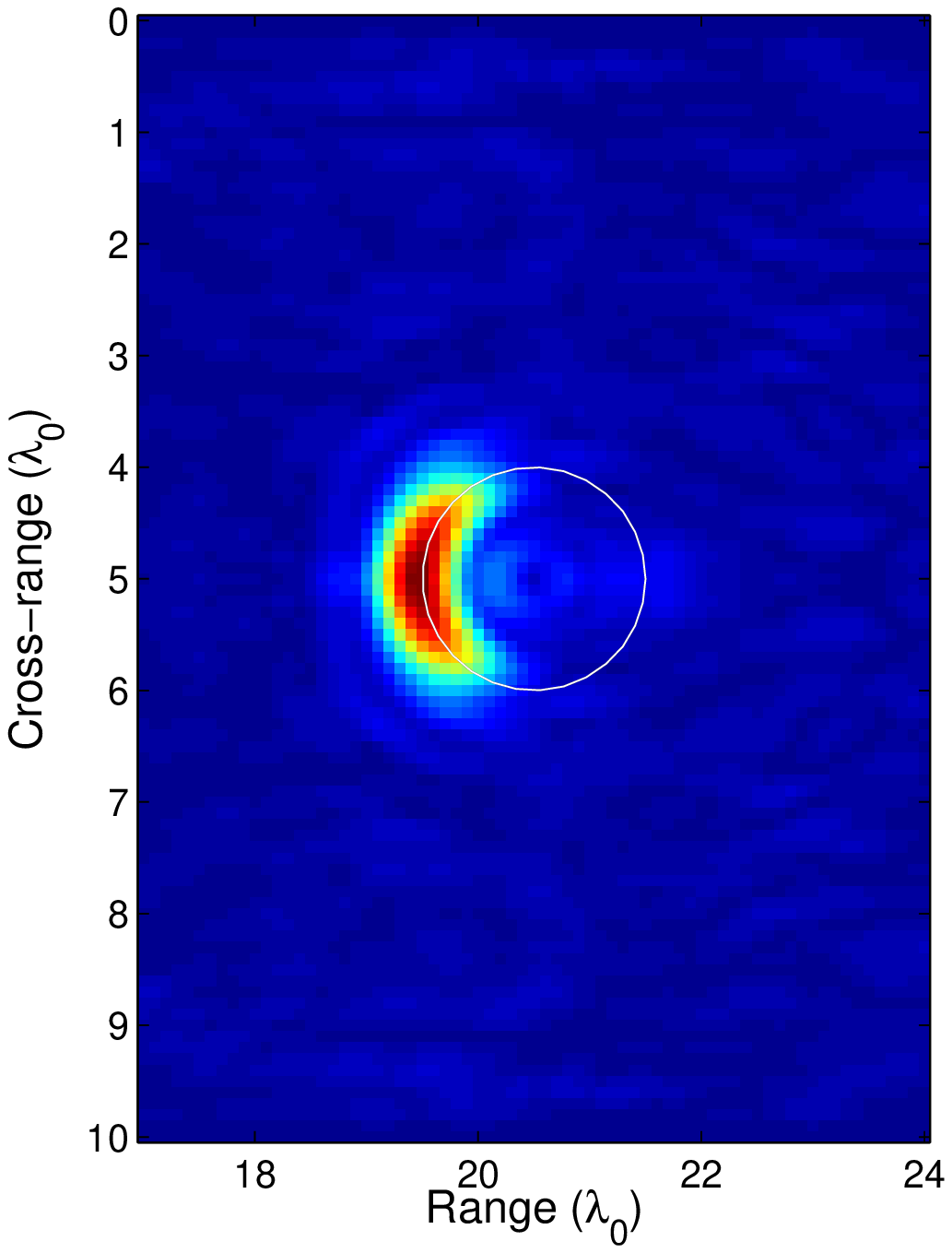}   
\includegraphics[width=0.22\textwidth]{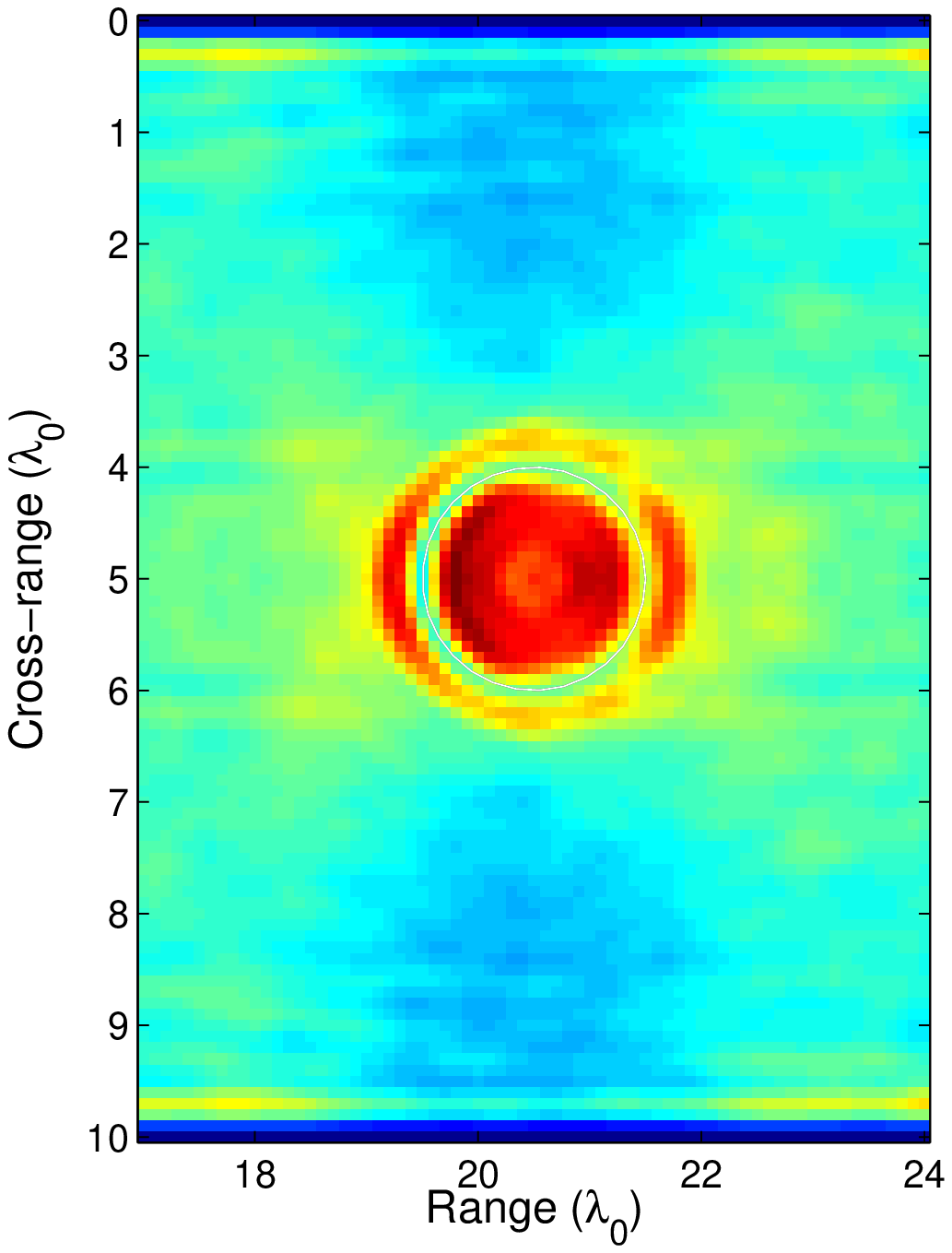} }
\subfloat[Threshold equal to $0.4$.]{\includegraphics[width=0.22\textwidth]{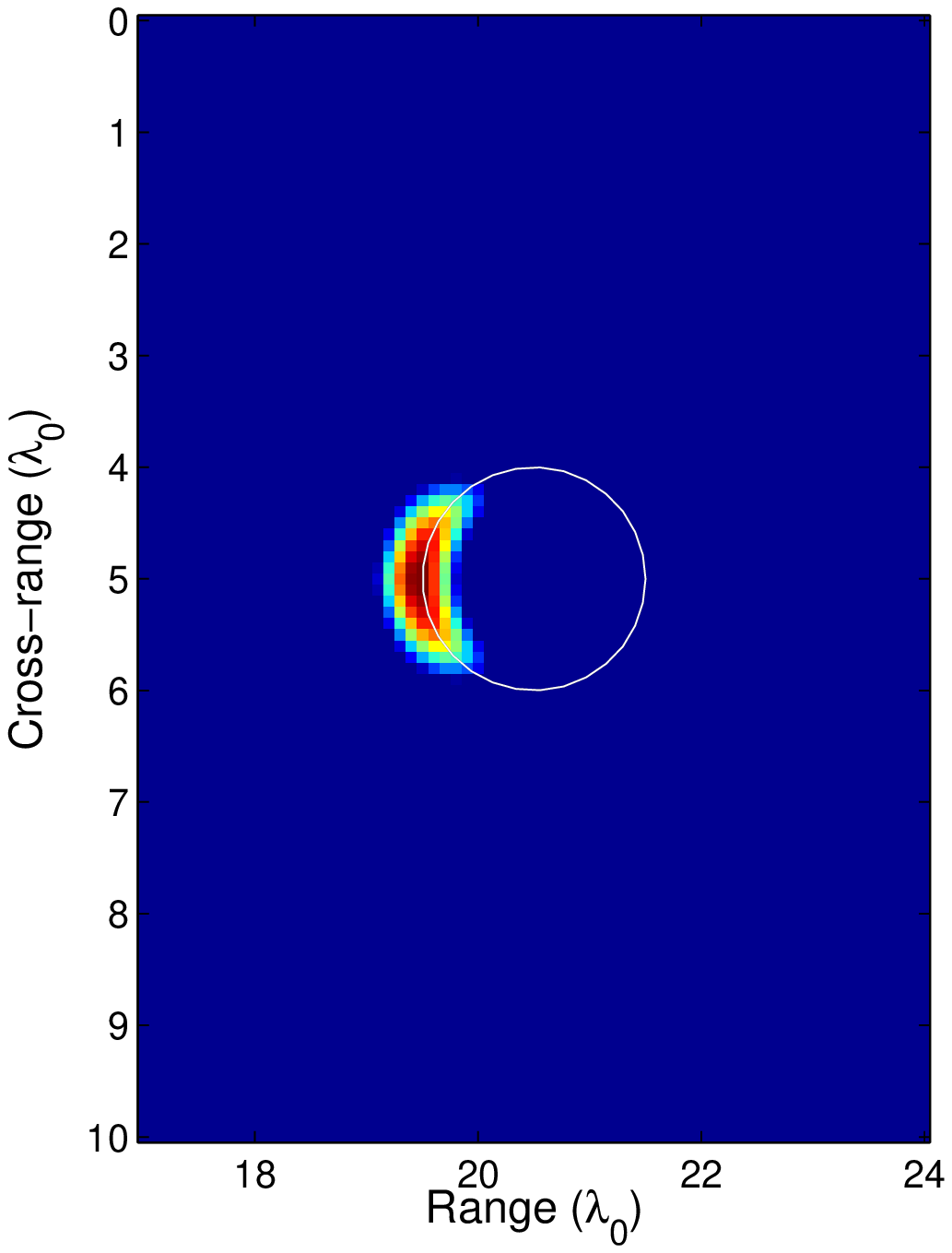}  
\includegraphics[width=0.22\textwidth]{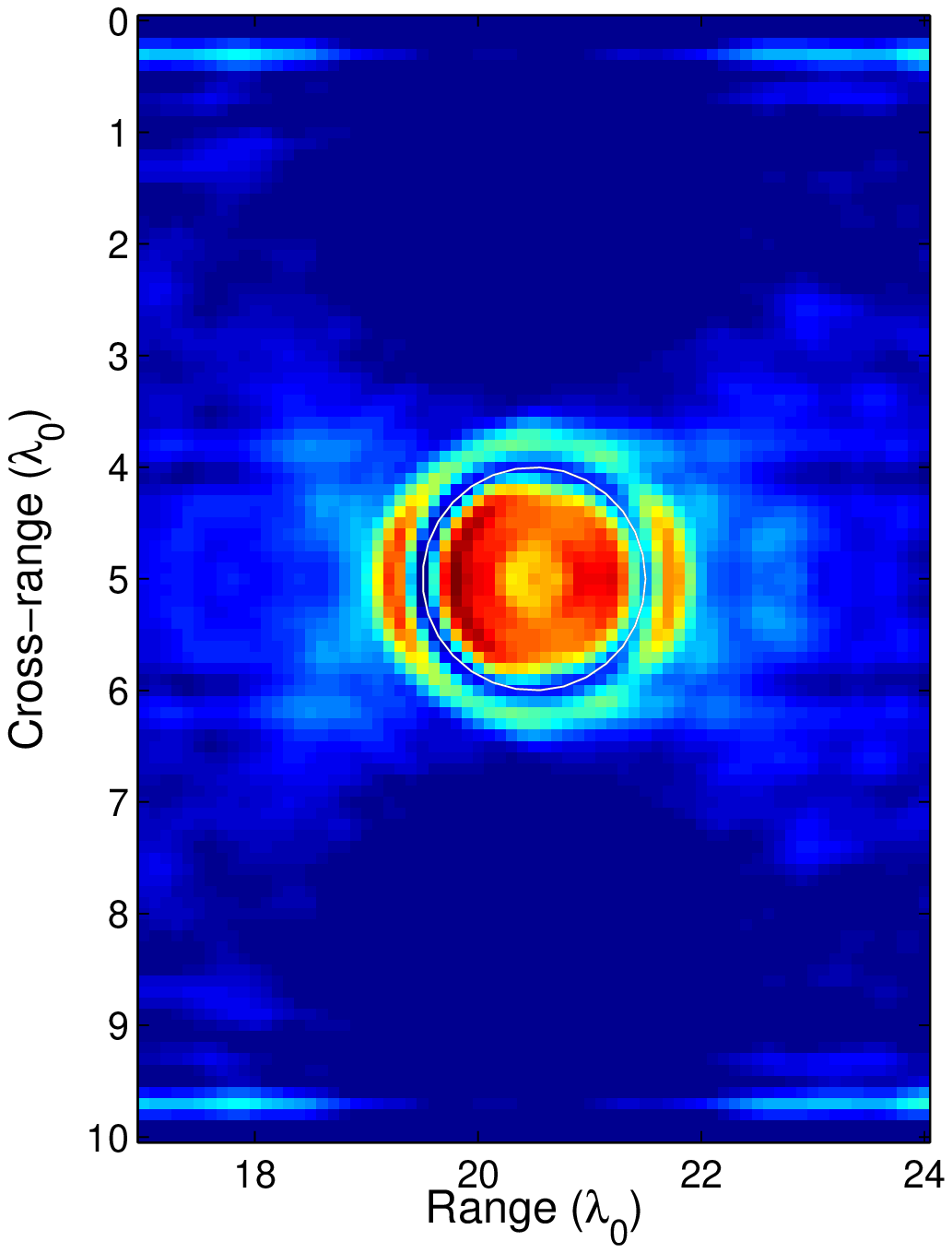}}
\caption{Imaging with $\Ia$ of a disc-shaped scatterer centered at $(z^*,x^*)=(20.5,5)\lambda_0$,  
for $k_c=0.9733 k_0$, $k_0 = \pi/10$ and $B=0.92f_c$. 
(a) Data are back propagated with the Green's function for the open ended (left subplot) and the 
terminating waveguide (right subplot), where we do not use thresholding. 
(b) Same setup as in (a) but we use thresholding  with $\ell = 0.4$.}
\label{fig:IKMt_d_c}
\end{figure}

The array imaging setup is similar to the one used in the previous subsection, with the exception that now 
our vertical array is placed at $\za=2\lambda_0$ and has a pitch $h=\lambda_0/4$. 
First, we consider the case of the semi-infinite strip, i.e. $\Omega = (-\infty,R)\times (0,D)$, where 
now the terminating vertical boundary is located at $R=28\lambda_0$ and
a disc-shaped scatterer of diameter $b=2\lambda_0$ is centered at $(z^\ast,x^\ast) = (20.5,5)\lambda_0$. 
A Neumann condition is imposed on the circular boundary of the scatterer.
In the right subplot of \cref{fig:IKMt_d_c}(a) (second image in the panel) we plot the modulus of $\Ia$ 
normalized by its maximum value. As one may immediately verify, even though  the SNR of the image 
is a bit low, the location, size and shape of the scatterer are fully recovered.
For the image in the left subplot (first image in the panel) we pretend that we are not aware of the fact that the
waveguide has a closed end, and we back propagate the same data with the ``wrong" Green's function, 
i.e. the one for the open-ended waveguide. We implement this by replacing in \cref{eq:Iamf}  
the terms $\GRm$, $\GRn$ by $\Gom$, $\Gon$, respectively, i.e., by the Fourier coefficients of the Green's function 
for the infinite waveguide (see \cref{eq:Gopen}) with respect to the orthonormal basis $\{X_n\}_{n=1}^\infty$ of $L^2(0,D)$. 
As a result, only the left part of the scatterer is recovered.  
In an attempt to improve the SNR of these images we plot in \cref{fig:IKMt_d_c}(b) the corresponding images 
after thresholding with $\ell=0.4$.

\begin{figure}[ht]
\centering
\subfloat[No threshold used.]{
\includegraphics[width=0.22\linewidth]{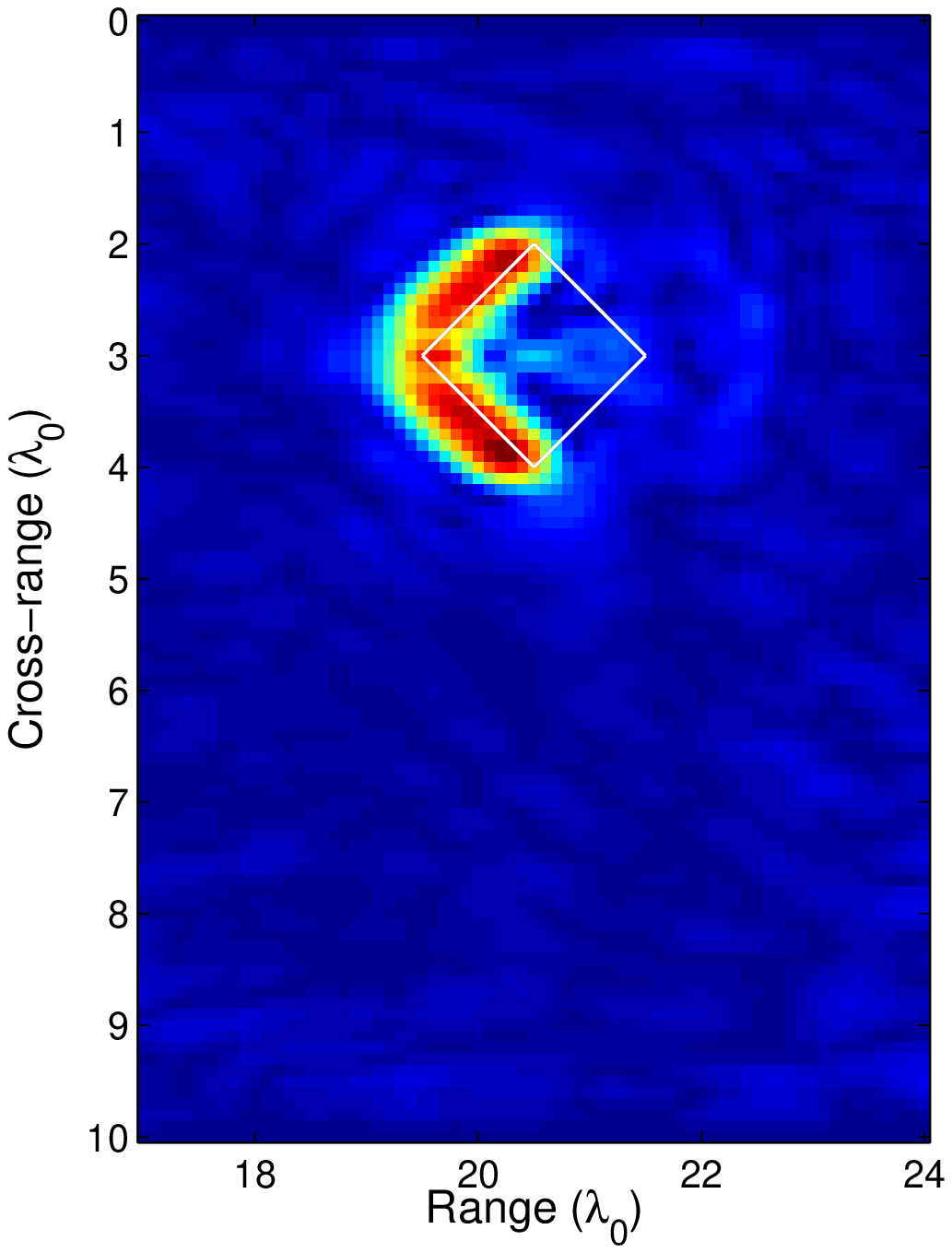}   
\includegraphics[width=0.22\textwidth]{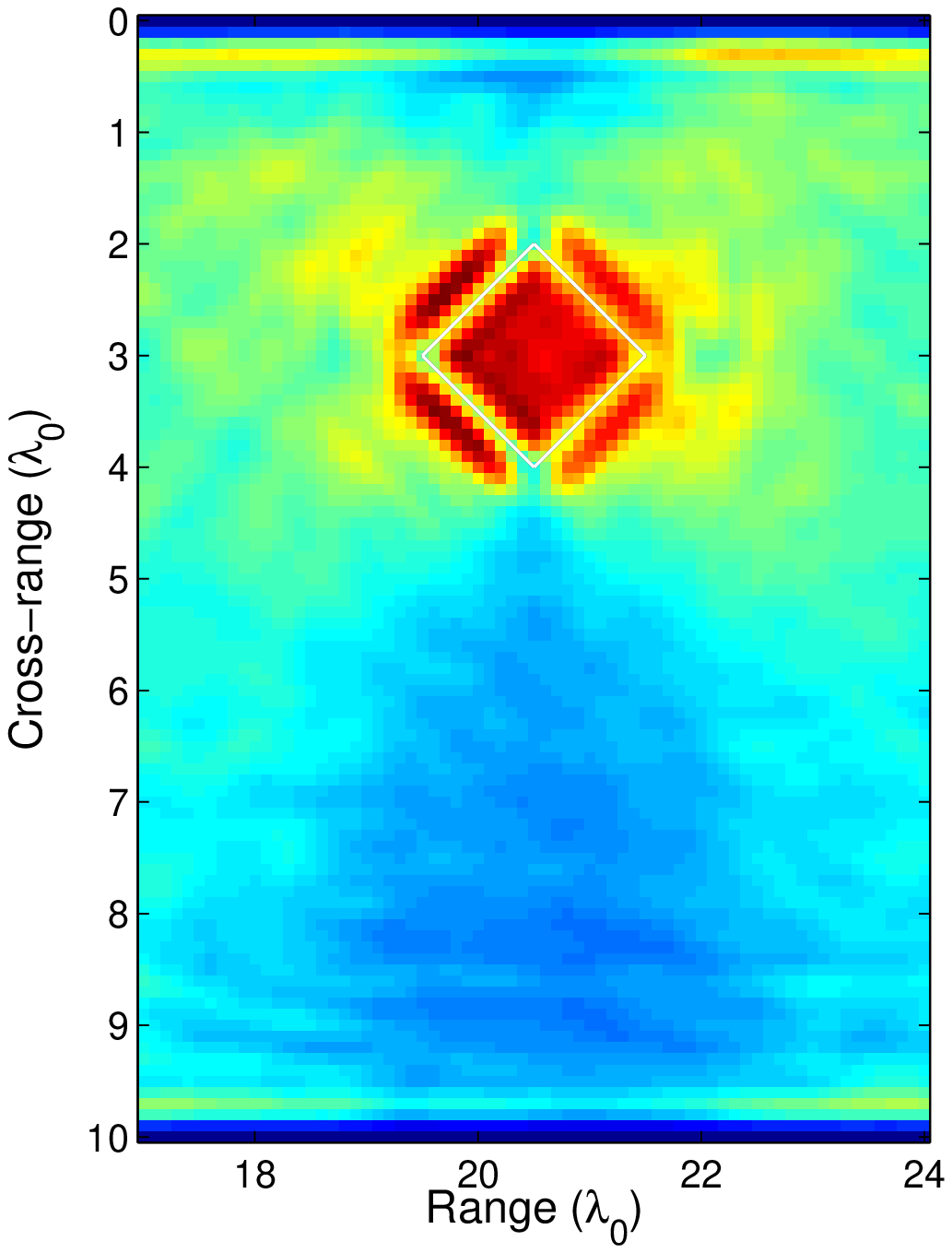} }
\subfloat[Threshold equal to $0.4$.]{\includegraphics[width=0.22\textwidth]{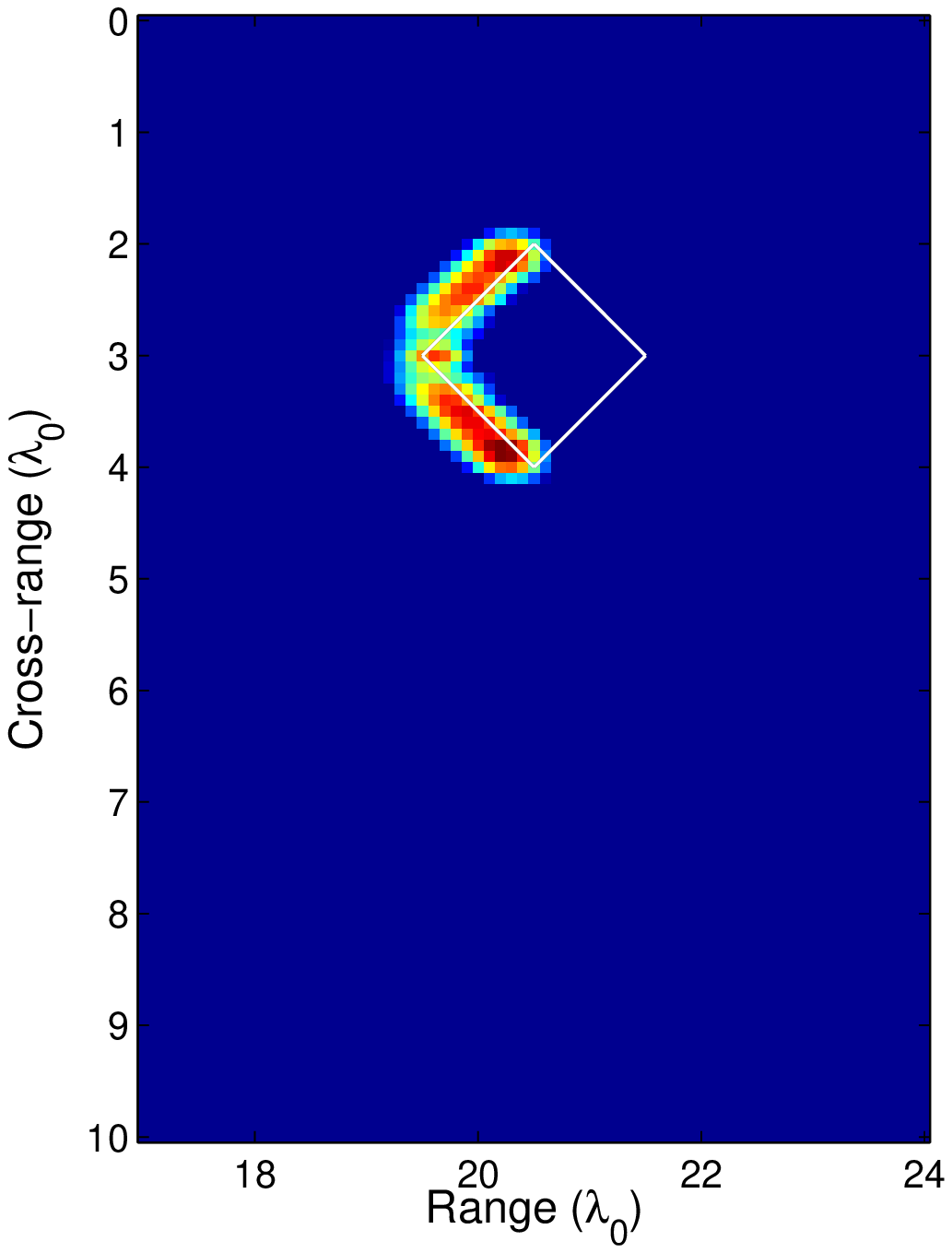}  
\includegraphics[width=0.22\textwidth]{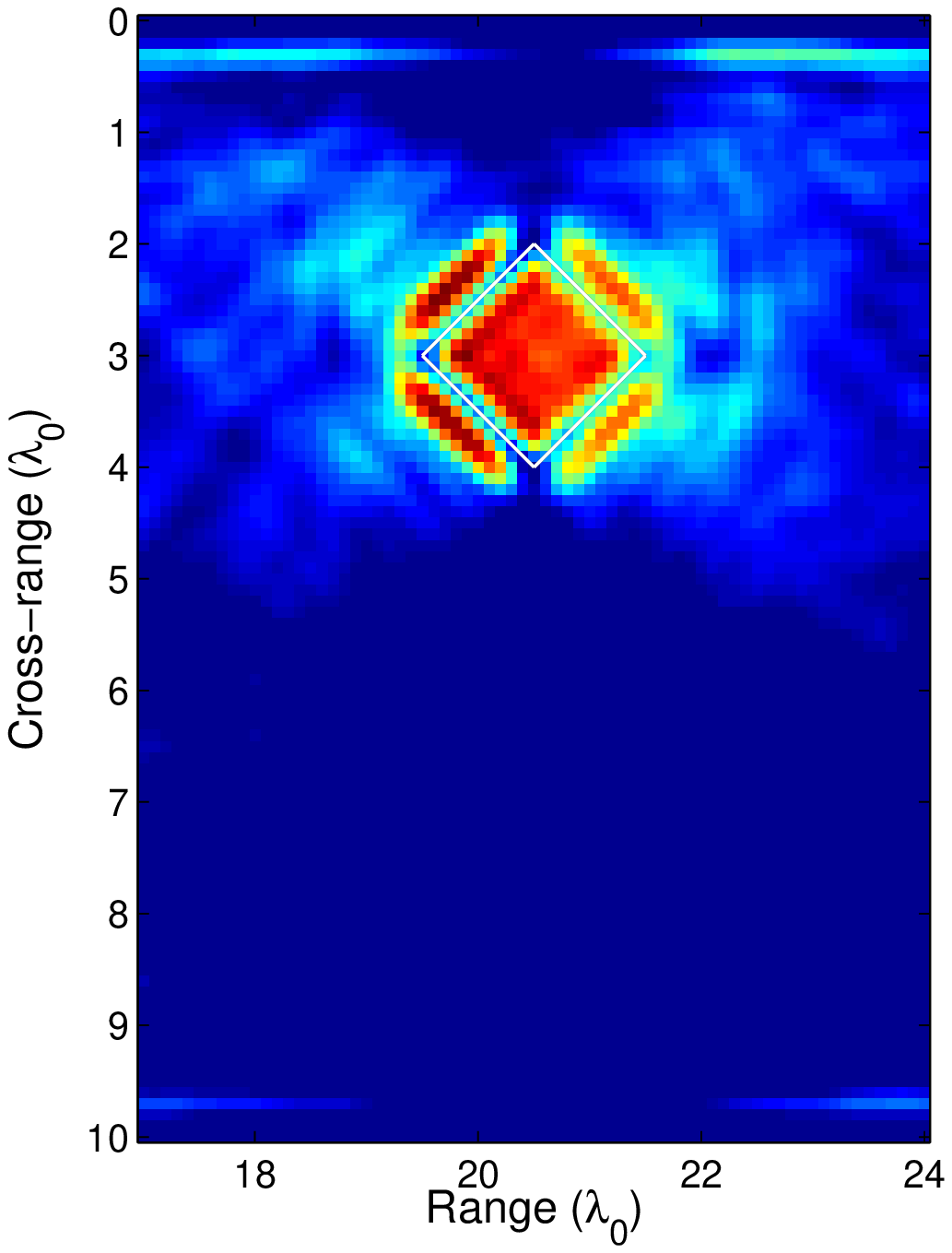}}
\caption{Imaging with $\Ia$ of a rhombus-shaped scatterer centered at $(z^*,x^*)=(20.5,3)\lambda_0$,  
for $k_c=0.9733 k_0$, $k_0 = \pi/10$ and $B=0.92f_c$. 
(a) Data are back propagated with the Green's function for the open ended (left subplot) and the 
terminating waveguide (right subplot), where we do not use thresholding. 
(b) Same setup as in (a) but we use thresholding  with $\ell = 0.4$.
}
\label{fig:IKMa_r_off}
\end{figure}

As a second example, we place in the previously described waveguide a rhombus-shaped scatterer of diameter 
$b=2\lambda_0$, centered at $(z^\ast,x^\ast) = (20.5, 3)\lambda_0$. 
\cref{fig:IKMa_r_off} is the analogous of  \cref{fig:IKMt_d_c}. 
As before, on the left subplot of each subfigure we present the image obtained when we back propagate our data with  
the Green's function for the open-ended waveguide; again we observe that only the left part of the scatterer can be 
reconstructed. When we use the correct Green's function, the corresponding images on the right subplots of each subfigure
exhibit a good reconstruction of the scatterer. 

\begin{figure}[ht]
\centering
\includegraphics[width=0.32\textwidth]{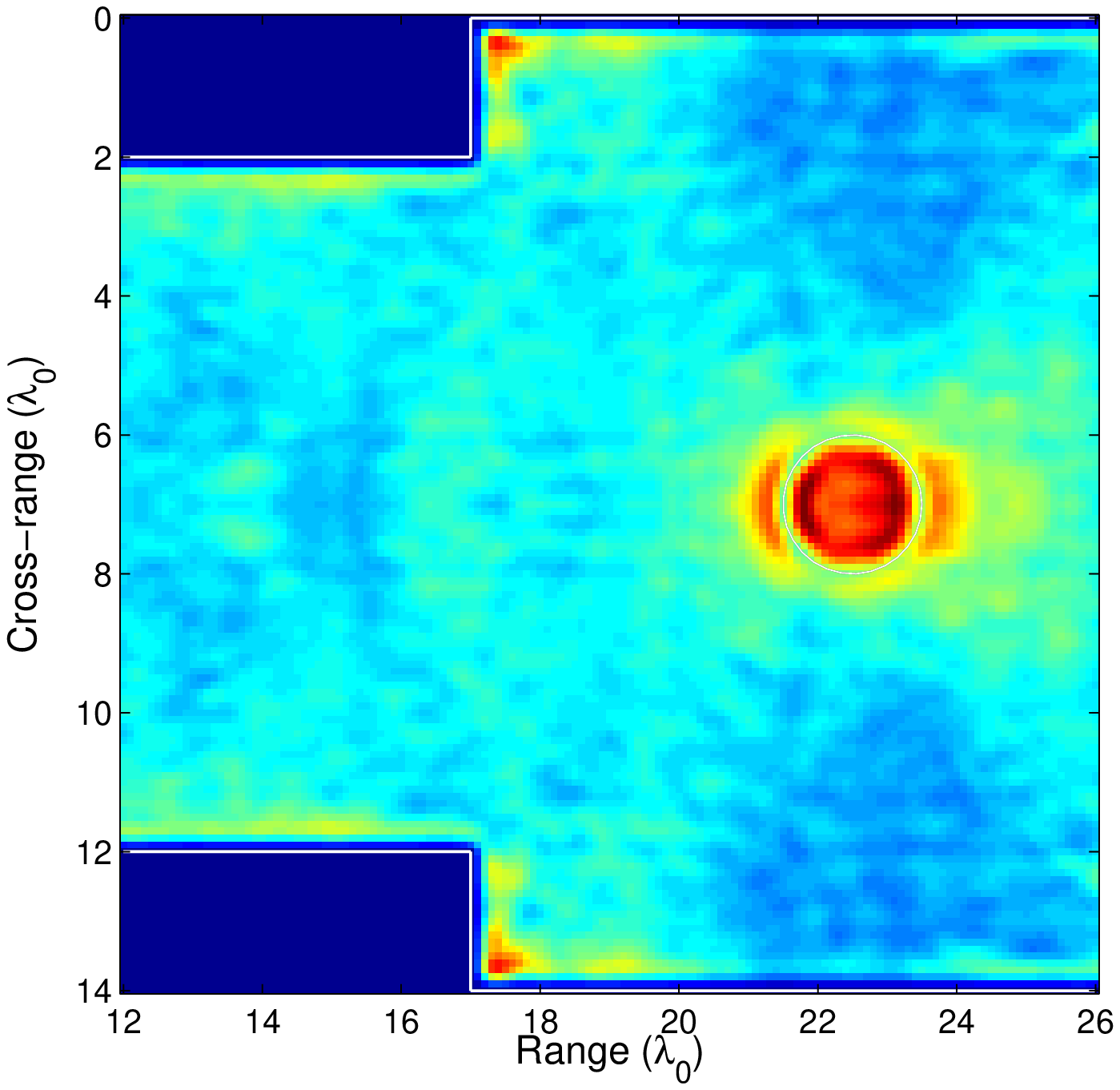}  
\includegraphics[width=0.32\textwidth]{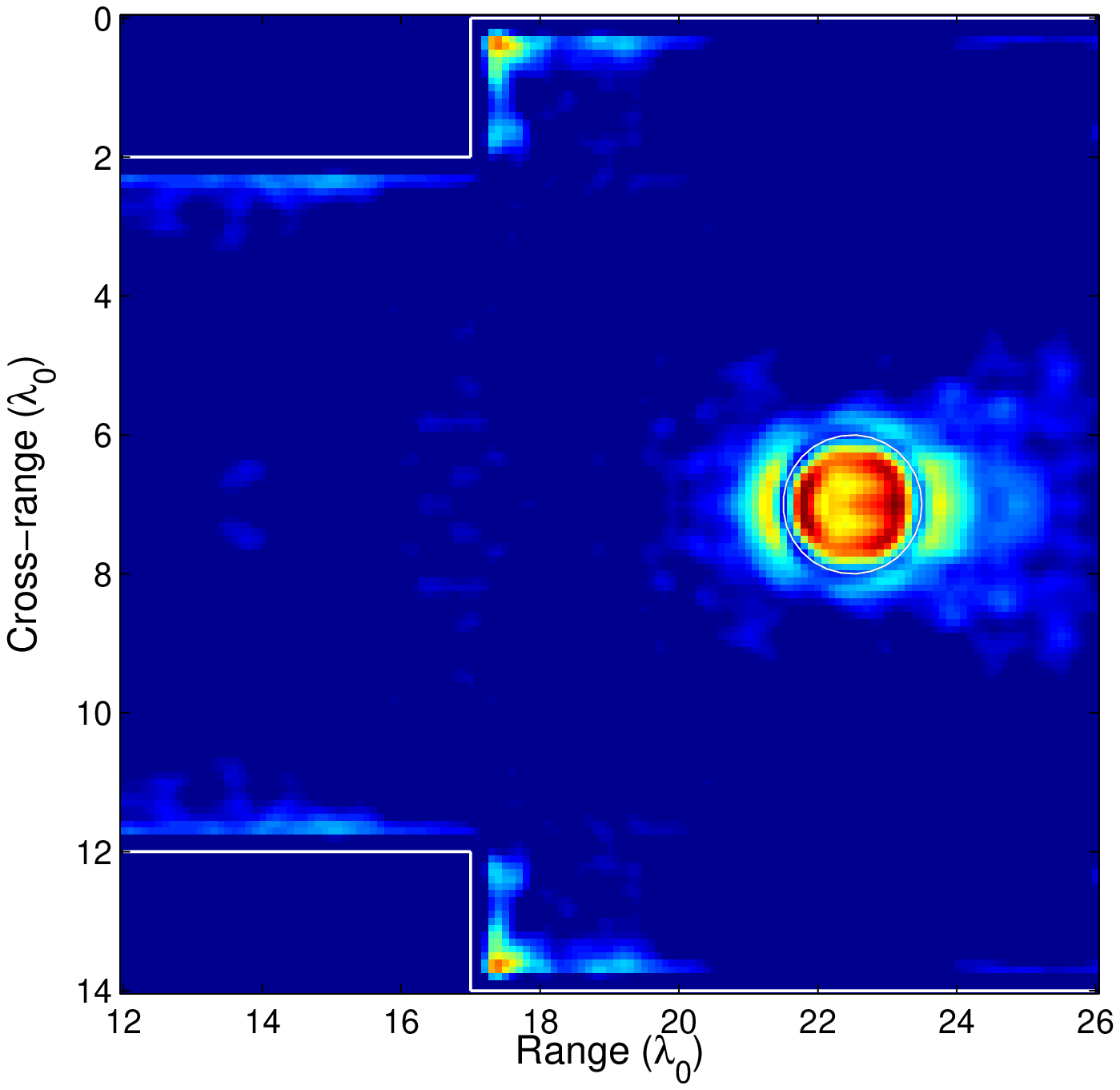}  
\caption{ Imaging with $\Ia$ for a disc scatterer centered at $(z^*,x^*)=(22.5,7)\lambda_0$,  
for $k_c=0.9733 k_0$, $k_0 = \pi/10$ and $B=0.92f_c$. 
On the left we use no threshold, while on the  right we have a threshold  $\ell=0.4$.}
\label{fig:IKMa_d_b}
\end{figure}

Finally, to demonstrate the robustness and the generality of our imaging approach we consider a waveguide 
$\Omega$  with a more complex geometry. Specifically, the waveguide has constant width in the cross-range direction 
equal to $10\lambda_0$ until $z=17\lambda_0$ and, then, it expands vertically by $2\lambda_0$ on both sides and 
keeps a  new constant width of $14\lambda_0$ until it is terminated by a vertical boundary located 
at $z=28\lambda_0$.  The geometry of part of the waveguide is depicted in the imaging results shown in 
\cref{fig:IKMa_d_b}. 
A disc-shaped scatterer with diameter $b=2\lambda_0$ is centered at $(z^\ast,x^\ast) = (22.5,7)\lambda_0$
and is  depicted in \cref{fig:IKMa_d_b} with a white continuous line.

For this waveguide geometry, we do not have an analytic expression for the Green's function $\Gh(\yb^s,\xb_s)$,
which is needed to form the image, hence we compute it numerically. To be more precise, $\Gh(\yb^s,\xb_s)$ 
is obtained by solving the wave equation in $\Omega$ in the absence of the scatterer for all sources' locations $\xb_s$, 
$s=1,\ldots,N$ and the solution is stored  for all search points $\yb^s$ in the imaging window. 
The computations are performed in the time domain and we use FFT to transform the data in the frequency 
domain.

The imaging results are shown in \cref{fig:IKMa_d_b} where we plot on the left the normalized modulus of 
$\Ia$ without a threshold, and on the right using a threshold $\ell=0.4$. 
The reconstruction is successful since it provides good estimates for the size and shape of the reflector. 
\subsection{Imaging with partial aperture}
We consider now the more challenging problem of imaging a reflector with an array that does not span the entire 
vertical cross-section of the waveguide. As we have described in \cref{alg:Active}, our imaging method requires 
the evaluation of the functional $\Ia$ in each point of the search domain. Recall that in the case of multiple 
frequencies $\Ia$ is given in \cref{eq:Iamf} and let us remark that this expression applies for any array aperture size. 
What alters is the way we construct 
the $M_l \times M_l$ modal projected matrix $\Qq$, which in the case of a partial-aperture array uses the trigonometric 
polynomials $s_j$, $j=1,\ldots,M_l$,  as in \cref{{eq:Qpartial},{eq:Smatrx}} that  account for the partial array aperture through 
the eigenvectors of the array matrix $A_{\rm arr}$.

\begin{figure}[ht]
\centering
\begin{minipage}{0.22\linewidth}
\begin{center}
\hspace*{0.3cm} $|\mathcal{A}| = 0.75D$ \\
\includegraphics[width=\linewidth]{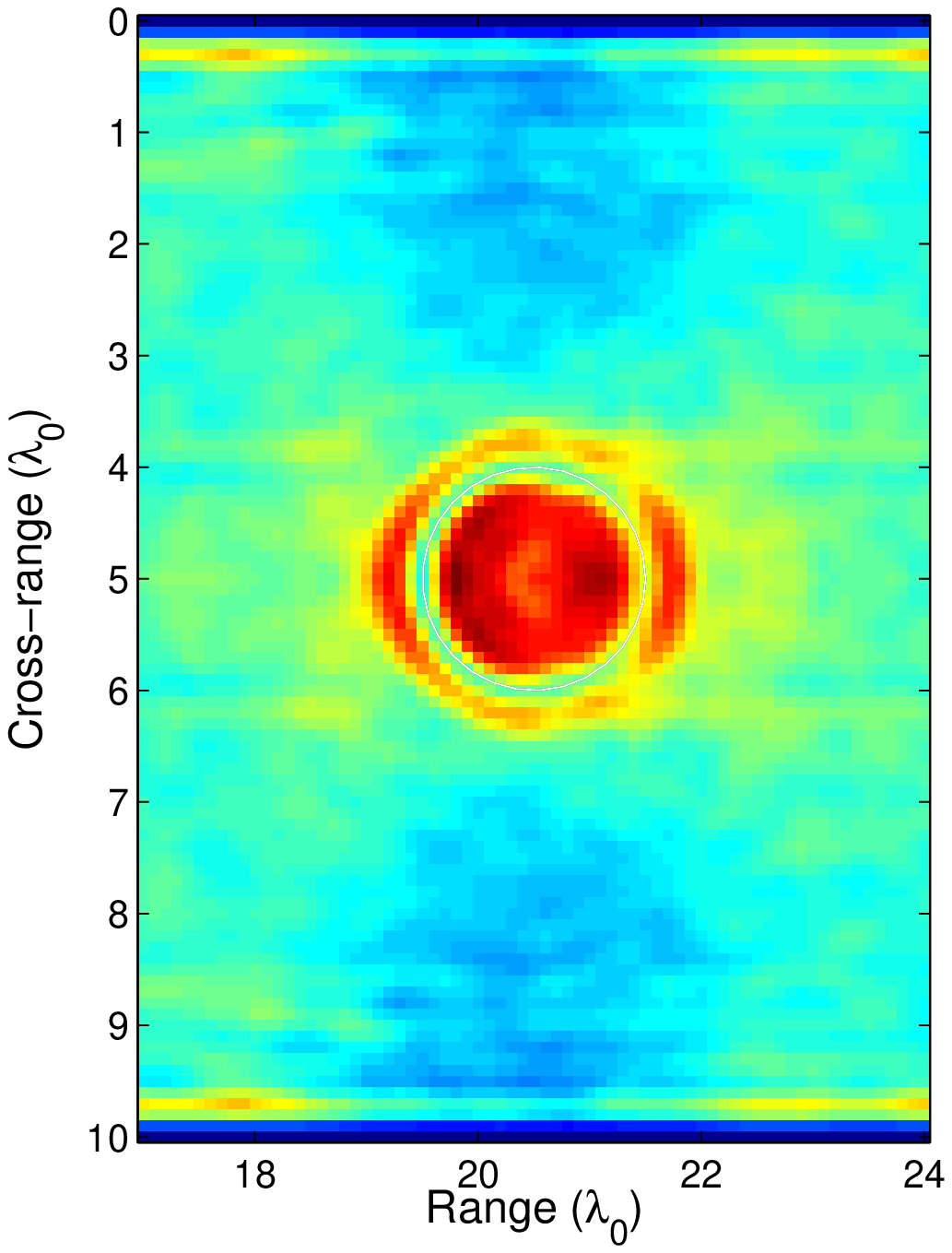}  \\
\includegraphics[width=\linewidth]{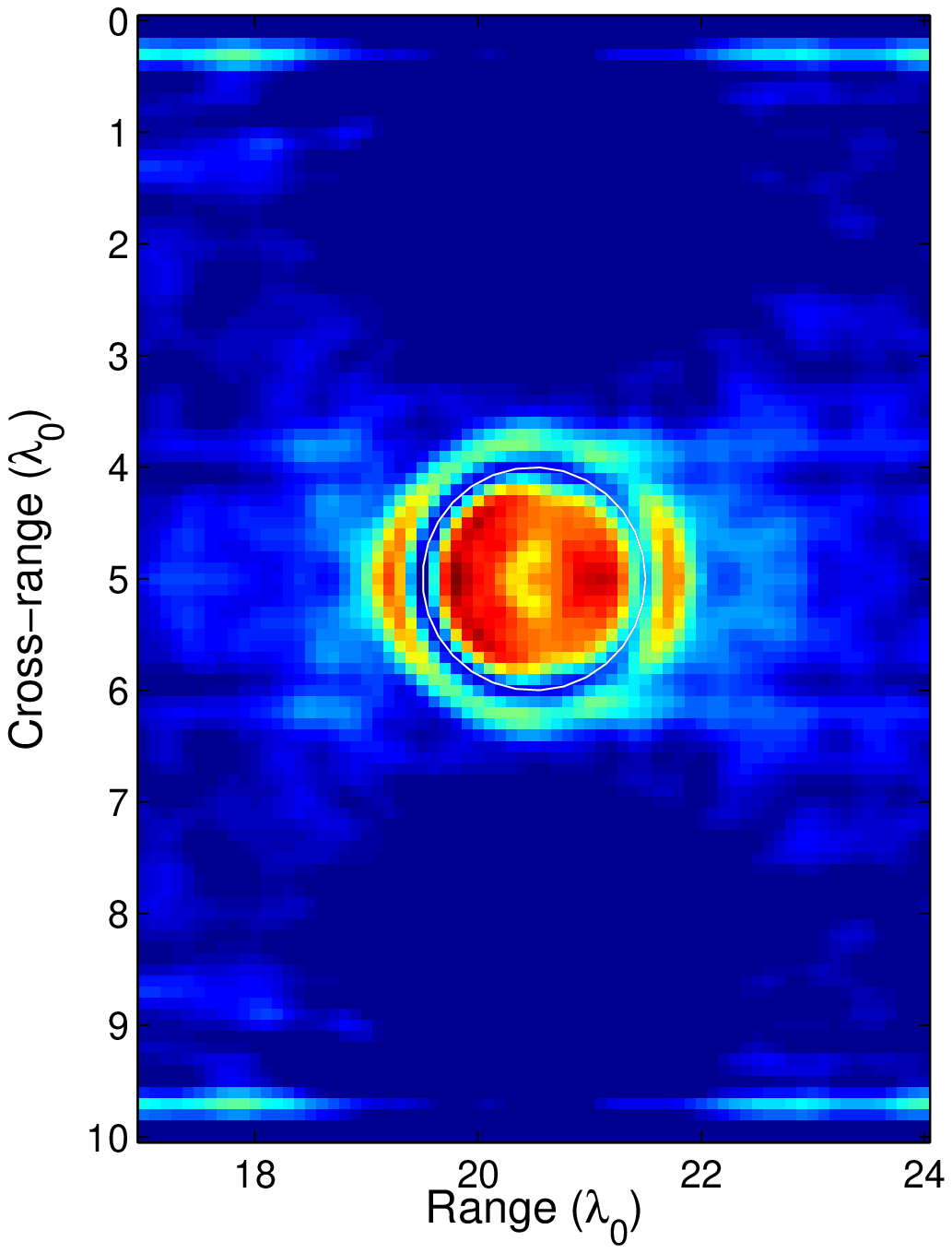}  
\end{center}
\end{minipage}
\begin{minipage}{0.22\linewidth}
\begin{center}
\hspace*{0.3cm} $|\mathcal{A}| = 0.5D$ \\
\includegraphics[width=\linewidth]{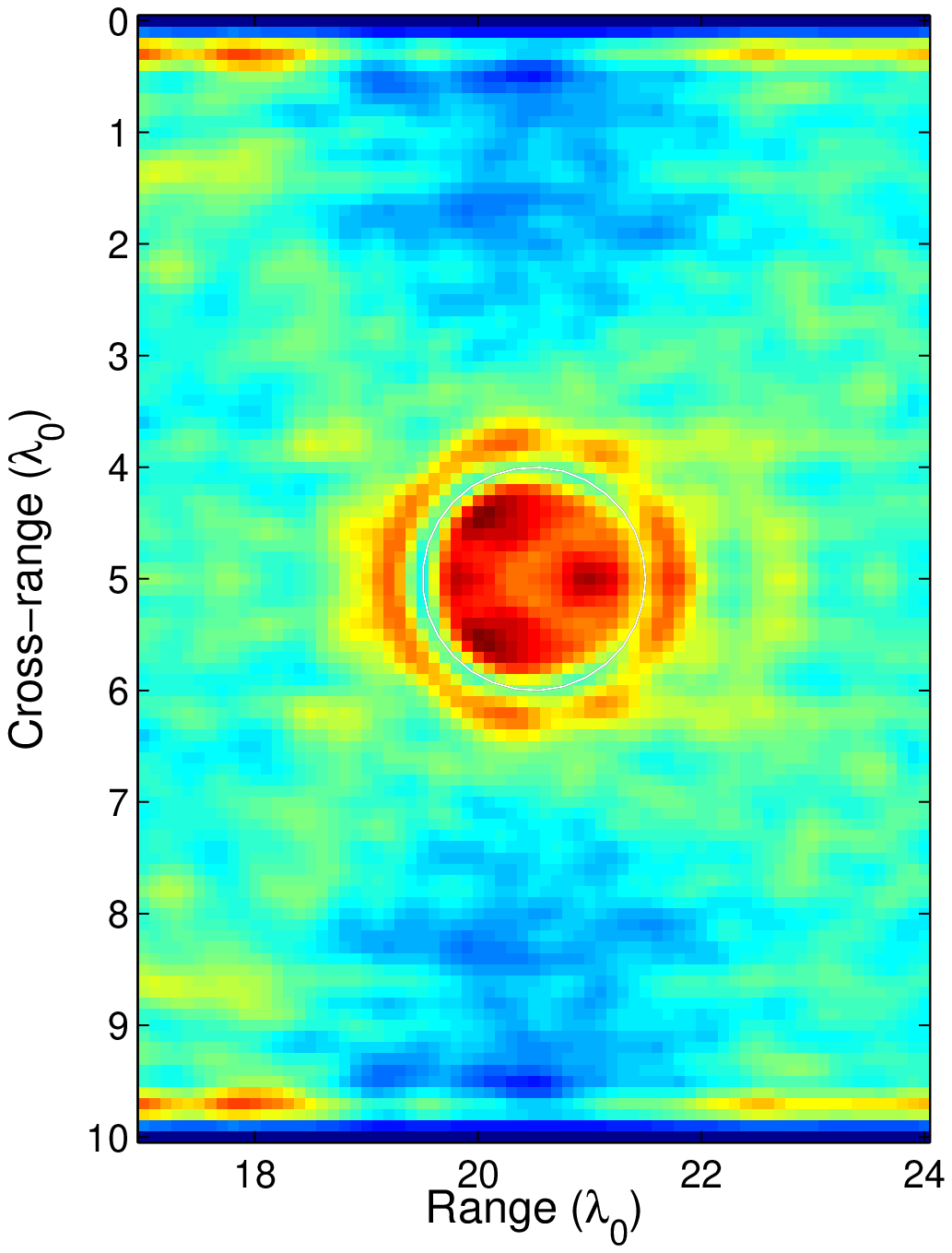}  \\
\includegraphics[width=\linewidth]{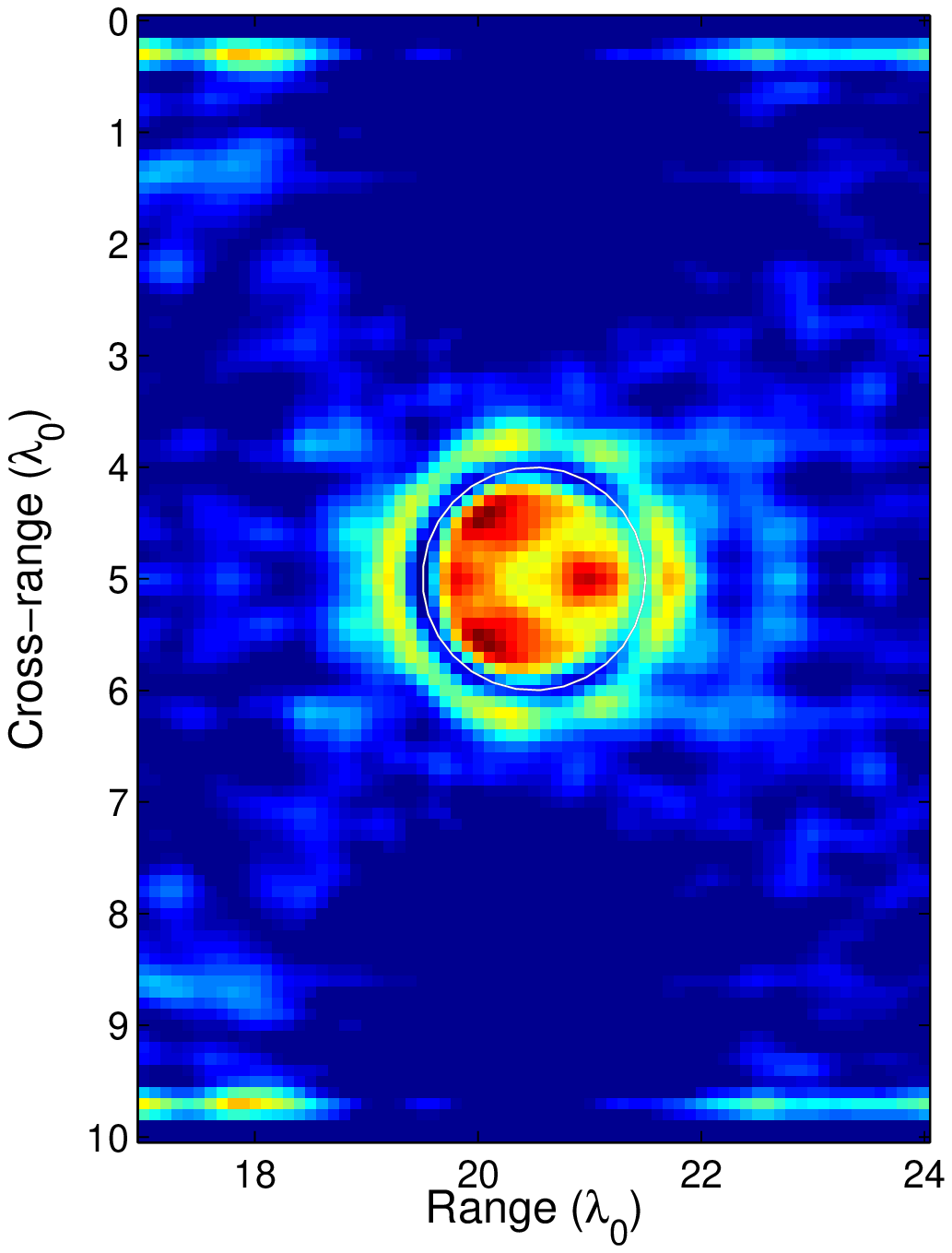}  
\end{center}
\end{minipage}
\begin{minipage}{0.22\linewidth}
\begin{center}
\hspace*{0.3cm} $|\mathcal{A}| = 0.25D$ \\
\includegraphics[width=\linewidth]{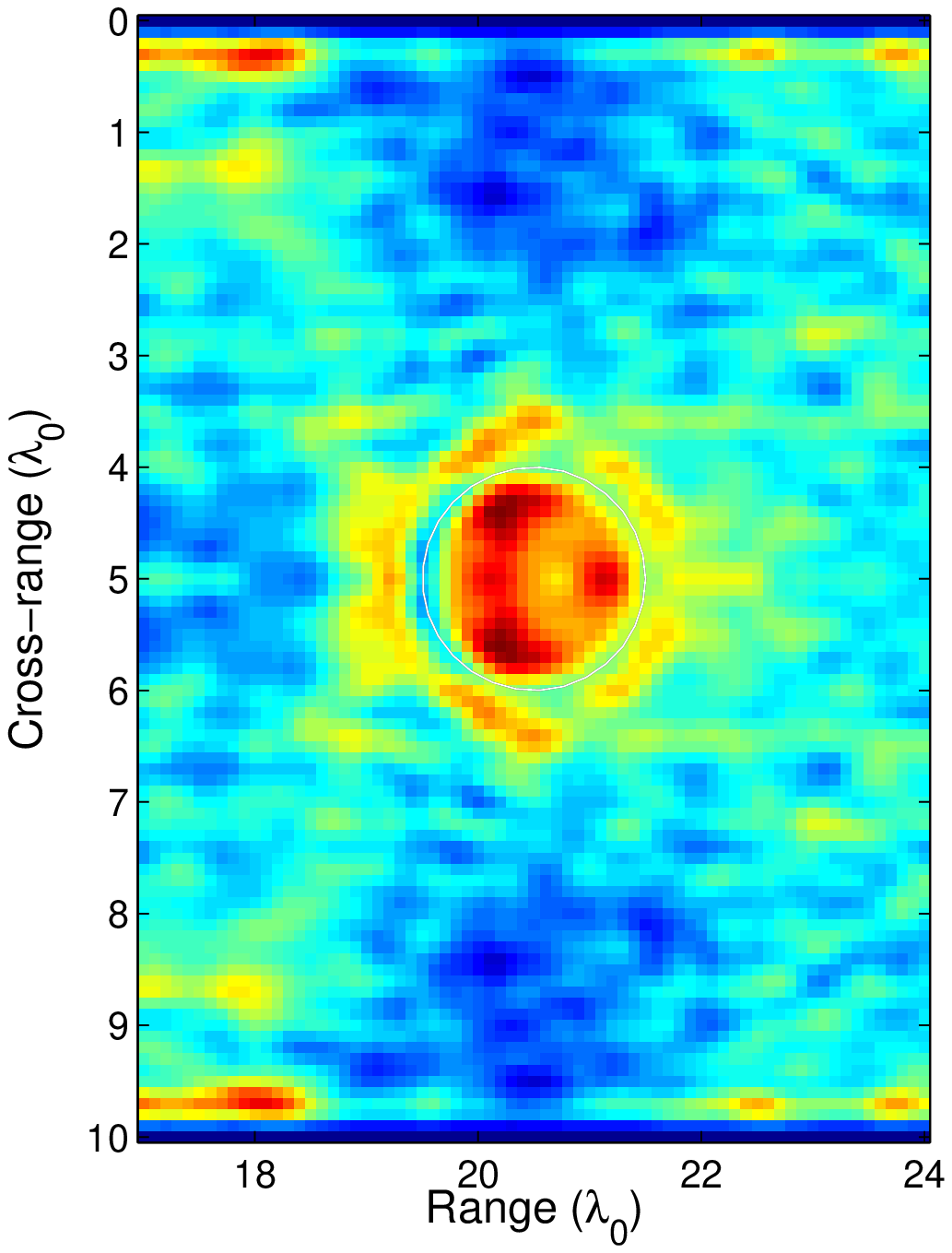}  \\
\includegraphics[width=\linewidth]{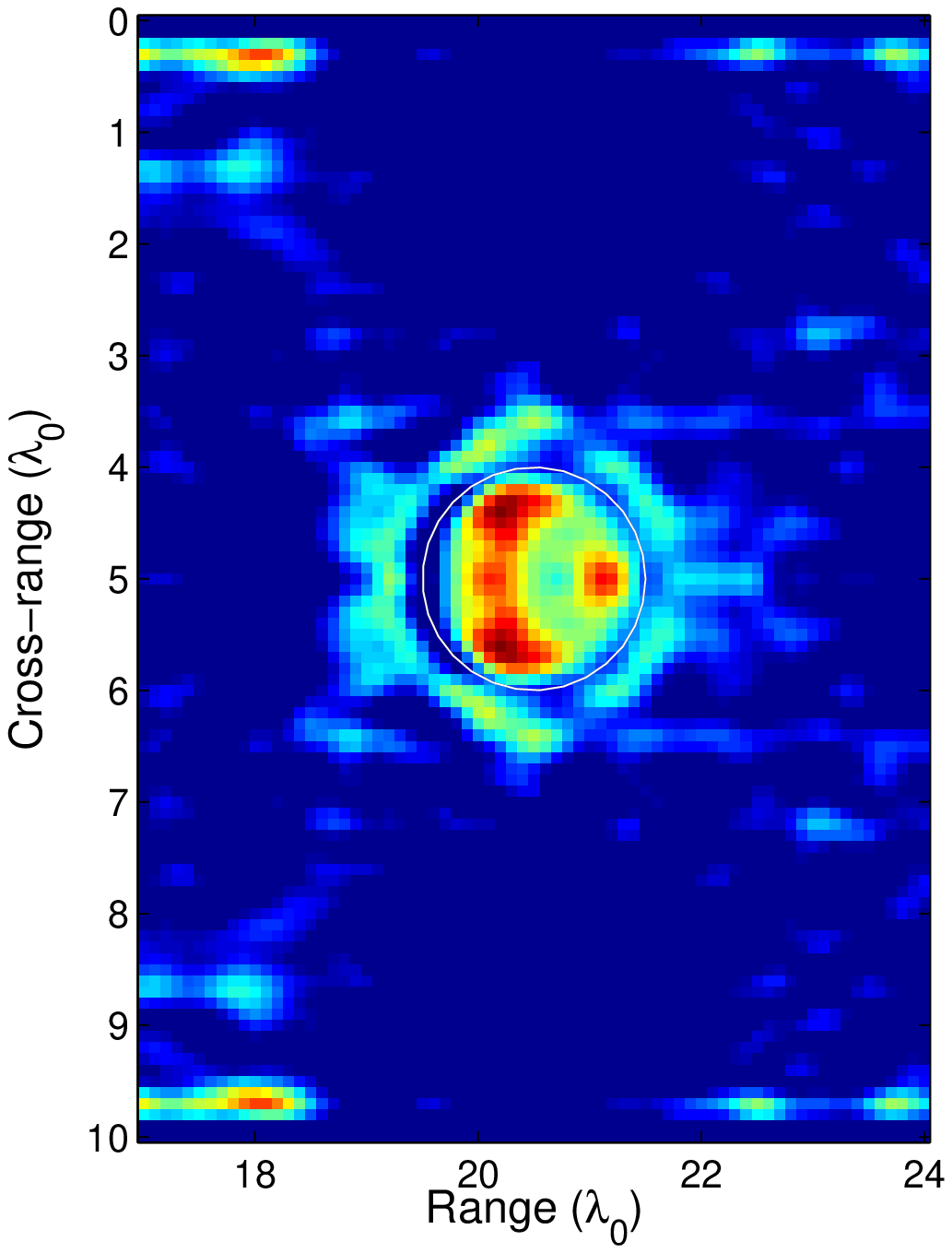}  
\end{center}
\end{minipage}
\caption{From left to right: Imaging with $\Ia$ for a disc scatterer centered at $(z^*,x^*)=(21.5,5)\lambda_0$, for different array apertures $|\mathcal{A}| = 0.75D,\ 0.5D$ and $0.25D$, for 
$k_c=0.9733 k_0$, $k_0 = \pi/10$ and $B=0.92f_c$. 
On the he top row plots we use no threshold, while for the plots in the bottom row we have a threshold  $\ell=0.4$. }
\label{fig:IKMa_d_partial}
\end{figure}
\begin{figure}[ht]
\centering
\begin{minipage}{0.32\linewidth}
\begin{center}
\hspace*{0.3cm} $|\mathcal{A}| = 0.75D$ \\
\includegraphics[width=\linewidth]{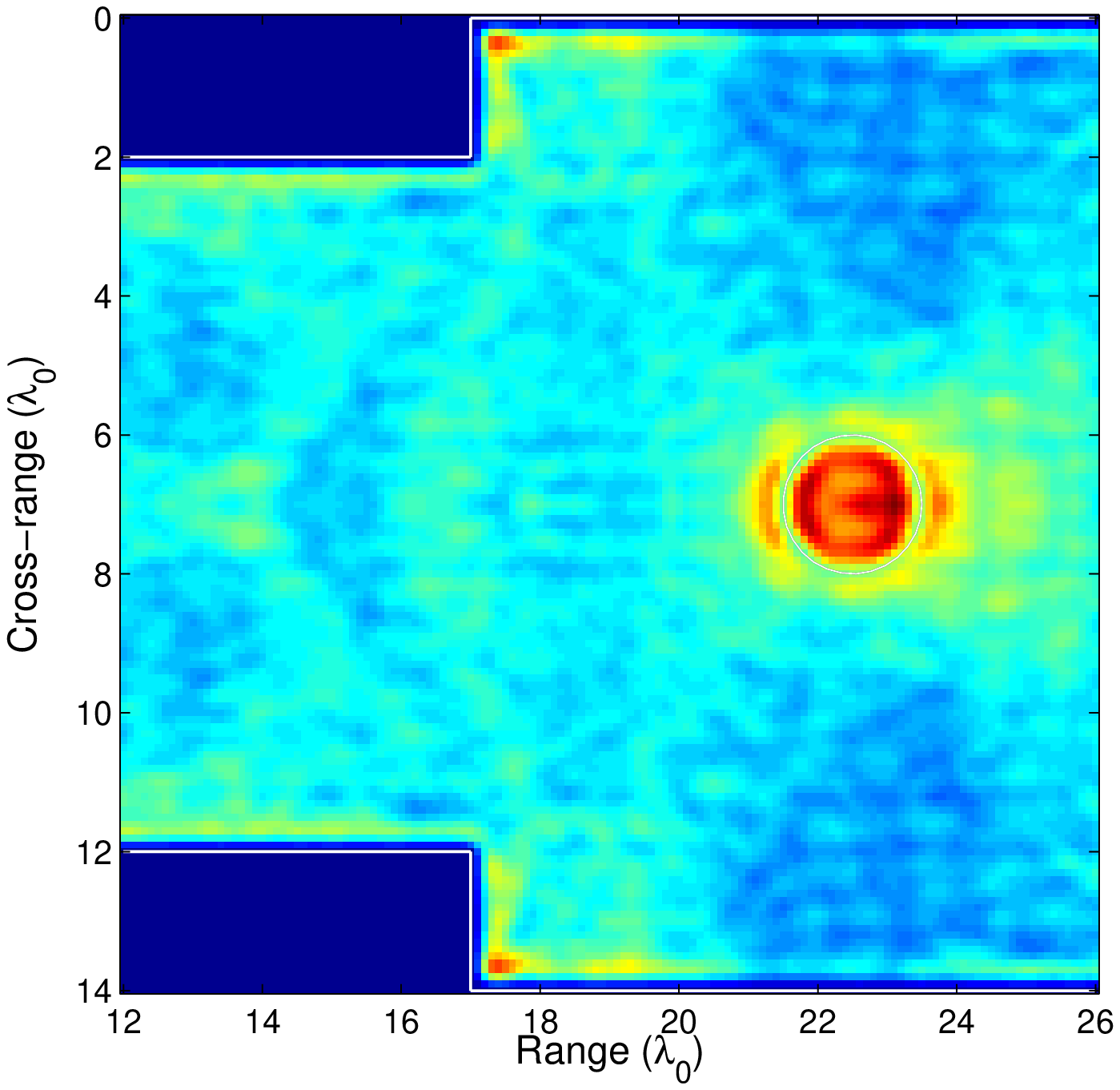}  \\
\includegraphics[width=\linewidth]{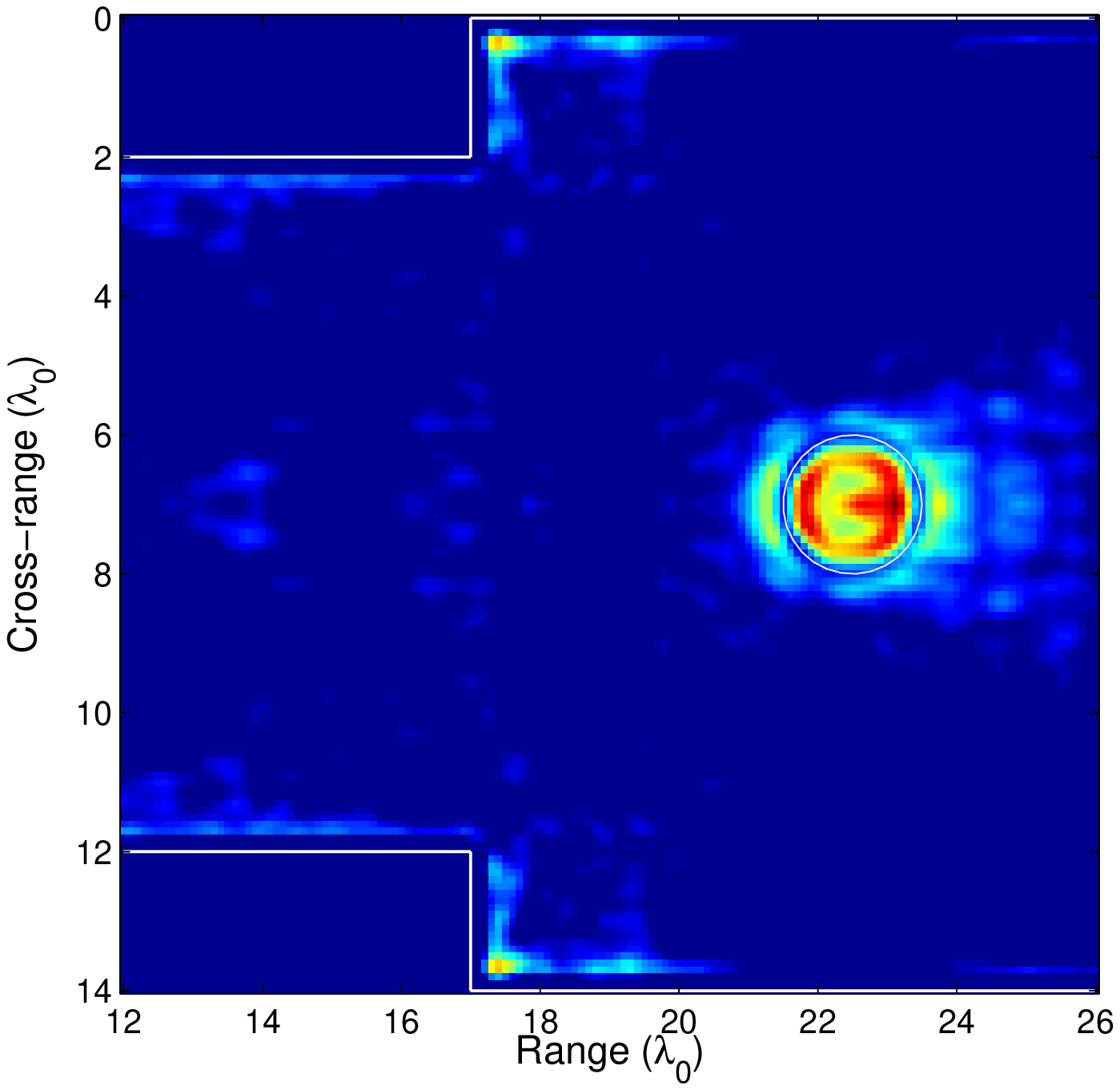}
\end{center}
\end{minipage}
\begin{minipage}{0.32\linewidth}
\begin{center}
\hspace*{0.3cm} $|\mathcal{A}| = 0.5D$ \\
\includegraphics[width=\linewidth]{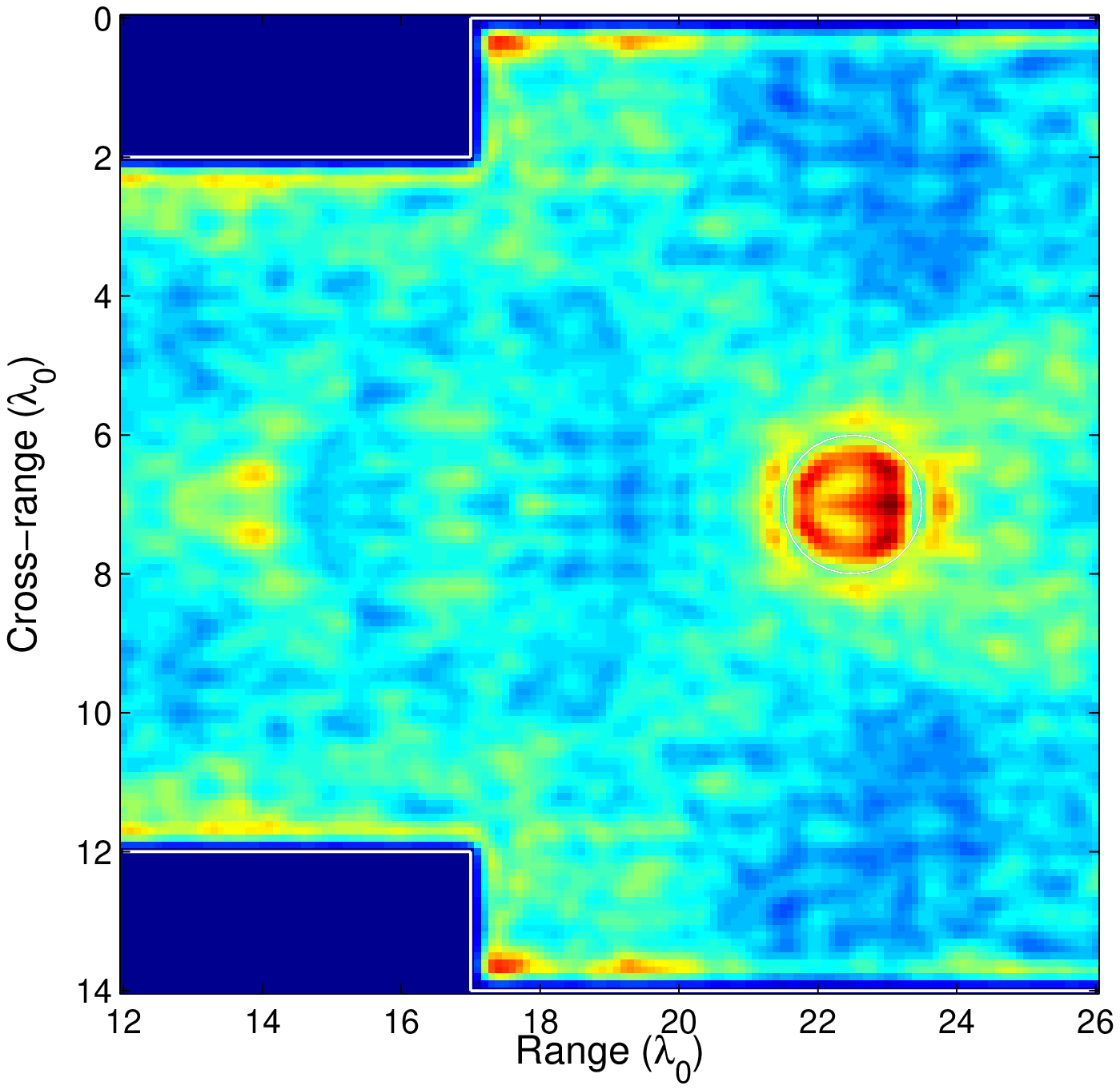}  \\
\includegraphics[width=\linewidth]{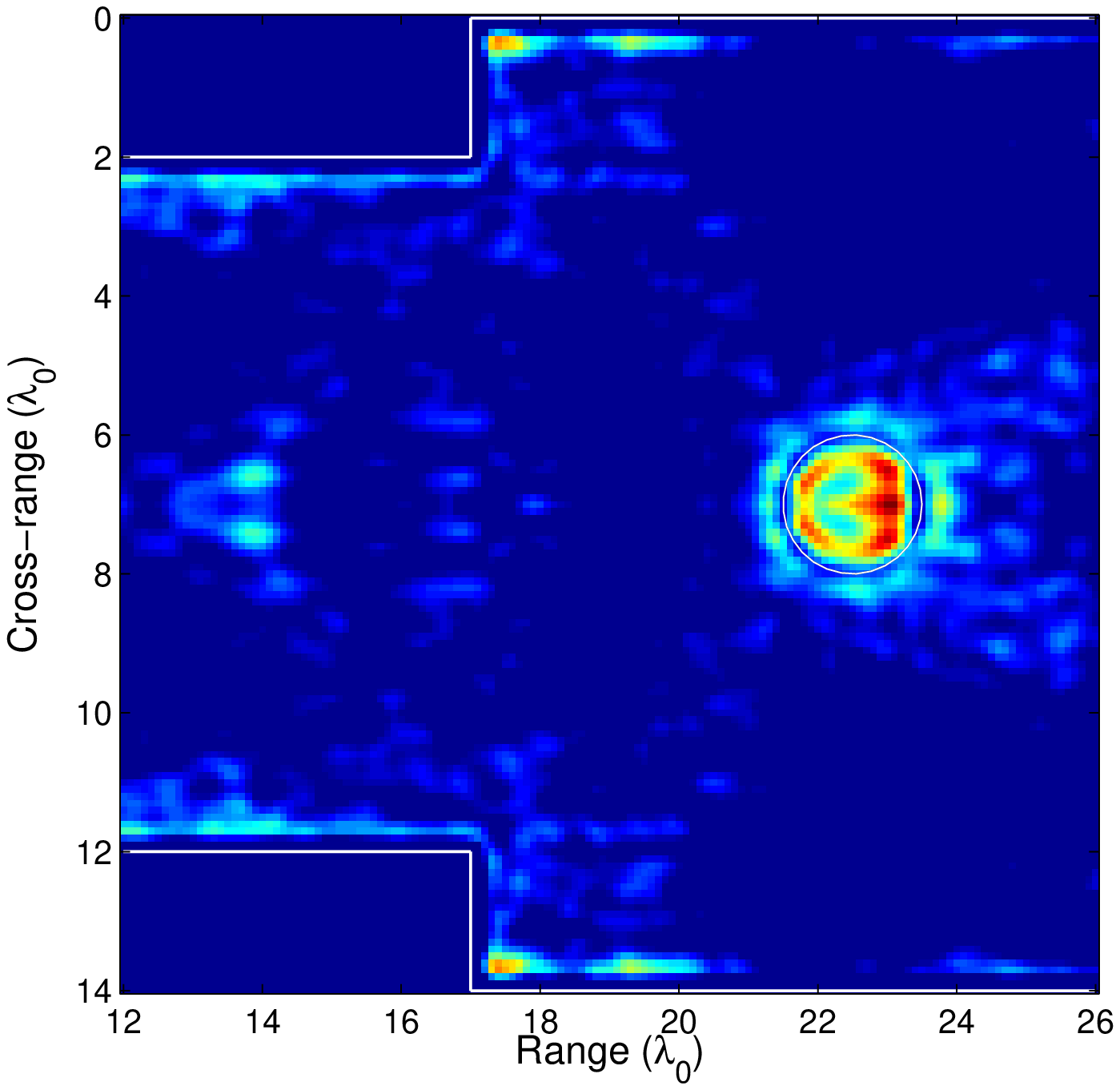}  
\end{center}
\end{minipage}
\begin{minipage}{0.32\linewidth}
\begin{center}
\hspace*{0.3cm} $|\mathcal{A}| = 0.25D$ \\
\includegraphics[width=\linewidth]{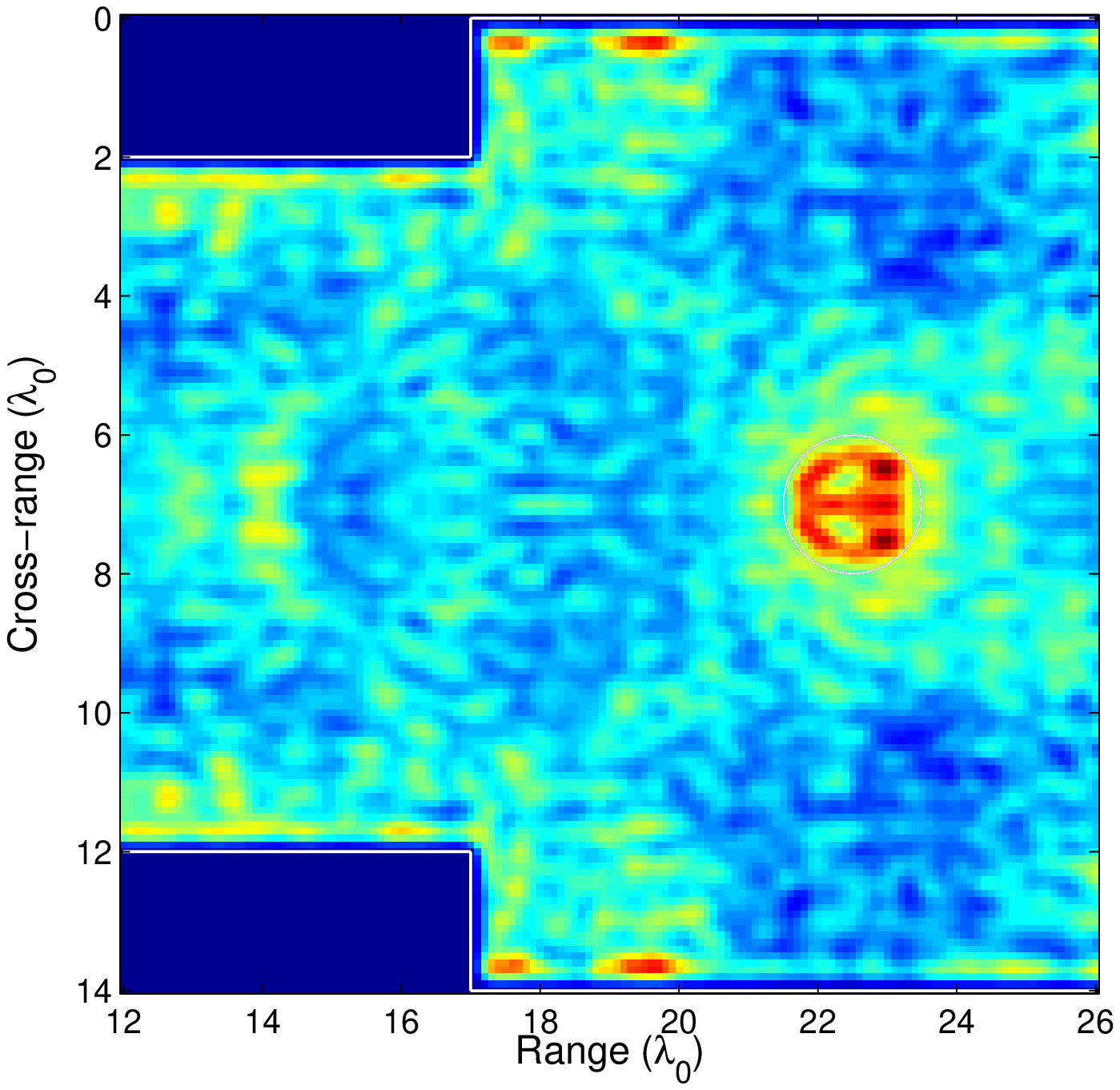}  \\
\includegraphics[width=\linewidth]{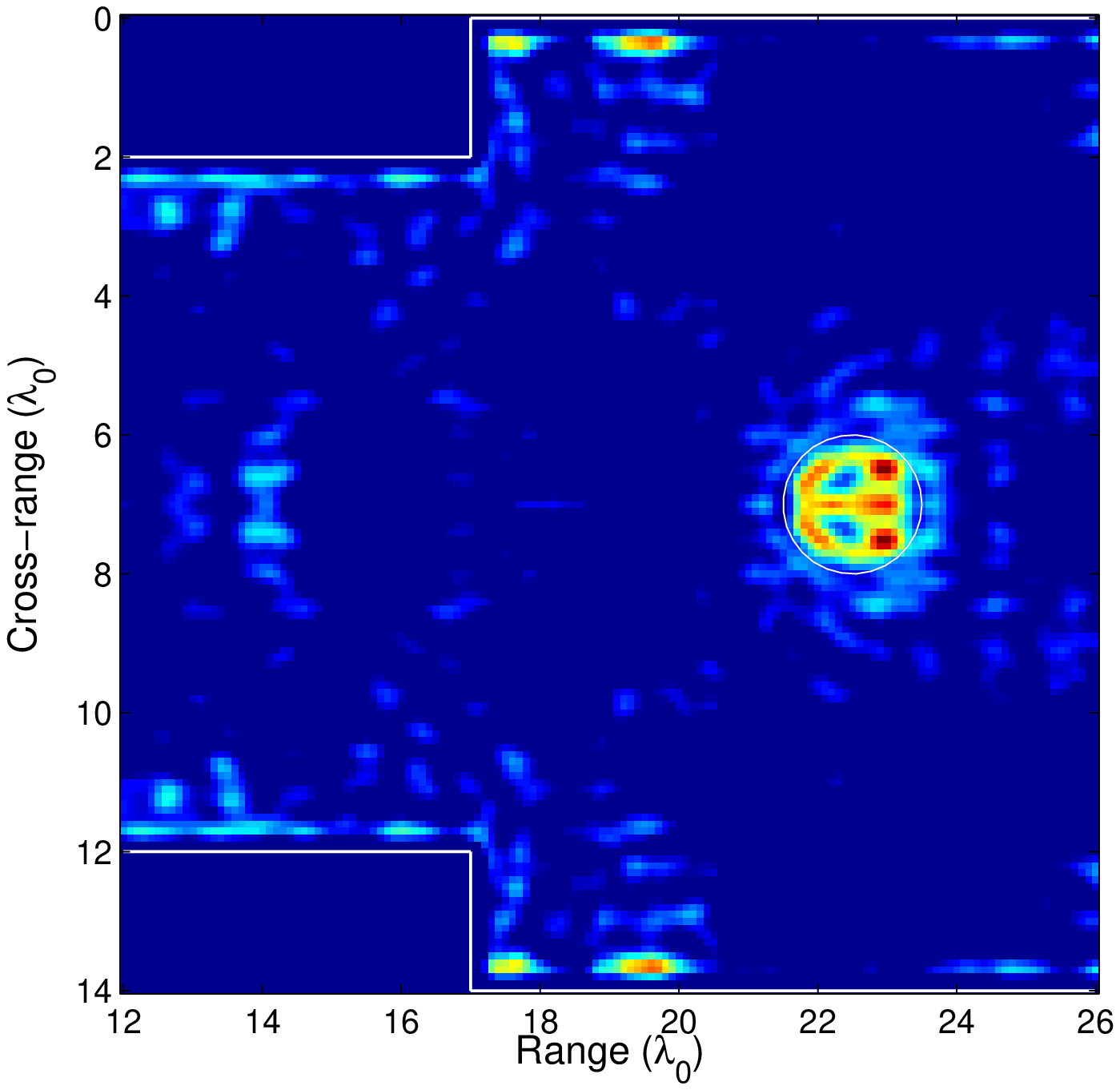}  
\end{center}
\end{minipage}
\caption{From left to right: Imaging with $\Ia$ for a disc scatterer centered at $(z^*,x^*)=(22.5,7)\lambda_0$, for different array apertures $|\mathcal{A}| = 0.75D,\ 0.5D$ and $0.25D$,  
 for $k_c=0.9733 k_0$, $k_0 = \pi/10$ and $B=0.92f_c$. 
On the top row plots we use no threshold, while for the plots in the bottom row we have a threshold  $\ell=0.4$. }
\label{fig:IKMa_d_b_partial}
\end{figure}

We show in \cref{fig:IKMa_d_partial,fig:IKMa_d_b_partial} imaging results obtained for the same 
configurations as in \cref{fig:IKMt_d_c,fig:IKMa_d_b}, respectively. The difference is that here we consider 
array apertures $|\mathcal{A}| = 0.75D,\ 0.5D$ and $0.25D$ where $D$ is the total width of the waveguide
in the cross-range direction. As illustrated in these figures the image quality deteriorates as the array aperture decreases 
but rather moderately. Indeed, comparing these images with the corresponding ones in \cref{fig:IKMt_d_c}, one may confirm
that the images for $|\mathcal{A}| = 0.75D$ are almost indistinguishable from the full aperture ones, and they are
still quite good for $|\mathcal{A}| = 0.25D$! 


\begin{figure}[ht]
\begin{center}
\begin{minipage}{0.22\linewidth}
\begin{center}
$|\mathcal{A}| = D $ \\
\includegraphics[width=\linewidth]{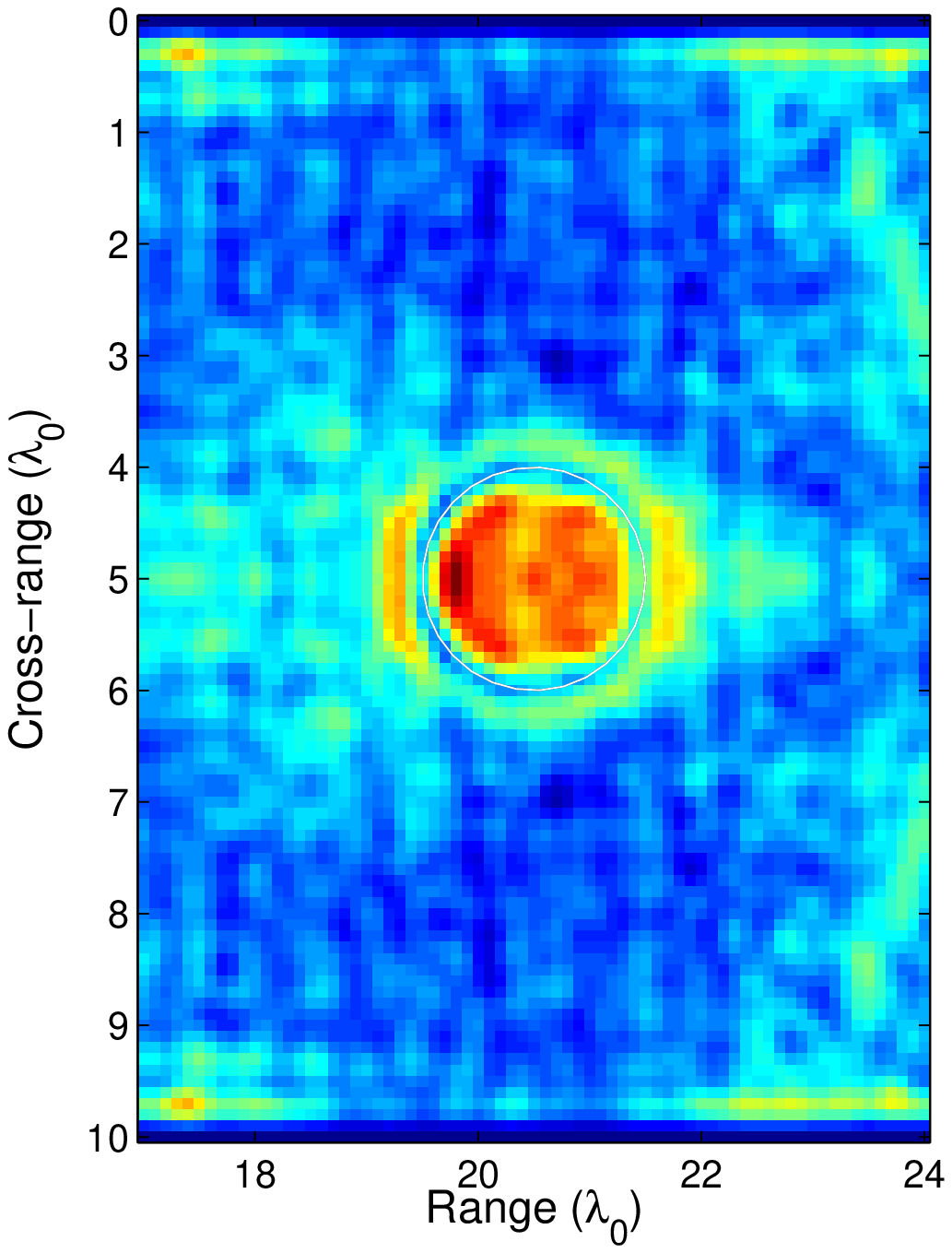}  \\
\includegraphics[width=\linewidth]{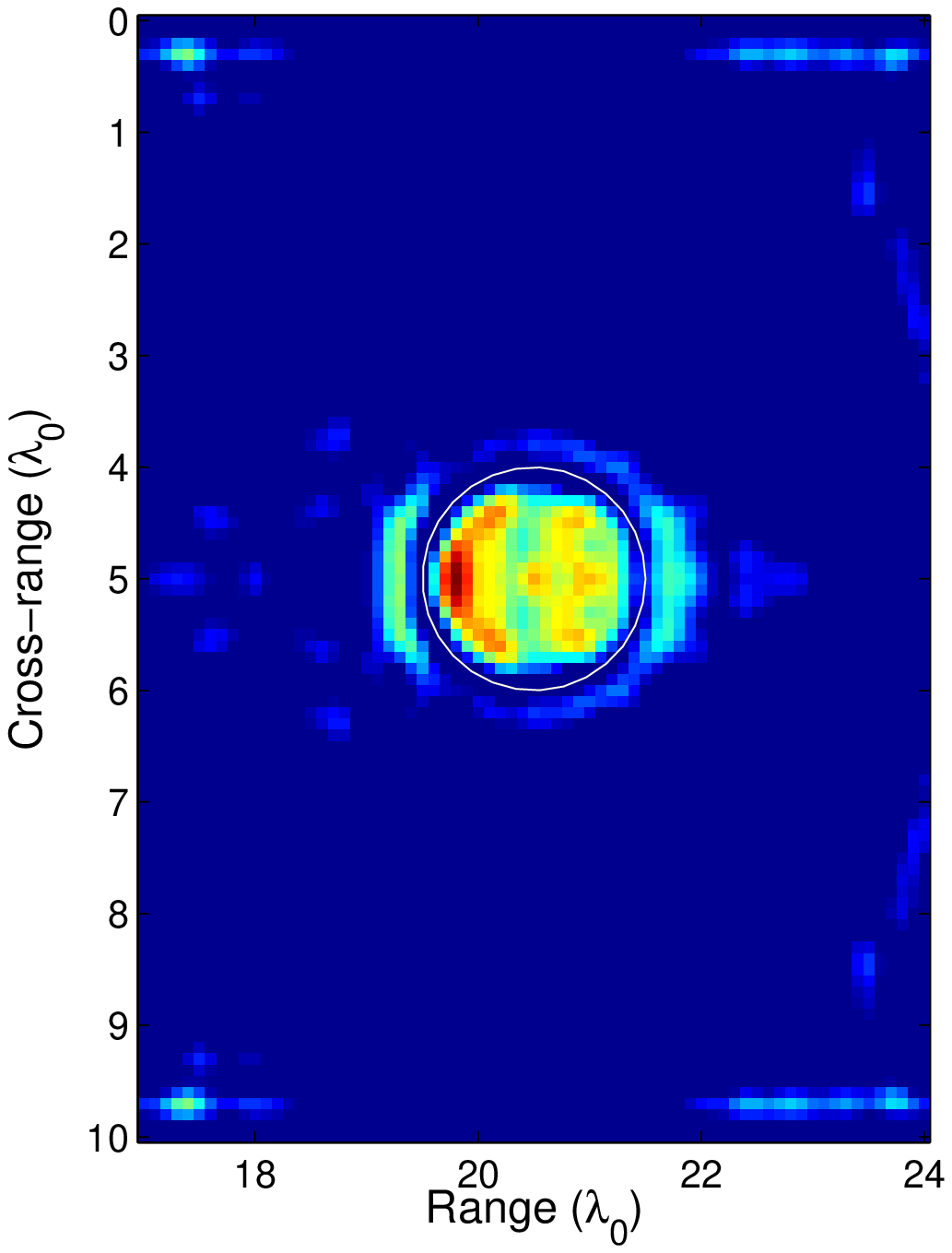}
\end{center}
\end{minipage}
\begin{minipage}{0.22\linewidth}
\begin{center}
$|\mathcal{A}| = 0.75D $ \\
\includegraphics[width=\linewidth]{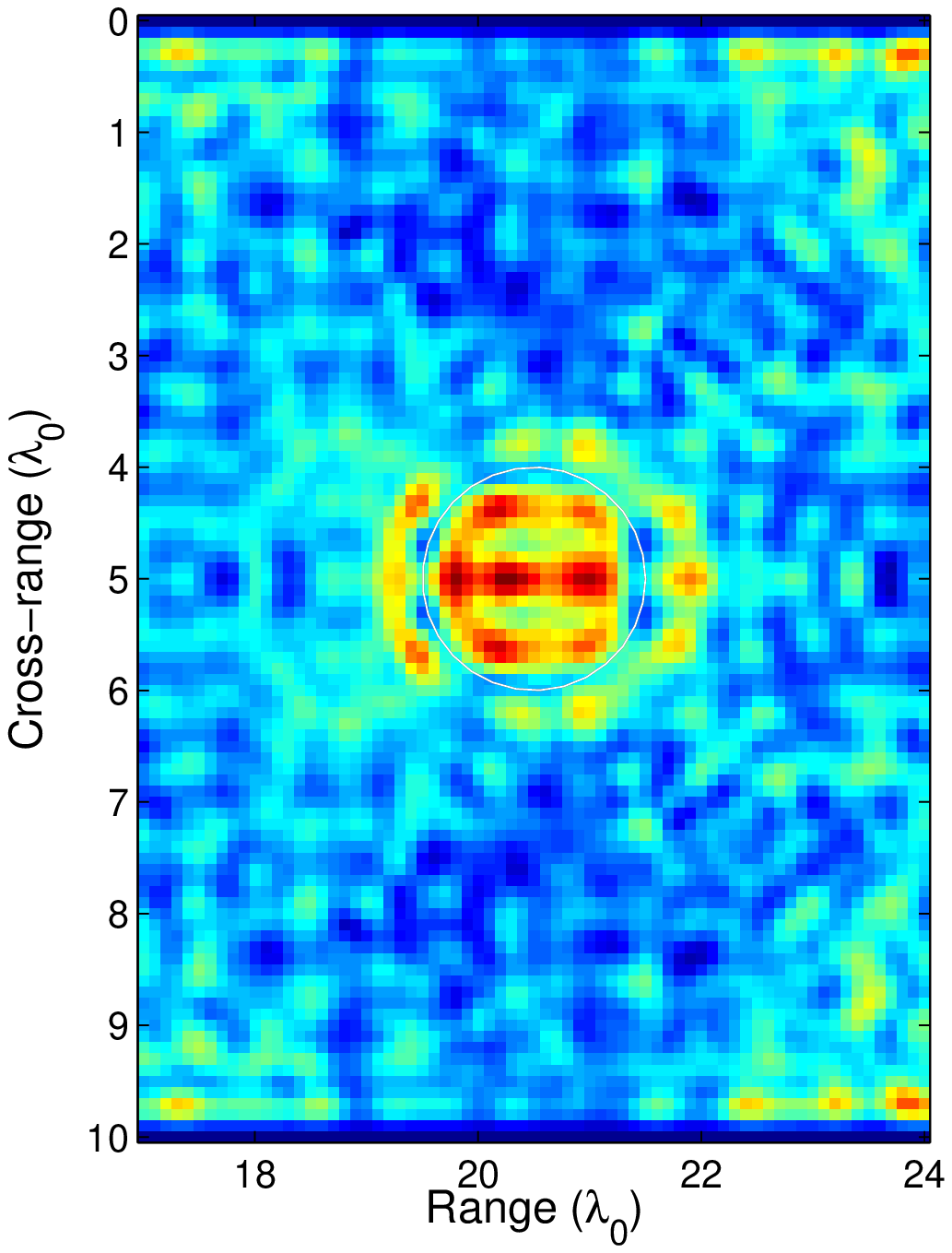}  \\
\includegraphics[width=\linewidth]{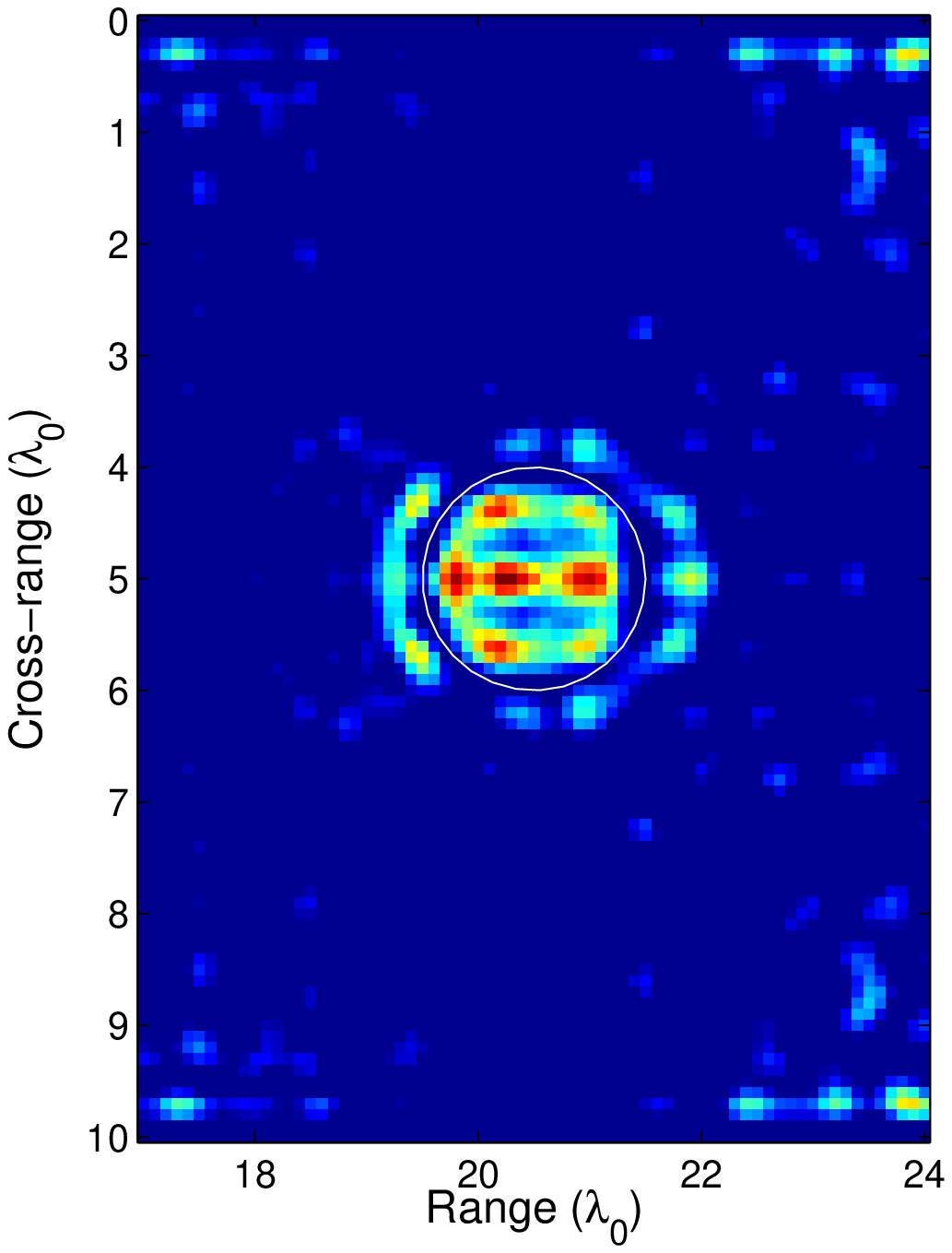}
\end{center}
\end{minipage}
\begin{minipage}{0.22\linewidth}
\begin{center}
$|\mathcal{A}| = 0.5D $ \\
\includegraphics[width=\linewidth]{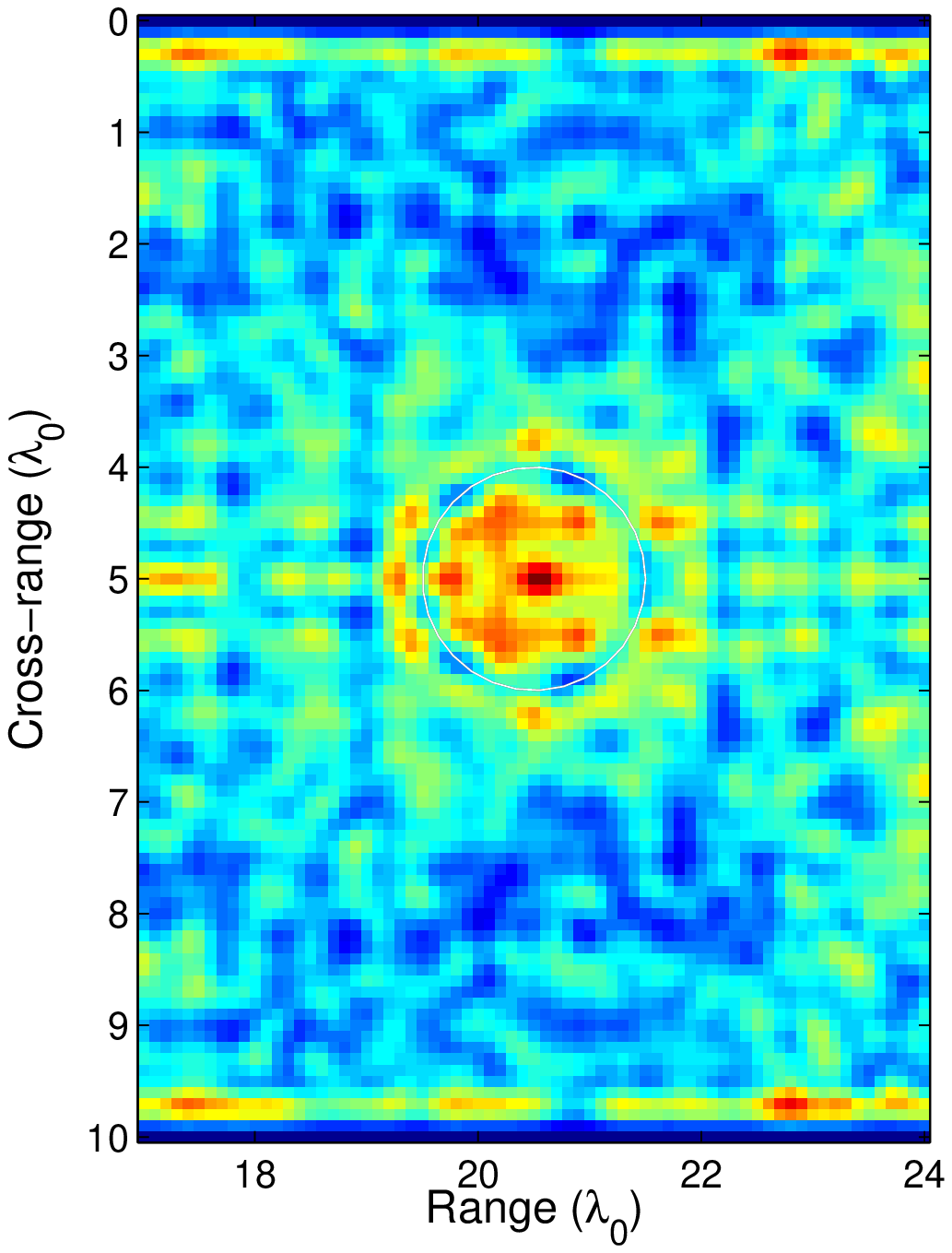}  \\
\includegraphics[width=\linewidth]{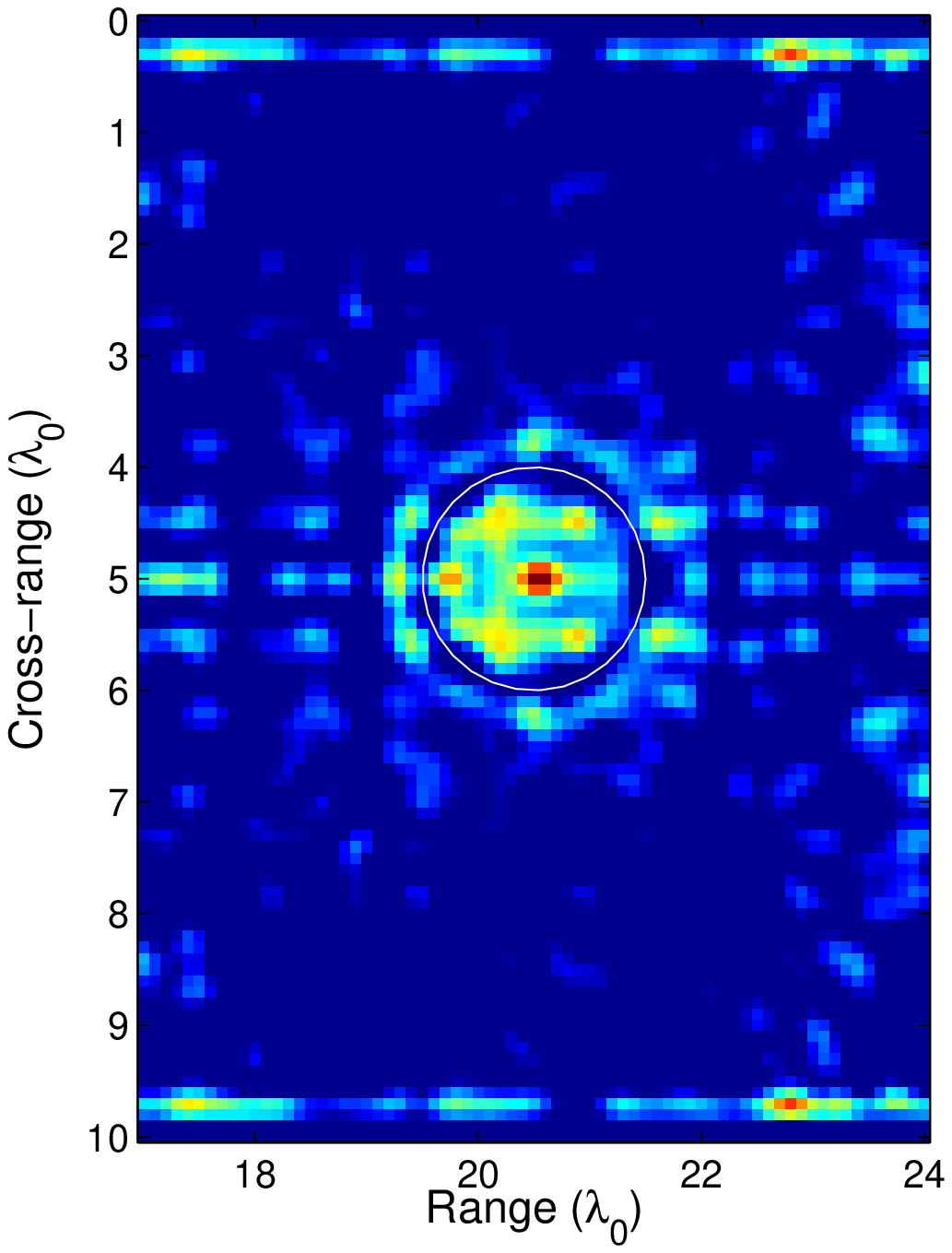}
\end{center}
\end{minipage}
\end{center}
\caption{From left to right: Imaging with $\Ia$ for a disc scatterer centered at $(z^*,x^*)=(20.5,5)\lambda_0$, using a synthetic aperture array with length $|\mathcal{A}| = D,\ 0.75D$ and $0.5D$, for $k_c=0.9733 k_0$, $k_0 = \pi/10$ and $B=0.92f_c$. 
On the  top row plots we use no threshold, while for the plots in the bottom row we have a threshold  $\ell=0.4$. }
\label{fig:IKMa_d_SAR}
\end{figure}

To synopsize, our numerical results indicate that the imaging method based on $\Ia$ can be used for reconstructing extended scatterers that are located in terminating waveguides  of complex geometry. The data used is the usual array response matrix which may cover 
only part of the vertical cross-section of the waveguide. The array response matrix is then projected on the propagating modes in an adequate way using the trigonometric polynomials on the array aperture as in \cref{{eq:Qpartial},{eq:Smatrx}}. We note that the same procedure can be followed for synthetic aperture data collected by a single transmit/receive element. In the latter case the data consist only of the diagonal entries of the array response matrix. We have numerically observed that the image resolution remains the same in this case while the SNR is worse; this is expected since the number of measurements is reduced to $N$ for the synthetic aperture instead of $N^2$ that are 
tabulated in the array response matrix. As an example, we show in \cref{fig:IKMa_d_SAR} full and partial aperture imaging results for the same imaging configuration as in  \cref{fig:IKMa_d_partial} but with a synthetic aperture that is formed with a single transmit/receive element. 

Note that to form the image with $\Ia$ we need the Green's function in the semi-infinite waveguide, which can be computed numerically 
assuming that the geometry and background velocity in the waveguide are known.  This is necessary for complex geometries and/or propagation media in which case it is not possible to derive an analytical expression for the Green's function. 
We have also assessed the performance of the imaging method with fully non-linear scattering data. 
Let us also mention that we expect this imaging method to be very robust to additive uncorrelated measurement noise 
since $\Ia$ is the average of the corresponding single frequency images over the available bandwidth. 

\subsection{Imaging in a three-dimensional terminating waveguide}
\begin{figure}[ht]
 \centering
\includegraphics[width=0.75\linewidth]{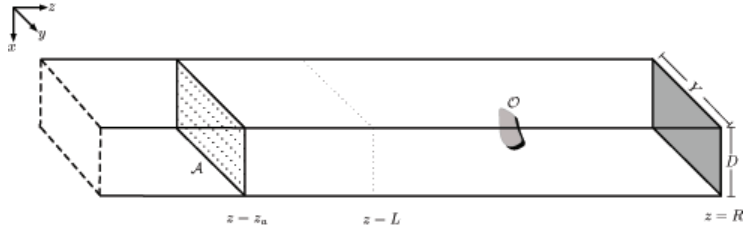}
\caption{Schematic representation of the imaging setup in a three-dimensional waveguide. }
  \label{fig:setup-3d}
\end{figure}

We finally consider the problem of imaging an extended reflector in a three-dimensional terminating waveguide with a bounded rectangular cross-section. The imaging setup is illustrated in \cref{fig:setup-3d}. We denote as before, with $z$ the range variable and with $x$, $y$ the two cross-range variables. The vertical cross-section of the waveguide (xy-plane) is the rectangle $(0,D) \times (0,Y)$  
and the terminating boundary is at $z=R$. Homogeneous Dirichlet boundary conditions are imposed on all of the waveguide's boundaries.

For a homogeneous waveguide with a simple geometry as the one depicted 
in \cref{fig:setup-3d}, the analytic expression for the Green's 
function in the waveguide may be retrieved in straightforward way from 
the analogous two-dimensional expressions. Consequently, the 
linearized scattered acoustic field may be computed on the array of 
receivers $\mathcal{A}$ that span the bounded cross-section of the 
waveguide. Imaging is performed by the functional  $\Ia$ as in 
\cref{eq:Ia} with the projected response matrix $\widehat{\mathbb{Q}}$ 
defined by adequately modifying \cref{eq:sct_respmatr_proj} so that 
the integrals are taken over the two-dimensional array aperture.

\begin{figure}[ht]
\begin{minipage}{0.3\linewidth}           
\includegraphics[width=\linewidth]{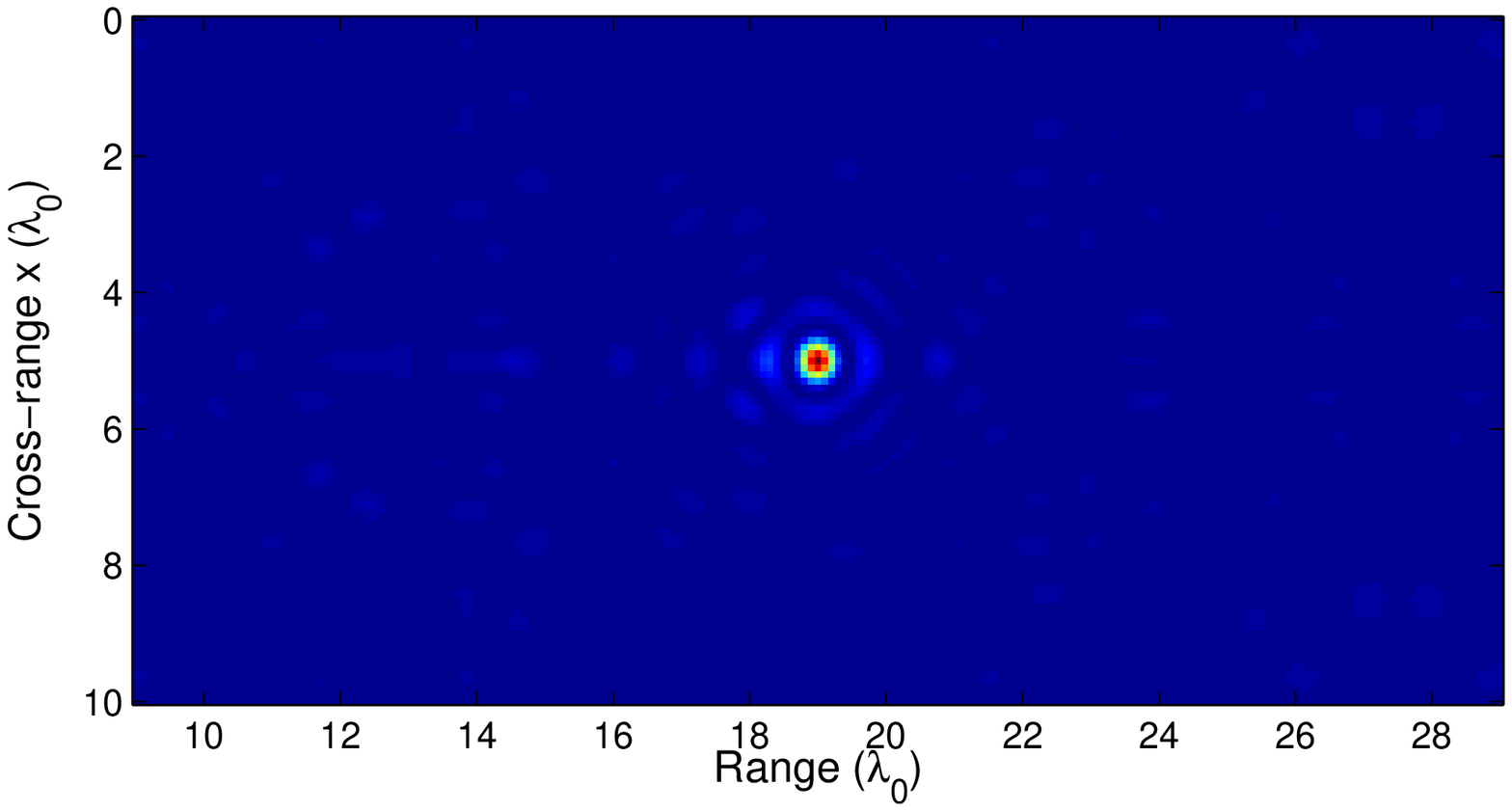} \\
\includegraphics[width=\linewidth]{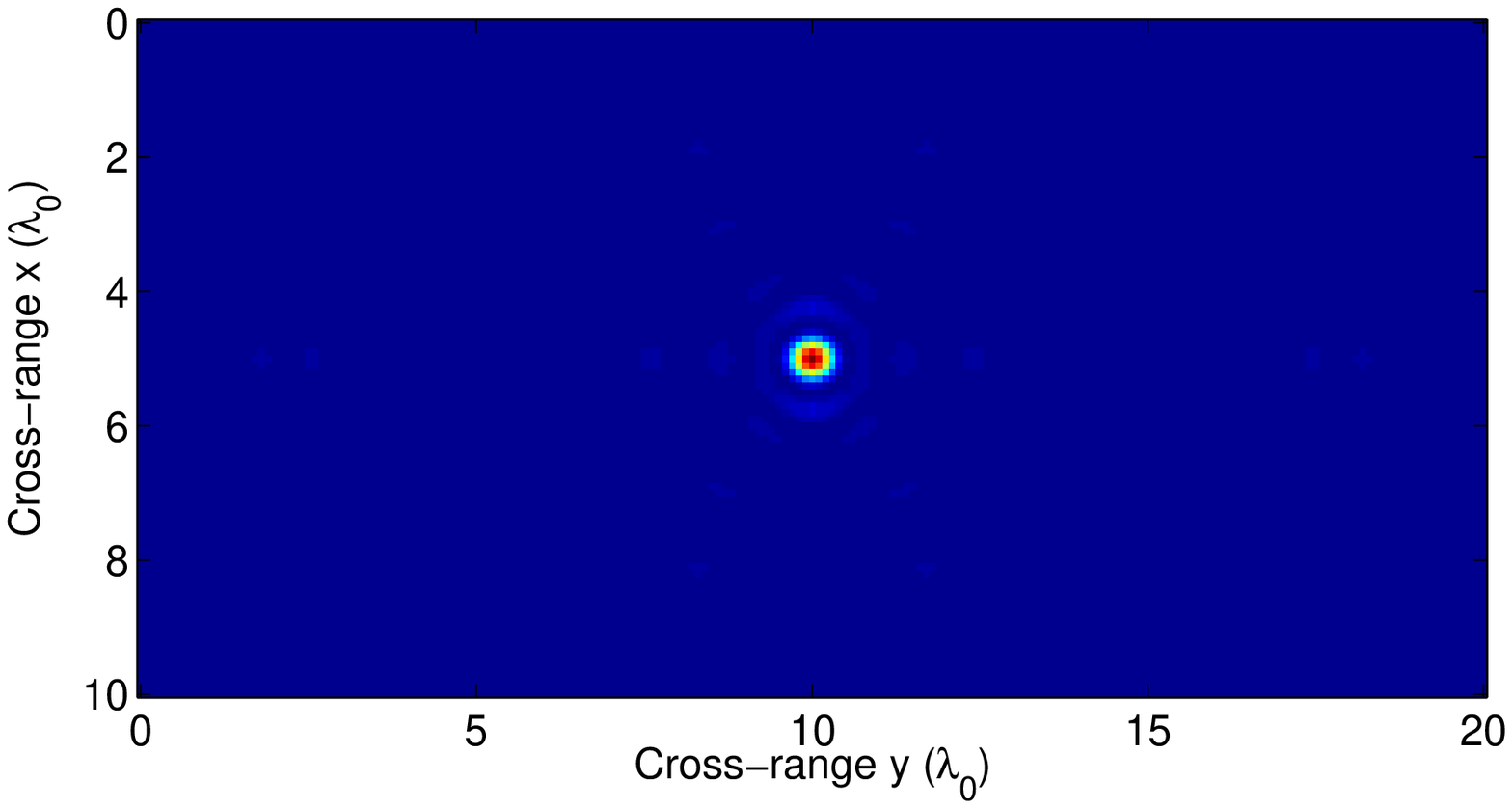} 
 \end{minipage}
 \begin{minipage}{0.282\linewidth}
\includegraphics[width=\linewidth]{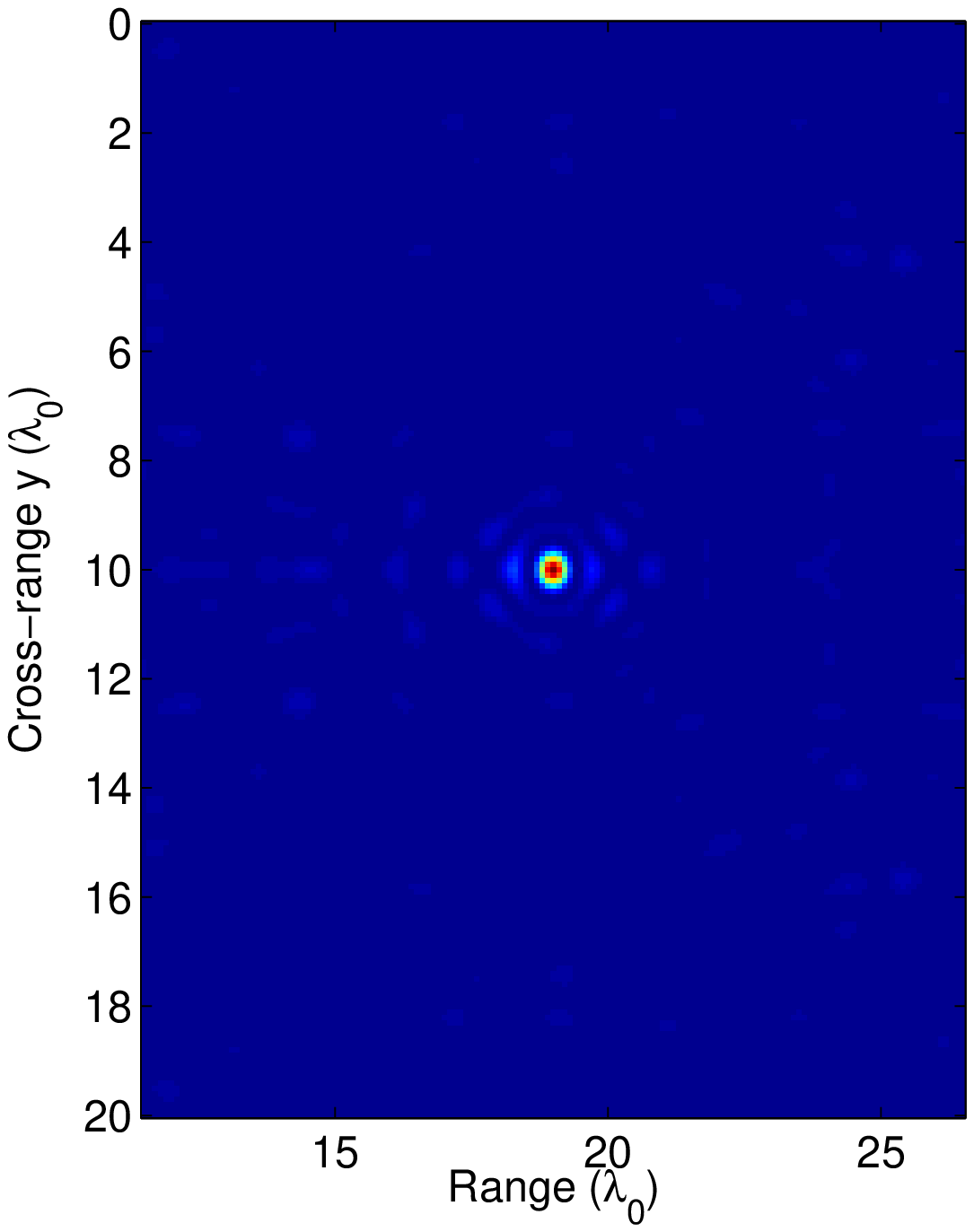}
\end{minipage}
\begin{minipage}{0.37\linewidth}
\includegraphics[width=\linewidth]{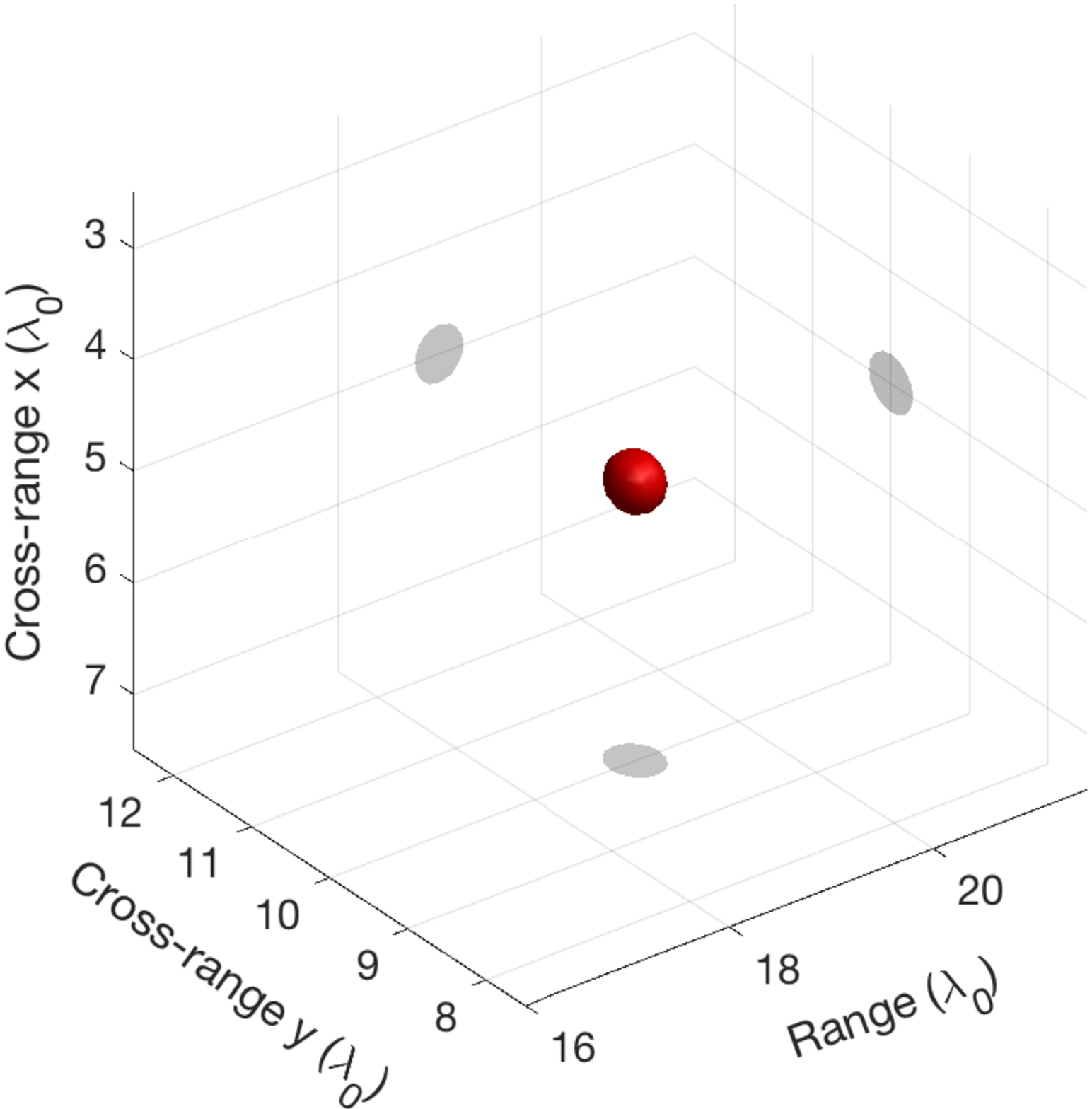}
\end{minipage}
\caption{Modulus of $\Ia$ for the $zx$-plane (top left), $yx$-plane (bottom left) and $zy$-plane (middle), for a single frequency 
$k = 0.973k_0$, $k_0 = \pi/10$, for a point reflector placed at $\xb^\ast = (19,5,10)~\lambda_0$. On the right plot we show the three dimensional reconstruction of the point reflector.}
  \label{fig:3dpoint}
\end{figure}

Without giving the details of the computations we present as a proof of concept in the following figures some preliminary results that illustrate how this imaging methodology performs in the three-dimensional case.  In \cref{fig:3dpoint} we show the reconstruction for a point reflector located at $\xb^\ast = (19,5,10)\,\lambda_0$. The vertical cross-section has size $[0,10\lambda_0] \times [0,20\lambda_0]$ and the terminating boundary is placed at $z=28\lambda_0$.
This is a single frequency result for $k = 0.973k_0$ ($k_0 = \pi/10$), and essentially depicts the point spread function of $\Ia$ in three dimensions.  We observe that the resolution is $\lambda/2$ in all directions as expected from our resolution analysis.

\begin{figure}[ht]
 \centering
 \begin{minipage}{0.31\linewidth}           
\includegraphics[width=\linewidth]{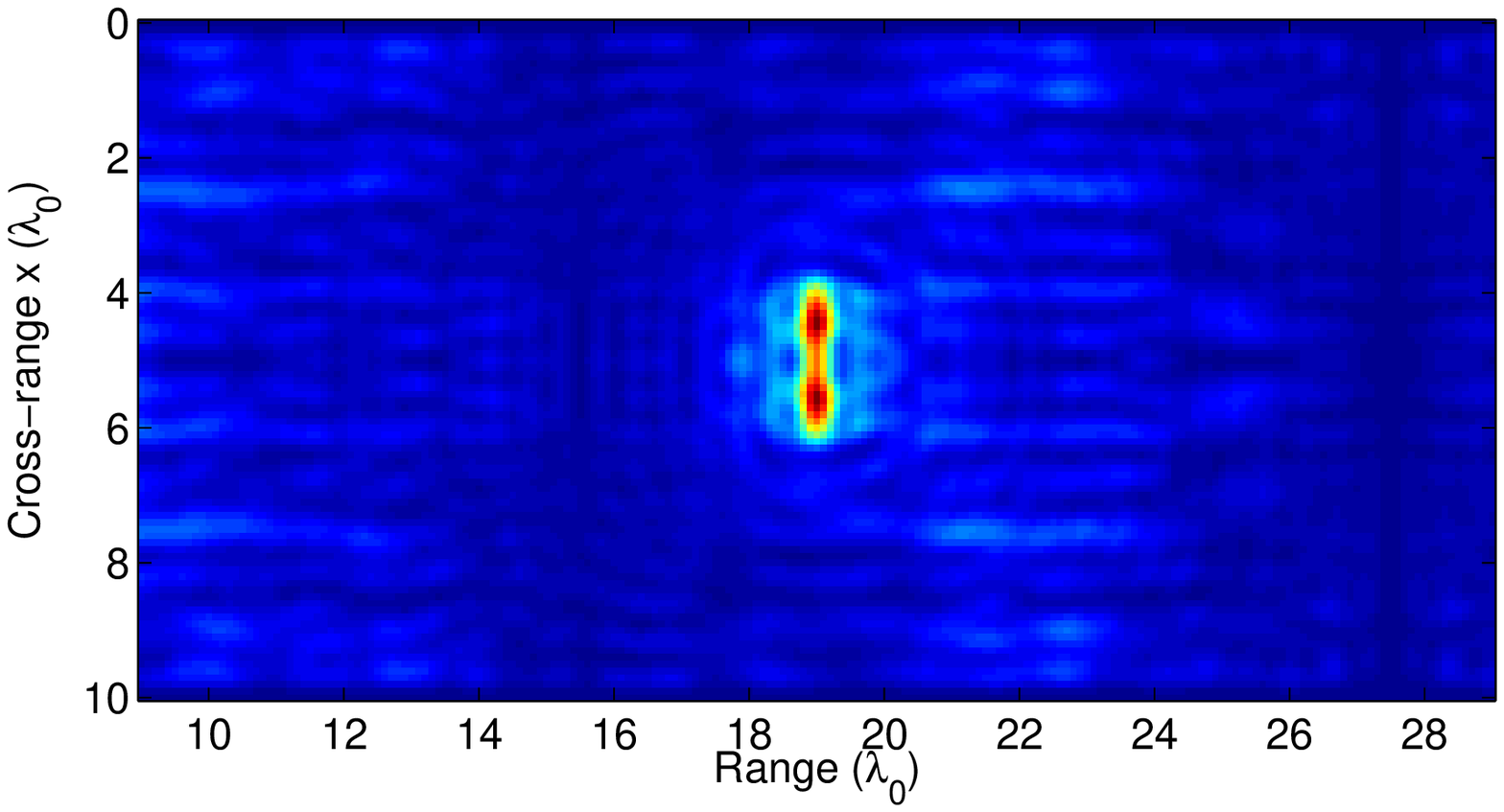} \\
\includegraphics[width=\linewidth]{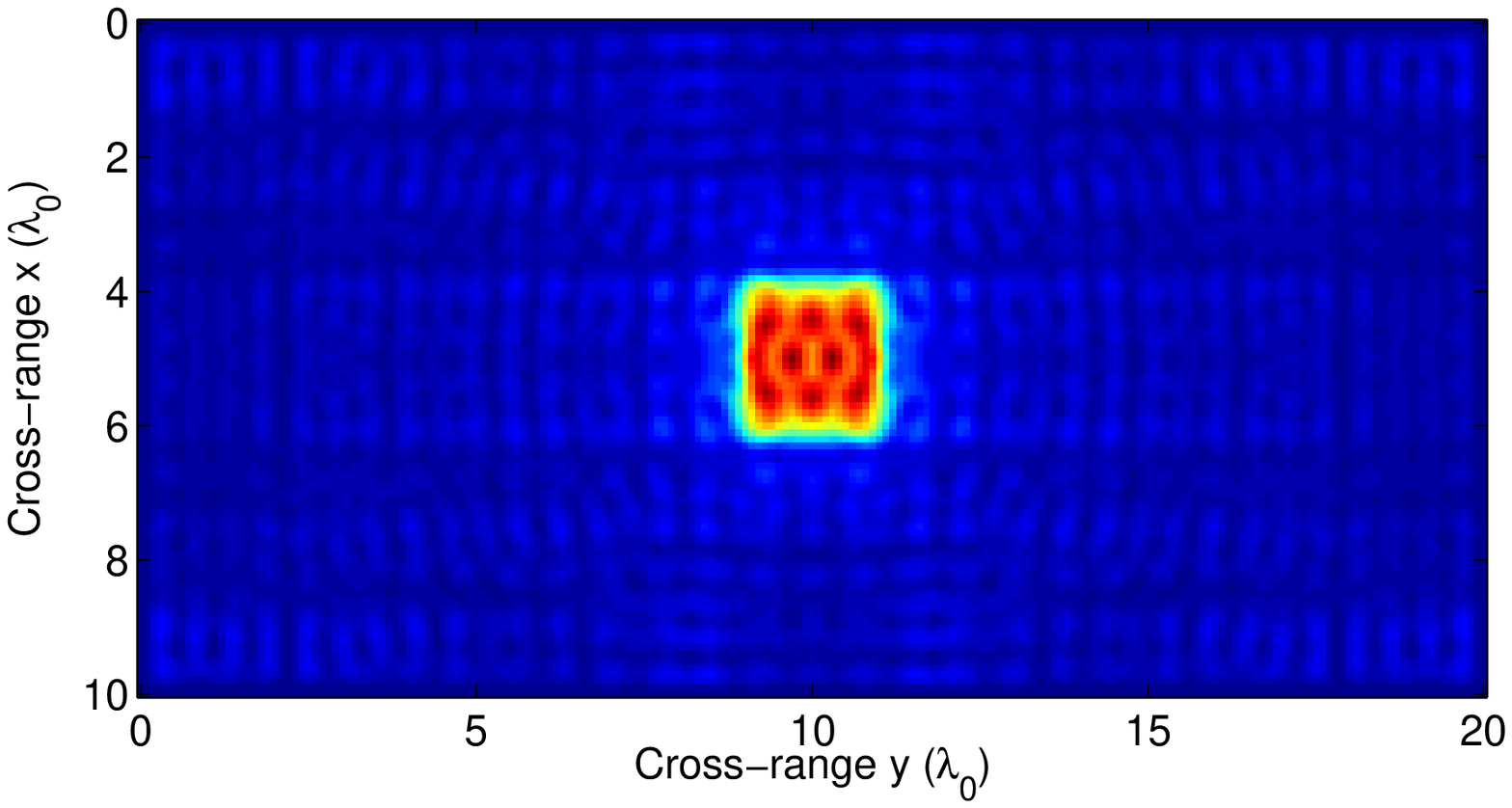} 
 \end{minipage}
\begin{minipage}{0.282\linewidth}
\includegraphics[width=\linewidth]{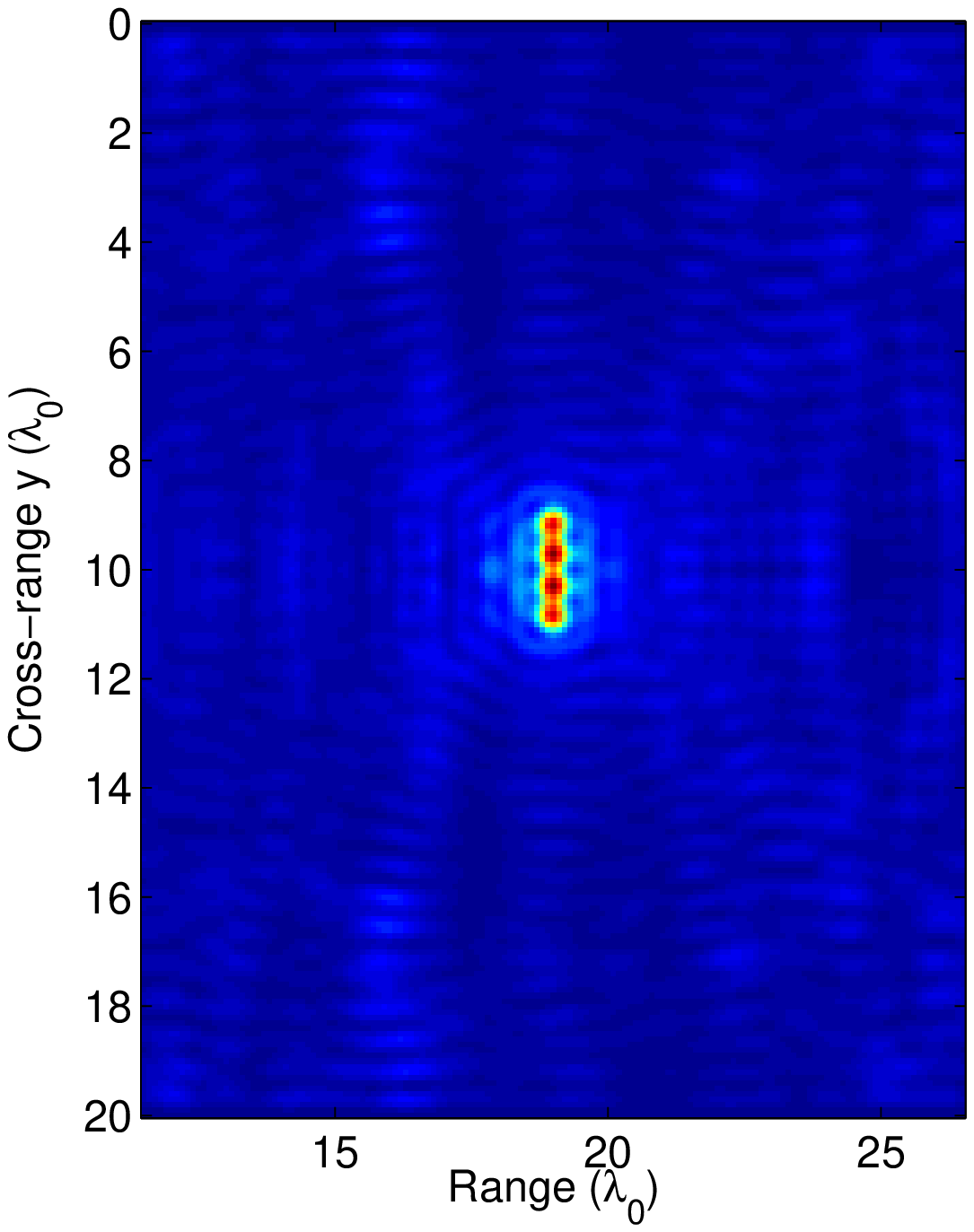}
\end{minipage}
\begin{minipage}{0.35\linewidth}
\includegraphics[width=\linewidth]{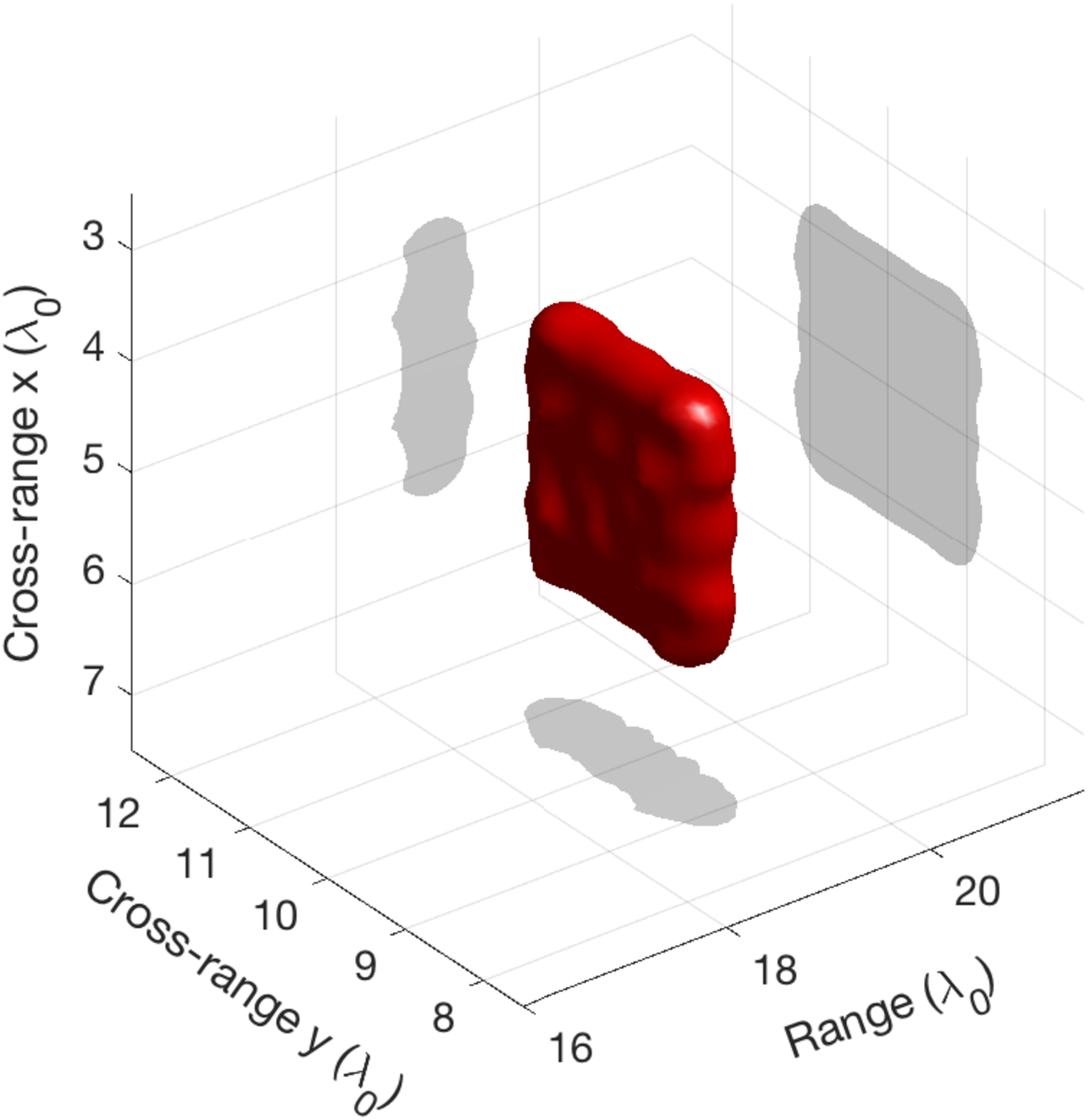}
\end{minipage}
\caption{Modulus of $\Ia$ for the $zx$-plane (top left), $yx$-plane (bottom left) and $zy$-plane (middle), for a single frequency $k = 0.973k_0$ for a square reflector $[9,11]\lambda_0 \times[4,6]\lambda_0$ placed at $z=19\lambda_0$. On the right plot we show the three dimensional reconstruction of the square reflector.}
  \label{fig:3dsquare}
\end{figure}

In Figure \ref{fig:3dsquare} we display the modulus of $\Ia$ for a square-shaped screen reflector. We observe that the reconstructions are very good and the shape of the reflector can be retrieved with accuracy. 

\begin{figure}[ht]
 \centering
\includegraphics[width=.35\linewidth]{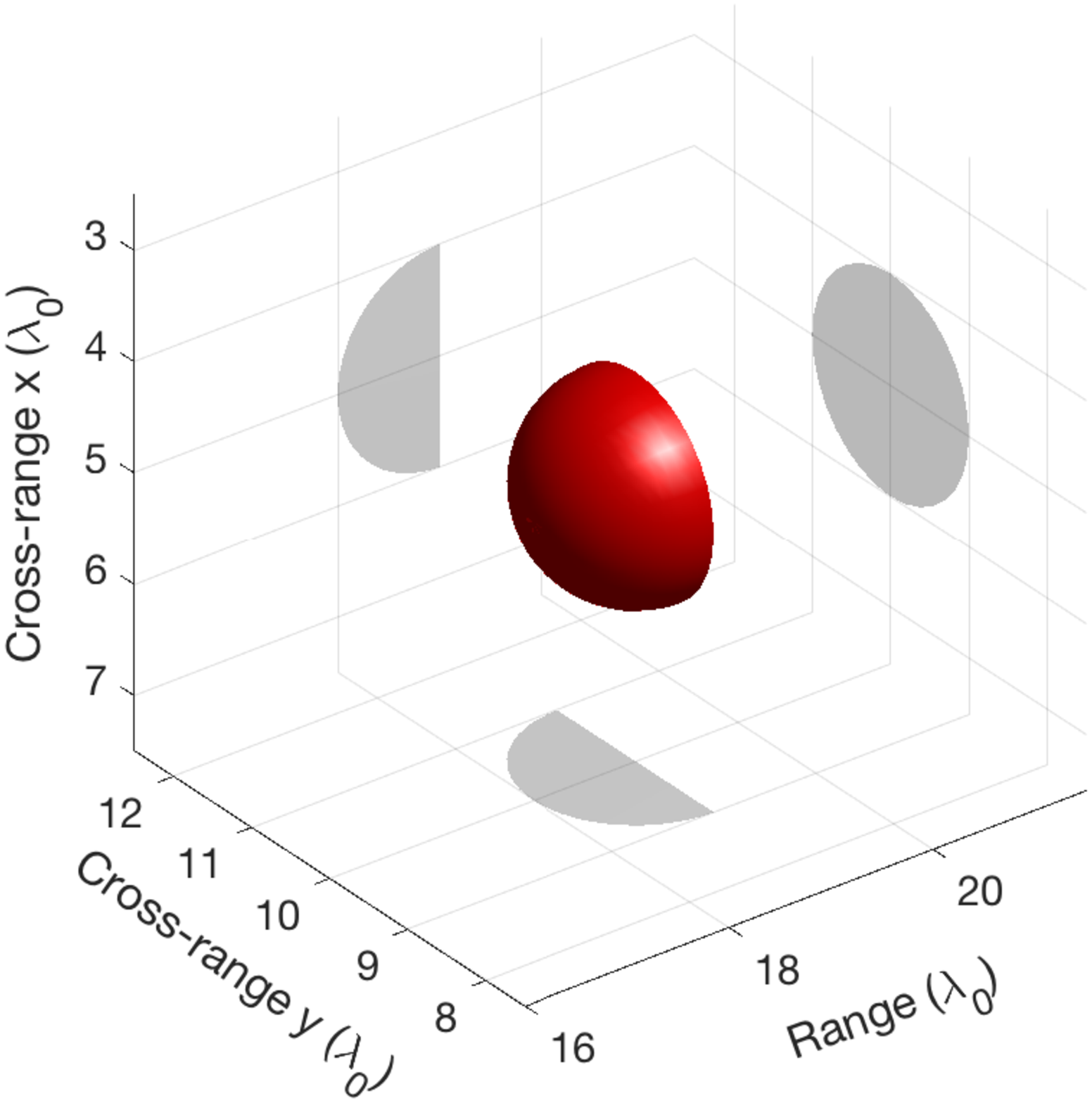}
\includegraphics[width=.35\linewidth]{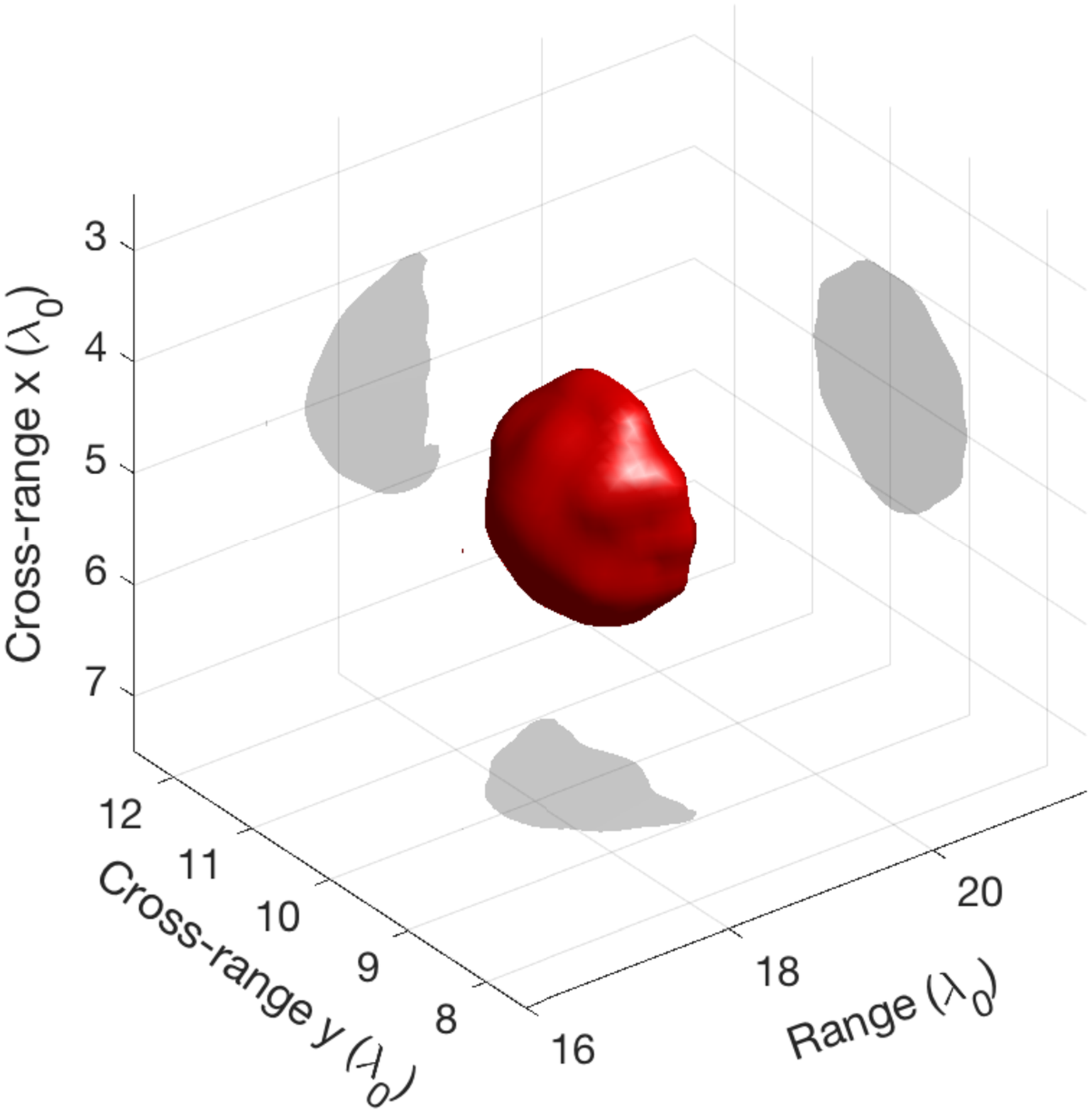}
\caption{Imaging a hemisphere with diameter $b=2\lambda_0$, centered at $\xb^\ast = (19,5,10)\lambda_0$. The true reflector is shown on the left. The modulus of $\Ia$ for a single frequency $k_c=0.9733 k_0$ is on the right plot. }
  \label{fig:3dhemi}
\end{figure}

A more challenging example is considered in \cref{fig:3dhemi} where we seek to reconstruct a hemisphere with diameter $b=2\lambda_0$, centered at $\xb^\ast = (19,5,10)\lambda_0$. The reconstruction shown on the right plot is very close to the true reflector's geometry shown on the left plot. These preliminary three-dimensional results are very promising.   Of course, more experiments with full wave scattered data and noise should be carried out to fully assess the performance of the method in three-dimensions. Also, we leave for future work the adequate modification of the imaging functional for the partial-aperture case in three dimensions.
%
%
%
%
%
\section*{Conclusions}
We considered the problem of imaging extended reflectors in terminating waveguides $\Omega \subset {\mathbb R}^2$ that consist of two subdomains: the semi-infinite strip 
$\Omega_{L^-} = (-\infty,L) \times (0,D)$ and a bounded domain $\Omega_{L^+}$.  We also assume that the medium is 
homogeneous in the semi-infinite strip $\Omega_{L^-}$ while it can be inhomogeneous in $\Omega_{L^+}$ which may also contain the reflector to be imaged. 
This formalism allows us to image reflectors in waveguides with complex geometries. We introduce an imaging functional that relies on the back-propagation of the modal projection of the array response matrix adequately defined so as to account for the array aperture. Our analysis shows that the resolution of the image is half a wavelength corresponding to the central frequency while the signal-to-noise ratio depends on the bandwidth. We observe a net improvement in the reconstructions compared to the infinite waveguide case  and recover the reflector's location, size and shape with very good accuracy. This is intuitively expected since in the terminating waveguide we benefit from the reflections (multiple-scattering paths) that bounce off the terminating boundary of the waveguide providing views of the reflector that are not available in the infinite waveguide. Our numerical results illustrate the robustness of the method for different array apertures ranging from full to one fourth of the waveguide's depth. We also obtain good reconstructions for synthetic array data obtained with a single transmit/receive element. Although the methodology was presented here in the two-dimensional case, the extension of the methodology to a three-dimensional waveguide with bounded cross-section does not present any conceptual difficulties.  
%
%
%
%
%
\section*{Acknowledgments}
Part of this material is based upon work supported by the National Science Foundation under Grant No. DMS-1439786 while the authors were in 
residence at the Institute for Computational and Experimental Research in Mathematics (ICERM) in Providence, RI, during the Fall 2017 semester.
The inspiring ICERM/BrownUniversity research environment and its kind hospitality 
are warmly acknowledged.
S. Papadimitropoulos and D. Mitsoudis also acknowledge support from IACM/FORTH that covered part of their expenses for  
participating in ICERM's  semester program. The work of C. Tsogka was partially supported by AFOSR FA9550-17-1-0238. 
%
%
%
%
%
\appendix
%
%
%
%
%
%
\section{The Kirchhoff-Helmholtz identity in a  terminating waveguide} \label{sec:KHpr}
Let $\Omega= \Omega_{L^-} \cup \Omega_{L^+}$ be the terminating waveguide described in 
\cref{sec:Form}, see \cref{fig:WG_setup}.
We assume that all the inhomogeneities of the medium are contained in $\Omega_{L^+}$ and that the wave speed
is constant in  $\Omega_{L^-}$. Let also $\{\mu_n,X_n\}_{n=1}^\infty$ be the eigenvalues and eigenvectors
of the negative Dirichlet Laplacian $-d^2 / dx^2$ in $(0,D)$ given in \cref{eq:eigs}.
We will consistently assume that the constant wavenumber, denoted by $k$, satisfies in $\Omega_{L^-}$:
$$
   \mu_M < k^2 < \mu_{M+1}, \quad \mbox{ for some index }M,
$$
and let $\beta_n$ denote the horizontal wavenumbers in $\Omega_{L^-}$ that are given in \cref{eq:betas}.
Hence $M$ is the number of {\it propagating modes} in $\Omega_{L^-}$.

For any $z_a < L$, let  $\mathcal{A} = \{(z_a,x):\ 0\le x\le D\}$ be the coresponding cross-section to the range direction, and 
let $\Omega_{\mathcal{A}} $ denote the bounded domain $\Omega \cap \{(z,x)\in{\mathbb R}^2: \; z\ge z_a\}$.
Note that here we are using the symbol $\mathcal{A}$  to denote an artificial boundary whereas in previous 
sections it denoted the array. Then, the eigenpairs $\{\mu_n,X_n\}_n$ allow us to define a
{\it Dirichlet-to-Neumann (DtN)} map, denoted by $T$, such that for each function $u$ in suitable 
function space
\begin{equation}   \label{eq:DtNop}
    T u(z,x) := \sum_{n=1}^{\infty} i \beta_n u_n(z) X_n(x) = T_1 u(z,x) + T_2 u(z,x),
\end{equation}
where
\begin{align}
  T_1 u(z,x) &=  i \sum_{n=1}^{M} \sqrt{k^2 - \mu_n}\, u_n(z) X_n(x) ,  \label{eq:DtNop_pr}\\
  T_2 u(z,x) &=  -\sum_{n=M+1}^{\infty} \sqrt{\mu_n - k^2}\, u_n(z) X_n(x),       \label{eq:DtNop_ev}
\end{align} 
and
\begin{equation}      \label{eq:Fourcoef}
     u_n(z) := \int_0^D  u(z,x)\, X_n(x) \, dx,
\end{equation}
are the Fourier coefficients of $u$ with respect to the orthonormal basis $\{X_n\}_{n=1,2,\ldots}$.
\begin{remark}~\\[-5pt]
\begin{enumerate}
\item On the artificial boundary ${\mathcal A}$ we may define the following
norms of fractional order:
$$
  \|u\|_{X^s({\mathcal A})} := 
  \left(\sum_{n=1}^{\infty} (\mu_n)^s |u_n (z_a)|^2\right)^{1/2} < \infty,
$$
The spaces $X^s({\mathcal A})$,  $s\geq 0$,
are then defined as the domain of $(-d^2/dy^2)^{s/2}$, while the space
of negative order $X^{-s}({\mathcal A})$ may be identified with the dual of
$X^s({\mathcal A})$. The notation is adopted from \cite{BG_99}.

The function space $X^s({\mathcal A})$ coincides with 
$H^s({\mathcal A})$ for $0<s< 1/2$. For $s=1/2$,
$X^{1/2}({\mathcal A})$ may be identified with $H_{00}^{1/2}({\mathcal A})$, the subspace of
functions of $H^{1/2}({\mathcal A})$ which when extended by zero belong to 
$H^{1/2}(\partial\Omega_{\mathcal{A}})$.
For $1/2<s\leq 1$, $X^{s}({\mathcal A}) = \,\Hnods({\mathcal A})$, (see \cite{LM_1972,BG_99}).
Then $T$ is a bounded linear operator from $X^{1/2}({\mathcal A})$ to $X^{-1/2}({\mathcal A})$. 
\item It is easy to show the following properties of the DtN operator. First,
         \begin{equation}     \label{eq:DtN_prop1}
             \int_{\mathcal{A}} Tu \, v = \int_{\mathcal{A}} Tv \, u,
         \end{equation} 
         and second,  letting 
         \begin{equation}      \label{eq:DtNopstar}
             \begin{array}{ll}
             & T^\ast u(z,x) = T_1^\ast u(z,x) + T_2 u(z,x), \\
             \mbox{\quad where \quad} &
             T_1^\ast u(z,x) = - i \sum_{n=1}^{\infty} \sqrt{k^2 - \mu_n}\, u_n(z) X_n(x),
             \end{array}
         \end{equation}
         it holds that
         \begin{equation}     \label{eq:DtN_prop2}
             \overline{Tu} = T^\ast\overline{u}.
         \end{equation}          
\end{enumerate}
\end{remark}

Now let $\Gh(\cdot,\xb_i)$ denote the Green's function for the Helmholtz operator with Dirichlet
conditions on the boundary $\partial\Omega$  due to a point source located at  $\xb_i = (z_i,x_i) \in \Omega_{L_+}$
for a fixed single frequency. (Here we consider a single frequency so when we refer to the Green's function we omit 
writing dependence on frequency.) Thus $\Gh(\cdot,\xb_i)$ solves the problem
\begin{align}   
    &-\Delta \Gh(\cdot,\xb_i) - k^2 \eta(\cdot)\, \Gh(\cdot,\xb_i)  = \delta(\cdot - \xb_i) \text{ in }\Omega_{\mathcal{A}}, 
    \label{eq:Greensbvp1} \\
    &\Gh(\cdot,\xb_i) = 0  \text{ on } \partial\Omega_{\mathcal{A}} \setminus \mathcal{A},   \label{eq:Greensbvp2} \\
    &\partial_\nu\Gh(\cdot,\xb_i)  = T\Gh(\cdot,\xb_i)  \text{ on } \mathcal{A} ,                    \label{eq:Greensbvp3}
\end{align} 
where $\nu$ is the outward unit normal on ${\mathcal A}$ and the last boundary condition, which is imposed
on the artificial boundary ${\mathcal A}$, accounts for the radiation condition.

In the following proposition we prove a reciprocity  relation for the Green's function.
%
%
\begin{proposition}   \label{pr:reciprocity}
For any $\xb_1,\xb_2\in\Omega_{\mathcal A}$ it holds that
\begin{equation}      \label{eq:Greens_recipr}
   \Gh(\xb_1,\xb_2) = \Gh(\xb_2,\xb_1).
\end{equation}
\end{proposition}
{\it Proof}. Let $\xb_i\in\Omega_{\mathcal A}$, $i = 1,2$. Since $\Gh(\cdot,\xb_i)$ satisfies \cref{eq:Greensbvp1}
we have for every $\yb = (z , x) \in\Omega_{\mathcal A}$ that
\begin{eqnarray*}
  & \Delta \Gh(\yb,\xb_2) + k^2 \eta(\yb) \,\Gh(\yb,\xb_2)  = -\delta(\yb - \xb_2),\\
  & \Delta \Gh(\yb,\xb_1) + k^2 \eta(\yb) \,\Gh(\yb,\xb_1)  = -\delta(\yb - \xb_1).
\end{eqnarray*}
We multiply the first equation by $\Gh(\yb,\xb_1)$, the second by $\Gh(\yb,\xb_2)$, subtract and 
integrate the resulting equation over $\Omega_{\mathcal{A}}$ to obtain that
\begin{align*}
   &\int_{\Omega_{\mathcal{A}}} \Big(\Delta \Gh(\yb,\xb_2) \,\Gh(\yb,\xb_1)
                           -  \Gh(\yb,\xb_2) \,\Delta\Gh(\yb,\xb_1)\Big)  \\
   &=  \int_{\Omega_{\mathcal{A}}} \Big(\delta(\yb - \xb_1)\, \Gh(\yb,\xb_2) - \delta(\yb - \xb_2) \,\Gh(\yb,\xb_1)\Big) \\
   &\Rightarrow  \int_{\Omega_{\mathcal{A}}} \Big(\Delta \Gh(\yb,\xb_2) \Gh(\yb,\xb_1)
                           -  \Gh(\yb,\xb_2) \Delta \Gh(\yb,\xb_1)\Big)  
   = \Gh(\xb_1,\xb_2) - \Gh(\xb_2,\xb_1).
\end{align*}
Using the second Green's identity, and the Dirichlet boundary conditions \cref{eq:Greensbvp2}, the equation above may be written as
\begin{align*}
     &\Gh(\xb_1,\xb_2) - \Gh(\xb_2,\xb_1) 
     = \int_{\mathcal{A}} \Big(\frac{\partial\Gh}{\partial\nu}(\yb,\xb_2) \Gh(\yb,\xb_1)
                           -  \Gh(\yb,\xb_2) \frac{\partial\Gh}{\partial\nu}(\yb,\xb_1)\Big) \\               
    &\stackrel{(\ref{eq:Greensbvp3})}{=} 
    \int_{\mathcal{A}} 
                 \Big(T\Gh((z_a,x),\xb_2) \, \Gh((z_a,x),\xb_1)  -  \Gh((z_a,x),\xb_2)\, T\Gh((z_a,x),\xb_1) \Big)\, dx
    \stackrel{(\ref{eq:DtN_prop1})}{=} 0   .
\end{align*}                               
Hence $\Gh(\xb_1,\xb_2) - \Gh(\xb_2,\xb_1)  = 0$. \hfill $\Box$

Now we are in a position to prove the following Kirchhoff-Helmholtz identity.
%
%
\begin{proposition}[Kirchhoff-Helmholtz identity]   \label{pr:KHidentity}
Let $\xb_1,\xb_2\in\Omega_{\mathcal A}$. Then 
\begin{equation}      \label{eq:KHid1}
        \Gh(\xb_1,\xb_2) - \ov{\Gh(\xb_1,\xb_2)} = 
        \int_{\mathcal{A}} \Big(\ov{\Gh(\yb,\xb_1)} \nabla\Gh(\yb,\xb_2) 
                           -  \Gh(\yb,\xb_2) \ov{\nabla\Gh(\yb,\xb_1)}~\Big)\cdot \bi{\nu} \, dx.
\end{equation}
Moreover,
\begin{equation}      \label{eq:KHid2}
       \Gh(\xb_1,\xb_2) - \ov{\Gh(\xb_1,\xb_2)} = 
      2i \sum_{n=1}^{M} \beta_n\, \ov{\Gh_n(z_a,\xb_1)}\,  \Gh_n(z_a,\xb_2),
\end{equation}
where the Fourier coefficients $\Gh_n(z_a,\cdot)$ are defined in \cref{eq:GRn}.
\end{proposition}
{\it Proof}. Since $\Gh(\cdot,\xb_1)$ solves \cref{eq:Greensbvp1}--\cref{eq:Greensbvp3} it is 
immediate to show that $\ov{\Gh(\cdot,\xb_1)}$ solves the problem
\begin{align}
    &-\Delta \ov{\Gh(\cdot,\xb_1)} - k^2 \eta(\cdot) \, \ov{\Gh(\cdot,\xb_1)}  = \delta(\cdot - \xb_1) \mbox{ in }\Omega_{\mathcal A}, 
    \label{eq:Gbvpc1}\\
    &\ov{\Gh(\cdot,\xb_1)} = 0  \mbox{ on } \partial\Omega_{\mathcal A} \setminus{\mathcal A},    \label{eq:Gbvpc2}\\
    &\partial_\nu\ov{\Gh(\cdot,\xb_1)}  = T^\ast\ov{\Gh(\cdot,\xb_1)}  \mbox{ on } \mathcal{A} .      \label{eq:Gbvpc3}
\end{align} 
Hence, for every $\yb = (z,x)\in\Omega_{\mathcal A}$ we have that
\begin{align*}
  & \Delta \Gh(\yb,\xb_2) + k^2\eta(\yb) \,\Gh(\yb,\xb_2)  = -\delta(\yb - \xb_2),\\
  & \Delta \ov{\Gh(\yb,\xb_1)} + k^2\eta(\yb) \,\ov{\Gh(\yb,\xb_1)}  = -\delta(\yb - \xb_1),
\end{align*}
Now, we multiply the first by $\ov{\Gh(\yb,\xb_1)}$, the second by $\Gh(\yb,\xb_2)$, subtract, and integrate
over $\Omega_{\mathcal{A}}$ to obtain that:
\begin{align*}
   &\int_{\Omega_{\mathcal{A}}} \Big(\Delta \Gh(\yb,\xb_2) \ov{\Gh(\yb,\xb_1)} 
                           -  \Gh(\yb,\xb_2) \Delta\ov{\Gh(\yb,\xb_1)}\Big)  \\
   & = \int_{\Omega_{\mathcal{A}}} \Big(\delta(\yb - \xb_1) \Gh(\yb,\xb_2) - \delta(\yb - \xb_2) \ov{\Gh(\yb,\xb_1)}\Big) 
    = \Gh(\xb_1,\xb_2) - \ov{\Gh(\xb_2,\xb_1)}.
\end{align*}
Then, from the reciprocity property \cref{eq:Greens_recipr} we get that
\begin{eqnarray*}
     \int_{\Omega_{\mathcal{A}}} \Big(\Delta \Gh(\yb,\xb_2) \ov{\Gh(\yb,\xb_1)} 
                           -  \Gh(\yb,\xb_2) \Delta\ov{\Gh(\yb,\xb_1)}\Big)  = 
     \Gh(\xb_1,\xb_2) - \ov{\Gh(\xb_1,\xb_2)}.
\end{eqnarray*}
Now, \cref{eq:KHid1} results using the second Green's identity and the boundary conditions  
\cref{eq:Greensbvp2},   \cref{eq:Gbvpc2}.

Since we have proven \cref{eq:KHid1}, namely
$$
    \Gh(\xb_1,\xb_2) - \ov{\Gh(\xb_2,\xb_1)} = 
    \int_{\mathcal{A}} \Big(\ov{\Gh(\yb,\xb_1)} \partial_\nu\Gh(\yb,\xb_2) 
                           -  \Gh(\yb,\xb_2) \partial_\nu\ov{\Gh(\yb,\xb_1)}~\Big) \, dx,
$$
the DtN conditions \cref{eq:Greensbvp3} and \cref{eq:Gbvpc3} allow us to write
\begin{align}
     &\Gh(\xb_1,\xb_2) - \ov{\Gh(\xb_1,\xb_2)} \nonumber\\
     &= \int_{\mathcal{A}} 
         \Big( \ov{\Gh((z_a,x),\xb_1)} T\Gh((z_a,x),\xb_2)
                    -  \Gh((z_a,x),\xb_2) T^\ast\ov{\Gh((z_a,x),\xb_1)}\Big)\,dx  \nonumber\\
   &\stackrel{\cref{eq:DtN_prop1}}{=} 
           \int_{\mathcal{A}}                  
                 \Big(T\ov{\Gh((z_a,x),\xb_1)}  -  T^\ast\ov{\Gh((z_a,x),\xb_1)}\Big)
                 \Gh((z_a,x),\xb_2)\,dx .        \label{eq:HK1_termwg} 
\end{align}                               
Therefore, in view of \cref{eq:DtNop_pr} and \cref{eq:DtNopstar}, we deduce that
\begin{align*}
   T\ov{\Gh((z_a,x),\xb_1)}  -  T^\ast\ov{\Gh((z_a,x),\xb_1)} 
   = 2i \sum_{n=1}^{M} \beta_n\, \ov{\Gh_n(z_a,\xb_1)}\, X_n(x).
\end{align*}
Inserting the above in \cref{eq:HK1_termwg} we conclude that
\begin{align*}
     \Gh(\xb_1,\xb_2) - \ov{\Gh(\xb_1,\xb_2)} &= 
     2i \int_0^D \sum_{n=1}^{M} \beta_n\, \ov{\Gh_n(z_a,\xb_1)}\, X_n(x)\,  \Gh(z_a,x,\xb_2)\,dx\\
     &= 
     2i \sum_{n=1}^{M} \beta_n\, \ov{\Gh_n(z_a,\xb_1)}\,  \int_0^D\Gh(z_a,x,\xb_2)\, X_n(x)\,dx,
\end{align*}
which completes the proof. \hfill $\Box$
%
%
%
%
%
%
\section{Derivation of $\GR$} \label{sec:GMI}
\begin{figure}[ht]
\centering
\includegraphics[width=0.84\textwidth]{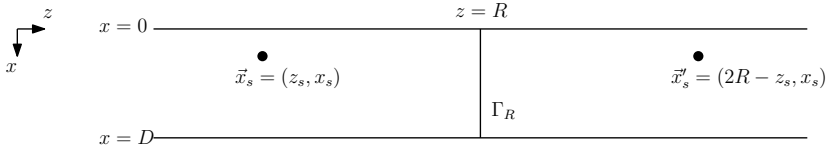} 
\caption{Two sources placed symmetrically with respect to $\Gamma_R$} 
\label{fig:WG_mirror}
\end{figure}

In this section we present the derivation of the Green's function $\GR$ for the Helmholtz operator when the 
terminating waveguide  $\Omega$ is a homogeneous ($\eta(\xb) = 1$) semi-infinite strip. Specifically, $\Omega = (-\infty,R)\times (0,D)$.
Then the Green's function $\GR(\cdot,\xb_s;\omega)$ due to a point source located at  $\xb_s = (z_s,x_s) \in \Omega$,
for a single frequency $\omega$,  solves the problem
\begin{align*}   
    -\Delta \Gh(\cdot,\xb_s;\omega) - k^2 \Gh(\cdot,\xb_s;\omega)  
    = \delta(\cdot - \xb_s) \quad&\mbox{ in }\Omega, \\ 
    \Gh(\cdot,\xb_s;\omega) = 0  &\mbox{ on } \partial\Omega, 
\end{align*}
In order to derive an analytic expression for $\GR(\cdot,\xb_s;\omega)$ we will use the method of images  
\cite{CH_89}. With reference to \cref{fig:WG_mirror}, we assume an infinite waveguide 
in the $z-$direction and we add a source at $\xb'_s$ that is symmetric to 
$\xb_s$ with respect to $\Gamma_{R}$, \textit{i.e.} $\xb'_s = (2R-z_s,x_s)$.
 
We then compute the field at a point $\yb = (z,x)\in\Omega$ as 
\begin{equation}    \label{eq:moimag}
   \GR(\yb,\xb_s;\omega) =\Go(\yb,\xb_s;\omega) - \Go(\yb,\xb'_s;\omega), 
\end{equation}
where $\Go(\yb,\xb_s;\omega)$ denotes the Green's function for an infinite waveguide.
A normal mode representation of $\Go(\cdot,\xb_s;\omega)$ reads, \cite{JKPS_04,PR_07},  
$$ 
   \Go(\xb,\xb_s;\omega)  = \frac{i}{2} \sum_{m=1}^{\infty}  \frac{\rme^{i\beta_m|z-z_s|}}{\beta_m}X_m(x)X_m(x_s),
$$ 
where $\{\mu_n,X_n\}_{n=1}^\infty$ are the eigenvalues and eigenvectors
of  $-d^2 / dx^2$ in $(0,D)$ given in \cref{eq:eigs},
and $\beta_n$ are the horizontal wavenumbers, see \cref{eq:betas}.

Then, \cref{eq:moimag} implies that  
\begin{align*} 
    \GR(\yb,\xb_s)  &  = \frac{i}{2} \sum_{m=1}^{\infty}
         \frac{\rme^{i\beta_m|z-z_s|}}{\beta_m}X_m(x)X_m(x_s) -\frac{i}{2} 
        \sum_{n=1}^{\infty}\frac{\rme^{i\beta_n|z-2R+z_s|}}{\beta_n}X_n(x)X_n(x_s) \\
    &= \frac{i}{2} \sum_{m=1}^{\infty}\frac{\rme^{i\beta_m|z-z_s|}-\rme^{i\beta_m |z+z_s-2R|}}{\beta_m}X_m(x)X_m(x_s),
\end{align*}
where $z< R$ and $0\le x\le D$.
Since $z,z_s < R$ it turns out that $z + z_s - 2R < 0$, hence 
\begin{align}     \label{eq:green2}
   \GR(\yb,\xb_s) 
    &= \frac{i}{2} \sum_{m=1}^{\infty} \frac{\rme^{i\beta_m|z-z_s|}-\rme^{-i\beta_m (z+z_s-2R)}}{\beta_m}X_m(x)X_m(x_s) \nonumber\\
    &= \left\{ \begin{array}{lc}
                  \displaystyle  \sum_{m=1}^\infty \frac{i}{2\beta_m}
                  \left( \rme^{i\beta_m(z-z_s)} - \rme^{-i\beta_m (z+z_s-2R)} \right)
                  X_m(x) X_m(x_s), &  z > z_s, \\
                  \displaystyle \sum_{m=1}^\infty\frac{i}{2\beta_m}\left( \rme^{-i\beta_m(z-z_s)} - \rme^{-i\beta_m (z+z_s-2R)}  \right) 
                  X_m(x)X_m(x_s), & z < z_s.   
    \end{array} \right.                                
\end{align}

Notice that 
\begin{align*}
 \rme^{i\beta_m(z - z_s)} - \rme^{-i\beta_m (z + z_s - 2R)}  &= 
  \rme^{i\beta_m(R - z_s)} \Bigl( \rme^{i\beta_m(z - R)} - \rme^{-i\beta_m (z - R)} \Bigr) \\
     &= -2i \,\rme^{i\beta_m(R - z_s)} \, \sin\beta_m(R - z) ,
\end{align*}
and, similarly,
\begin{align*}
\rme^{-i\beta_m(z - z_s)} - \rme^{-i\beta_m (z + z_s - 2R)}  =  -2i \,\rme^{i\beta_m(R - z)} \, \sin\beta_m(R - z_s).
\end{align*}

Therefore, (\ref{eq:green2}) may also be written as
\begin{align}    \label{eq:green3}
\GR(\yb,\xb_s) 
   = \left\{ \begin{array}{lc}
                  \displaystyle  \sum_{m=1}^\infty \frac{1}{\beta_m}\,
                  \rme^{i\beta_m (R - z_s)}  \, \sin\beta_m(R - z) \, X_m(x) \, X_m(x_s), &  z > z_s, \\
                  \displaystyle  \sum_{m=1}^\infty \frac{1}{\beta_m}\,
                  \rme^{i\beta_m (R - z)}  \, \sin\beta_m(R - z_s) \, X_m(x) \, X_m(x_s), & z < z_s .  
    \end{array} \right.                       
\end{align}
%
%
%
%
%
%
\clearpage
\bibliographystyle{siamplain}
\bibliography{biblio}
\end{document}